\documentclass{article}
\def\baseDir{./}
\def\cmakeRelDir{cmake/tests/release/}
\def\cmakeDir{\baseDir\cmakeRelDir}
\def\pythonRelDir{src/python_scripts/cgalpy_examples/}
\def\pythonDir{\baseDir\pythonRelDir}
\usepackage[T1]{fontenc}
\usepackage{lmodern}
\usepackage{slantsc}
\usepackage{amssymb}
\usepackage{amsfonts}
\usepackage[american]{babel}
\usepackage[latin1]{inputenc}
\usepackage{dsfont}
\usepackage{pifont}
\usepackage{placeins}
\usepackage{appendix}
\usepackage{hyphenat}
\usepackage{multirow}
\usepackage{multicol}
\usepackage{tabularx}
\usepackage{alltt}
\usepackage{listings}
\usepackage{mdwlist}
\usepackage{paralist}
\usepackage[caption=false,font=footnotesize]{subfig}

\usepackage[disable]{todonotes}
\usepackage{etex}
\usepackage{etoolbox}
\usepackage{xparse}
\usepackage{tikz}
\usepackage{forest}
\usetikzlibrary{calc,patterns,decorations.pathmorphing,decorations.markings,matrix,fit,decorations.pathreplacing,arrows,arrows.meta,automata,positioning,shapes,chains,spy}
\usetikzlibrary{intersections,through,backgrounds}
\usetikzlibrary{shadows,fadings,shadows.blur}
\usepackage{tikz-3dplot}
\tikzset{
  dots/.style args={#1per #2}{%
    line cap=round,
    dash pattern=on 0 off #2/#1
  },
  invisible/.style={opacity=0},
  only/.code args={<#1>#2}{\only<#1>{\pgfkeysalso{#2}}},
  alt/.code args={<#1>#2#3}{\alt<#1>{\pgfkeysalso{#2}}{\pgfkeysalso{#3}}},
  temporal/.code args={<#1>#2#3#4}{%
    \temporal<#1>{\pgfkeysalso{#2}}{\pgfkeysalso{#3}}{\pgfkeysalso{#4}}},
  point/.style={circle,inner sep=1.5pt,minimum size=1.5pt,draw,fill=#1},
  point/.default=red,
  big arrow/.style={
    decoration={markings,mark=at position 1 with {\arrow[scale=1.5,#1]{>}}},
    postaction={decorate},
    shorten >=0.4pt},
  big arrow/.default=black}
\tikzset{cross/.style={%
    cross out, draw=black, minimum size=2*(#1-\pgflinewidth), inner sep=0pt,%
    outer sep=0pt},%
  cross/.default={1pt}}
\usepackage{xspace}
\usepackage{wrapfig}
\usepackage[outercaption]{sidecap}
\usepackage{authblk}
\usepackage{amsthm}

\usepackage[pdfborder={0 0 0}]{hyperref} % last \usepackage
\def\calA{{\cal A}}
\def\calB{{\cal B}}
\def\calC{{\cal C}}

\def\calP{{\cal P}}

\def\calT{{\cal T}}

\newcommand{\R}{\ensuremath{\mathbb{R}}}               % R reelle Zahlen
\newcommand{\Rx}[1]{\ensuremath{\R \rule{0.3mm}{0mm}^{#1}}}

\newcommand{\Rtwo}{\Rx{2}\xspace}
\newcommand{\Rthree}{\Rx{3}\xspace}
\newcommand{\Rd}{\Rx{d}\xspace}

\newcommand{\cgalPackage}[1]{{\emph{#1}\index{CGAL package@\cgal{} package!#1@\emph{#1}}}}

\newcommand{\iiDandiiiDGeometryKernelPackage}{\cgalPackage{2D and 3D Geometry Kernel}}
\newcommand{\dDGeometryKernelPackage}{\cgalPackage{dD Geometry Kernel}}

\newcommand{\iiDConvexHullsandExtremePointsPackage}{\cgalPackage{2D Convex Hulls and Extreme Points}}
\newcommand{\iiiDConvexHullsPackage}{\cgalPackage{3D Convex Hulls}}

\newcommand{\iiDPolygonsPackage}{\cgalPackage{2D Polygons}}
\newcommand{\iiDRegularizedBooleanSetOperationsPackage}{\cgalPackage{2D Regularized Boolean Set-Operations}}

\newcommand{\iiDPolygonPartitioningPackage}{\cgalPackage{2D Polygon Partitioning}}

\newcommand{\iiDMinkowskiSumsPackage}{\cgalPackage{2D Minkowski Sums}}
\newcommand{\iiiDPolyhedralSurfacesPackage}{\cgalPackage{3D Polyhedral Surfaces}}

\newcommand{\iiDArrangementsPackage}{\cgalPackage{2D Arrangements}}

\newcommand{\iiDTriangulationsPackage}{\cgalPackage{2D Triangulations}}

\newcommand{\iiiDTriangulationsPackage}{\cgalPackage{3D Triangulations}}

\newcommand{\iiDPeriodicTriangulationsPackage}{\cgalPackage{2D Periodic Triangulations}}
\newcommand{\iiiDPeriodicTriangulationsPackage}{\cgalPackage{3D Periodic Triangulations}}
\newcommand{\iiDAlphaShapesPackage}{\cgalPackage{2D Alpha Shapes}}
\newcommand{\iiiDAlphaShapesPackage}{\cgalPackage{3D Alpha Shapes}}

\newcommand{\surfaceMeshPackage}{\cgalPackage{Surface Mesh}}

\newcommand{\PolygonMeshProcessing}{\cgalPackage{Polygon Mesh Processing}}

\newcommand{\dDSpatialSearchingPackage}{\cgalPackage{dD Spatial Searching}}

\newcommand{\HandlesandCirculatorsPackage}{\cgalPackage{Handles and Circulators}}

\newcommand{\ignore}[1]{}
\newcommand{\cgal}{{\sc Cgal}}
\newcommand{\cmake}{{\sc CMake}}

\newcommand{\disco}{{\sc DiscoPygal}}
\newcommand{\kdtree}{$k$D-tree}

\usepackage{relsize}

\def\ifmonospace{\ifdim\fontdimen3\font=0pt }

\def\C++{%
\ifmonospace%
    C++%
\else%
    C\kern-.1667em\raise.30ex\hbox{\smaller{++}}%
\fi%
\spacefactor1000 }

\def\Csharp{%
\ifmonospace%
    C\#%
\else%
    C\kern-.1667em\raise.30ex\hbox{\smaller{\#}}%
\fi%
\spacefactor1000 }

\definecolor{polyALightColor}{rgb}{0.7,0.8,1}
\definecolor{polyBLightColor}{rgb}{1,0.85,0.6}
\definecolor{polyCLightColor}{rgb}{0.85,1,0.3}
\definecolor{polyDLightColor}{rgb}{1,1,0.75}

\definecolor{polyAColor}{rgb}{0.5,0.7,1}
\definecolor{polyBColor}{rgb}{1,0.72,0.225}
\definecolor{polyCColor}{rgb}{0.8,0.9,0.2}
\definecolor{polyDColor}{rgb}{0.8,0.8,0.6}

\definecolor{polyADarkColor}{rgb}{0.2,0.4,0.67}
\definecolor{polyBDarkColor}{rgb}{0.67,0.4,0.12}
\definecolor{polyCDarkColor}{rgb}{0.53,0.6,0.13}
\definecolor{polyDDarkColor}{rgb}{0.53,0.53,0.4}

\definecolor{polyABDarkColor}{rgb}{0.5,0.35,0.4}

\definecolor{crossColor}{rgb}{1,0,0.5}

\newlength{\ccBaseWidth}
\lstdefinestyle{cppStyle}{%
  language=[11]{C++},
  identifierstyle=\color{red!50!black},
  escapechar=~~,
  moredelim=**[is][\only<+>{\color{red}}]{@}{@}}
\lstdefinestyle{cppNumberedStyle}{%
  style=cppStyle,
  numbers=left,
  numberstyle=\tiny\color{gray},
  numbersep=5pt}
\lstdefinestyle{cppFramedStyle}{%
  style=cppNumberedStyle,
  frame=single,
  frameround=tttt}
\lstdefinestyle{pythonStyle}{%
  language={Python},
  identifierstyle=\color{red!50!black},
  escapechar=~~}
\lstdefinestyle{pythonNumberedStyle}{%
  style=pythonStyle,
  numbers=left,
  numberstyle=\tiny\color{gray},
  numbersep=5pt}
\lstdefinestyle{pythonFramedStyle}{%
  style=pythonNumberedStyle,
  frame=single,
  frameround=tttt}
\NewDocumentCommand\framedInputPython{O{\scriptsize}O{1}om}{%
  \IfNoValueTF{#3}{
    \lstinputlisting[%
      escapechar=~~,
      numbers=left,
      numberstyle=\tiny\color{gray},
      numbersep=5pt,
      frame=single,
      frameround=tttt,
      language=Python,
      basicstyle=#1,
      firstline=#2,
      identifierstyle=\color{red!50!black}]{#4}}{%
    \lstinputlisting[%
      escapechar=~~,
      numbers=left,
      numberstyle=\tiny\color{gray},
      numbersep=5pt,
      frame=single,
      frameround=tttt,
      language=Python,
      basicstyle=#1,
      firstline=#2,
      lastline=#3,
      identifierstyle=\color{red!50!black}]{#4}}}
\lstdefinelanguage{Cmake}{
    morekeywords={},
    sensitive=false, % keywords are not case-sensitive
    morecomment=[l]{\#}, % l is for line comment
    morestring=[b]" % defines that strings are enclosed in double quotes
}
\lstdefinestyle{cmakeStyle}{%
  language={Cmake},
  identifierstyle=\color{red!50!black},
  escapechar=~~,
}
\lstdefinestyle{cmakeNumberedStyle}{%
  style=cmakeStyle,
  numbers=left,
  numberstyle=\tiny\color{gray},
  numbersep=5pt}
\lstdefinestyle{cmakeFramedStyle}{%
  style=cmakeNumberedStyle,
  frame=single,
  frameround=tttt}
\NewDocumentCommand\framedInputCmake{O{\scriptsize}O{1}om}{%
  \IfNoValueTF{#3}{
    \lstinputlisting[%
      escapechar=~~,
      numbers=left,
      numberstyle=\tiny\color{gray},
      numbersep=5pt,
      frame=single,
      frameround=tttt,
      language=Cmake,
      basicstyle=#1,
      firstline=#2,
      identifierstyle=\color{red!50!black}]{#4}}{%
    \lstinputlisting[%
      escapechar=~~,
      numbers=left,
      numberstyle=\tiny\color{gray},
      numbersep=5pt,
      frame=single,
      frameround=tttt,
      language=Python,
      basicstyle=#1,
      firstline=#2,
      lastline=#3,
      identifierstyle=\color{red!50!black}]{#4}}}
\def\myLstinline#1{\lstinline[columns=fixed]{#1}}
\def\bashLstinline#1{\lstinline[language={bash},columns=fixed]{#1}}
\def\pyLstinline#1{\lstinline[language={Python},columns=fixed]{#1}}
\def\cppLstinline#1{\lstinline[language={C++},columns=fixed]{#1}}
\def\cmakeLstinline#1{\lstinline[language={Cmake},columns=fixed]{#1}}

\usepackage{newfloat}
\DeclareFloatingEnvironment[
  fileext = lop,
  listname = {!htp},
  name = Listing
]{lstfloat}

\newsavebox{\draftbox}
\savebox{\draftbox}{\color{red}\huge $\times$}% create image for overlay

\newenvironment{draft}{\begin{lrbox}{0}\minipage[b]{\linewidth}}%
  {\endminipage\end{lrbox}%
  \leavevmode\rlap{\resizebox{\wd0}{\ht0}{\usebox\draftbox}}\usebox0}

\usepackage[margin=0.7in]{geometry}
\usepackage{algorithm}
\usepackage{algpseudocode}
\usepackage[export]{adjustbox}
\usepackage{algorithm}
\usepackage{algpseudocode}
\usepackage{amsthm}
\usepackage{wrapfig}
\usepackage{lipsum}
\usepackage{tikz,graphicx}
\usepackage[framemethod=tikz]{mdframed}

\newtheorem*{theorem*}{Theorem}

\definecolor{eficolor}{HTML}{EE4035}
\definecolor{nircolor}{HTML}{F37736}
\definecolor{dancolor}{HTML}{7BC043}
\definecolor{todocolor}{HTML}{0392CF}
\graphicspath{{figs/}}

\title{\cgal{} Made More Accessible}
\date{}
\author{Nir Goren, Efi Fogel, and Dan Halperin}
\begin{document}
\maketitle
% Tables
\tikzset{table/.style={
  matrix of nodes,
  row sep=-\pgflinewidth,
  column sep=-\pgflinewidth,
  nodes={rectangle,draw=black,align=left,anchor=center},
  nodes in empty cells,
  every even row/.style={nodes={fill=gray!20}},
  column 1/.style={nodes={text width=19.8em}},
  column 2/.style={nodes={text width=4.8em}},
  column 3/.style={nodes={text width=3.8em}},
  column 4/.style={nodes={text width=19.2em}},
  row 1/.style={nodes={minimum height=1.6em,anchor=center,align=center,fill=darkgray,text=white,font=\bfseries}}}}
\tikzset{tableBindingSelection/.style={table,
  nodes={minimum height=2.8em},
  column 1/.style={nodes={text width=20em}},
  column 2/.style={nodes={text width=4.6em}},
  column 3/.style={nodes={text width=3.8em}},
  column 4/.style={nodes={text width=19.2em}}}}
\tikzset{tableModuleFlags/.style={table,
  nodes={minimum height=1.6em},
  column 1/.style={nodes={text width=18.0em}},
  column 2/.style={nodes={text width=4.6em}},
  column 3/.style={nodes={text width=6.7em}},
  column 4/.style={nodes={text width=18.6em}}}}
\tikzset{tableModuleName/.style={table,
  nodes={minimum height=1.6em},
  column 1/.style={nodes={text width=20em}},
  column 2/.style={nodes={text width=16em}},
  column 3/.style={nodes={text width=7em}}}}
\tikzset{tableName/.style={table,nodes={minimum height=1.6em}}}

\begin{abstract}
We introduce Python bindings that enable the convenient, efficient, and reliable use of software modules of \cgal{} (The Computational Geometry Algorithms Library), which are written in C++, from within code written in Python. There are different tools that facilitate the creation of such bindings. We present a short study that compares three main tools, which leads to the tool of choice. 
The implementation of algorithms and data structures in computational geometry presents tremendous difficulties, such as obtaining robust software despite the use of (inexact) floating point arithmetic, found in standard hardware, and meticulous handling of all degenerate cases, which typically are in abundance. The code of \cgal{} extensively uses function and class templates in order to handle these difficulties, which implies that the programmer has to make many choices that are resolved during compile time (of the C++ modules). While bindings take effect at run time (of the Python code), the type of the C++ objects that are bound must be known when the bindings are generated, that is, when they are compiled. The types of the bound objects are instances (instantiated types) of C++ function and class templates. The number of object types that can potentially be bound, in implementation of generic computational-geometry algorithms, is enormous; thus, the generation of the bindings for all these types in advance is practically impossible. For example, the programmer needs to choose among a dozen types of curves (e.g., line segments, circular arcs, geodesic arcs on a sphere, or polycurves of any curve type) to yield a desired arrangement type; often there are several choices to make, resulting in a prohibitively large number of combinations. We present a system that rapidly generates bindings for desired object types according to user prescriptions, which enables the convenient use of any subset of bound object types concurrently. After many years, in which the usage of these packages was restricted to C++ experts, the introduction of the bindings made them easily accessible to newcomers and practitioners in non-computing fields, as we report in the paper.
Additional information can be found at \url{http://acg.cs.tau.ac.il/cgal-python-bindings}. The bindings software can be found at \url{https://bitbucket.org/taucgl/cgal-python-bindings}.
\end{abstract}

% =============================================================================
\section{Introduction}
\label{sec:intro}
% =============================================================================
\cgal{} (The Computational Geometry Algorithms Library) is an open source software project that provides access to efficient and reliable implementations of geometric algorithms in the form of a C++ library~\cite{cgal:eb-22}. \cgal{} has evolved through the years and now represents the state of the art in computational geometry software. \cgal{} is used in a diverse range of domains requiring geometric computation, such as computer graphics, scientific visualization, computer aided design and modeling, geographic information systems, molecular biology, and medical imaging. \cgal{} provides a large number of components that cover a wide range of areas. The bindings for most components are supported now. The principles for the bindings described in this paper applies to all packages of \cgal{} (and perhaps to other generic C++ libraries as well), and can be easily adopted to support the bindings of the few components that are still missing.

There is a large community of users of \cgal{} packages in academia and the industry, and these packages seem to be helpful in many diverse projects. In all the projects that we are aware of, the packages have been used by C++ experts or by users who have been strongly supported by such experts from the \cgal{} developer community, including GeometryFactory.\footnote{GeometryFactory is the company established to market and support the usage of \cgal{} packages.} In our experience, incorporating \cgal{} in education or in academic projects lacking C++ experts has been rugged. We attribute the difficulties that arise in such settings to the highly sophisticated usage of C++ in \cgal. The binding project that we describe in this paper is meant to mitigate these difficulties and make all \cgal{} packages more accessible to a wider audience of programmers. We already see, and report below on, first signs that the project is on a good trajectory toward achieving this goal.

The software modules of \cgal{} rigorously adhere to the generic programming paradigm---a discipline that consists of the gradual lifting of concrete algorithms abstracting over details, while retaining the algorithm semantics and efficiency. The software modules of \cgal{} also follow the exact geometric-computation paradigm, which simply amounts to ensuring that errors in predicate evaluations do not occur; it guarantees robustness of the applied algorithms.  As a result, the software of \cgal{} is robust despite the use of (inexact) floating point arithmetic, found in standard hardware, it is complete as it handles all degenerate cases, which typically are in abundance, and it is efficient---all at the same time.

Generic programming identifies an abstraction that consists of a formal hierarchy of polymorphic abstract (syntactic and semantic) requirements on data types, referred to as \emph{concepts}, and a set of classes that conform precisely to the specified requirements, referred to as \emph{models}~\cite{a-gps-99}. A hierarchy of related concepts can be viewed as a directed acyclic graph, where a node of the graph represents a concept and an arc represents a refinement relation. An arc directed from concept $A$ to concept $B$ indicates that concept $B$ refines concept $A$. A model of concept $B$ is inherently also a model of concept $A$. When a class template is instantiated, each one of its template parameters is substituted with a model of one or more concepts associated with the template parameter.

% -----------------------------------------------------------------------------
\subsection{Python Bindings for C++ Modules}
\label{ssec:intro:bind}
% -----------------------------------------------------------------------------
Python, like many other languages, was originally defined by a reference implementation. The original reference implementation of the Python programming language is CPython. To date, CPython is still the default and commonly used reference implementation of the language, although newcomers are emerging, see, e.g., PyPy.\footnote{PyPy is a replacement for CPython; see \url{https://www.pypy.org}.} As the Python language evolved, its definition has tightened and a well defined specification has materialized.

Bindings are essentially wrapper libraries that bridge two programming languages, so that a software module developed in one language can be used in code written in another language, exploiting (i) the strengths of both languages, and (ii) the availability of modules developed in the former language. C++ Python bindings enable (i) the invocation of C++ functions, and (ii)~the access to C++ variables from Python code, taking advantage of Python's quick development cycles and C++'s high performance.

Python is sufficiently efficient for many tasks; nevertheless low-level code written in Python tends to be too slow, largely because Python is dynamically typed. In particular, low-level computational loops are simply infeasible. In addition, the magnitude of existing and well-tested code in Fortran, C, or C++, is enormous. The Python/C API, the module that enables the usage of external modules developed in Fortran, C, or C++, from Python code, or embedding Python code in code written in other languages, is very rich.  Indeed, the Python/C API module of CPython exposes a vast amount of details of CPython internals. It enables the efficient exploitation of existing code in Fortran, C, or C++, as well as the replacement of critical sections written in Python when speed is essential. Wrapping existing code has traditionally been the domain of Python experts due to the steep learning curve, which characterizes its Python/C API. Although using such wrappers is possible without ever knowing their internals, providing such wrappers creates a sharp line between developers using Python and developers using C or C++ with the Python/C API. Several tools that facilitate the creation of such wrappers have been introduced; A sample of such tools are listed in Section~\ref{sec:comp}.

When running Python code that uses bindings for some C++ modules, one or more libraries that provide the bindings must be accessible.  A software module in C++ that adheres to the generic programming paradigm consists of function and class templates; these templates are instantiated at compile time of the binding libraries. In other words, the types of the C++ objects that are bound, that is, instances (instantiated types) of C++ function and class templates, must be known when the bindings are generated.

% -----------------------------------------------------------------------------
\subsection{\cmake}
\label{ssec:intro:cmake}
% -----------------------------------------------------------------------------
\cmake{}~\cite{mh-mc-08} is an open-source, cross-platform suite of tools designed to build, test, and package software. The suite of \cmake{} tools were created to address the need for cross-platform build environments for open-source projects. However, since it was conceived over two decades ago, it has grown and become powerful, rich, and flexible. It is now considered a modern tool; it is widely spread and used by major commercial software products for their build and test environments.

\cmake{}, for the most part, is a cross-platform build system generator.  It is used to control the software build process using simple platform and compiler independent configuration files, and it generates native build environments of the user choice. Users build a project by using \cmake{} to generate a build system for native tools on their platforms. \cmake{} supports an interpreted, functional, scripting language. The build process of a project is specified with platform-independent \cmake{} text files, named \myLstinline{CMakeLists.txt}, written in the \cmake{} language, included in several directories of the source tree of the project. For example, once the \cmake{} configuration files are in place for a certain project, \cmake{} can be used to generate \myLstinline{Makefile} files required by the \myLstinline{make} utility of the GNU suite to compile a project written in C++ on a Unix system using the \myLstinline{g++} compiler. The command-line interface of the cross-platform build system generator of \cmake{} is an executable called \myLstinline{cmake}. (It has a counterpart called \myLstinline{cmake-gui}, which includes a graphical user interface.) In order to generate native project build files, the user invokes \myLstinline{cmake} in a terminal and specifies the directory that contains the root \myLstinline{CMakeLists.txt} file.

A key feature of \cmake{} is the ability to (optionally) place compiler outputs (such as object files) outside the source tree. This enables multiple and concurrent builds from the same source tree and cross-compilation. It also keeps the source tree clean and enables the removal of entire build directories without impairing the source files.  We exploit this feature, and the scripting capabilities to conveniently generate the C++ Python bindings for several \cgal{} objects concurrently.

% -----------------------------------------------------------------------------
\subsection{Bindings for \cgal{} and Other Geometric Software}
\label{ssec:intro:cgal}
% -----------------------------------------------------------------------------
Instantiated types in \cgal{} are characterized by long instantiation chains of C++ class templates. An instantiated type is a template that has one or more template parameters and every parameter is substituted by another type, which is typically an instance of another template. While the number of models of most concepts is small, the number of potential types of objects that could be bound, and thus must be supported by the bindings, is enormous. Offering bindings for all these types in advance is practically impossible. Moreover, in some cases models need to be extended with types provided by the user, which may not be available when the bindings are generated.

\cgal{} uses \cmake~\cite{mh-mc-08} for the build process.  Since version~5.0, \cgal{} is header-only by default, which means that an application that depends on \cgal{} only depends on (i) the source code of \cgal{} that resides in header files, (the names of which typically end with the suffix \myLstinline{.hpp}), and (ii) native configuration files generated by \cmake{} (in particular, the execution of \myLstinline{cmake}). In other words, there is no need to compile any library before an application that depends on \cgal{} and written in C++ is built.\footnote{Some dependencies of \cgal{}   need to be installed in advance.} On the other hand, when running an application written in Python that depends on \cgal{}, the libraries that contain the bindings must be accessible. Typically, those bindings are generated ahead of time with little knowledge about the application itself. This would impose a severe limitation in cases like ours, where the number of objects to be bound is large. In cases where the types of the objects to be bound are closely tied with the application itself, generating bindings ahead of time is impractical. Our system enables easy and convenient (re)generation of bindings, thus alleviating the burden caused by these problems.

The Shapely Python package supports planar geometric algorithms provided by GEOS~\cite{geos-manual}. The common functionality exposed by Shapely and our bindings is small. The scikit-geometry Python package supports geometric algorithms derived from \cgal.\footnote{See \url{https://github.com/scikit-geometry/scikit-geometry}.} The API of scikit-geometry is detached from \cgal{}. While the API is Pythonic in its nature, the package exposes only a small fraction of the rich API of \cgal{}. C++ Python bindings are also available for a certain subset of \cgal{} packages as part of an experimental project that uses SWIG.\footnote{See \url{http://www.swig.org/}.}
While the project has been conceived more than a decade ago, it contains bindings only for a small set of type and function instances of a limited subset of \cgal{} packages.

% -----------------------------------------------------------------------------
\subsection{Contribution}
\label{ssec:intro:results}
% -----------------------------------------------------------------------------
We introduce C++ Python bindings that enable the convenient, efficient, and reliable use of \cgal{} software modules in Python code. There are different tools that facilitate the creation of C++ Python bindings. We present a short study that compares several tools, which leads to the method of choice. The binding themselves are implemented in C++ and their implementation exploits advanced features in C++, which we explicate. This results in elegant and compact code that is easy to maintain and extend with additional bindings.
We present a system that rapidly generates bindings of desired objects according to user prescriptions, which enables the convenient use of any subset of bound object types concurrently.  With our system it is possible to generate a single library that contains bindings for instances of different \cgal{} templates, e.g., an instance of the 2D arrangement class template and an instance of the 2D triangulation class template, or several libraries (which can be used concurrently), such that distinct libraries contain different instances of the same template.

% -----------------------------------------------------------------------------
\subsection{Outline}
\label{ssec:intro:outline}
% -----------------------------------------------------------------------------
The rest of this paper is organized as follows. A study that compares several methods for implementing C++ Python bindings is presented in Section~\ref{sec:comp}.
General instructions for rapidly generating bindings with our system are presented in Section~\ref{sec:gen}.
The description of the bindings generated by our system for various modules are given in Section~\ref{sec:modules}.
A sample of highlighted modern C++ techniques and idioms used in the binding implementation are described in Section~\ref{sec:code}.
The tight coupling between concepts and binding generation is described in Section~\ref{sec:concepts}.
Applications of the new bindings, which are geared toward making the \cgal{} code more accessible are presented in Section~\ref{sec:app}.

% -----------------------------------------------------------------------------
\subsection{Conventions}
\label{ssec:intro:conventions}
% -----------------------------------------------------------------------------
The paper is packed with code samples written either in the \cmake{}, Python, or C++ languages. We use several conventions to improve readability.  Code excerpt written in C++ is always numbered, while short code excerpts written in Python is sometimes preceded with \pyLstinline{>>>}. The C++ identifier \cppLstinline{py} is an alias for the namespace \cppLstinline{nanobind}.

% =============================================================================
\section{Python Bindings Tool Comparison}
\label{sec:comp}
% =============================================================================
Several tools that facilitate binding generation have been introduced throughout the years. The following is by no means an exhaustive list of such tools; however, it gives a taste of the rich possibilities one faces when there is a need to make C++ code accessible to Python developers. One goal that all these tools share is transforming Python/C API into a relatively more user-friendly interface. Most listed tools are designed to wrap C++ interfaces non-intrusively; that is, the C++ code of an entity does not change when it is wrapped.

% -----------------------------------------------------------------------------
\subsection{Binding Tools}
\label{ssec:comp:tools}
% -----------------------------------------------------------------------------
% ctypes
\pyLstinline{ctypes} is a foreign function module for Python that provides C compatible data
types.\footnote{See \url{https://docs.python.org/3/library/ctypes.html\#module-ctypes}.} It supports the loading of shared libraries and marshaling data between Python and C.\footnote{Marshaling is the process of transforming the memory representation of an object into a data format suitable for storage or transmission; see \url{https://en.wikipedia.org/wiki/Marshalling\_(computer\_science)}.} The \pyLstinline{ctypes} module is part of the Python standard library; thus, it can be used to wrap shared libraries in pure Python. Using the \pyLstinline{ctypes} module requires manual programming labour and detailed knowledge of the \pyLstinline{ctypes} interface. However, it does not require writing C code (in addition to existing C code that needs to be wrapped).

% CFFI
\pyLstinline{cffi} is another Python module that generates Python bindings (it stands for C Foreign Function Interface). It interacts with almost any C code from Python, based on C-like declarations that can often be copy-pasted from header files. It is compatible with both PyPy and CPython; Moreover, in the PyPy (see~\ref{ssec:intro:bind}) reference implementation it is a first class citizen (and provided by default), thus, exhibits outstanding performance. Compared to \pyLstinline{ctypes}, the \pyLstinline{cffi} module takes a more automated approach to generate Python bindings and it minimizes the extra bits of API that need to be mastered; thus, it scales better to larger projects than \pyLstinline{ctypes}. Like \pyLstinline{ctypes},
\pyLstinline{cffi} directly interfaces C libraries only. Wrapping C++ libraries with any one of these modules requires the provision of another layer of C wrapper around the C++ code.

% cppyy
\pyLstinline{cppyy} is yet another external module for generating bindings.\footnote{See \url{https://cppyy.readthedocs.io}.} It is an automatic, run-time, C++ Python bindings generator, for calling C++ from Python and Python from C++.\footnote{See \url{https://cppyy.readthedocs.io}.} Run-time generation enables detailed specialization for higher performance, lazy loading for reduced memory use in large scale projects, Python-side cross-inheritance and callbacks for working with C++ frameworks, run-time template instantiation, automatic object downcasting, exception mapping, and interactive exploration of C++ libraries. \pyLstinline{cppyy} delivers this without any language extensions, intermediate languages, or the need for boiler-plate hand-written code. For design and performance, see~\cite{ld-hppcn-16}, albeit that the \pyLstinline{cppyy} performance has been vastly improved since.  Like \pyLstinline{cffi}, \pyLstinline{cppyy} was designed to ease binding development for the Python programmer, minimizing the need for extra C++ code; it is also compatible with both CPython and PyPy.

% Cython
A completely different approach is adopted by Cython~\cite{bbcds-cbbw-11}, which refers to a programming language and to an optimising static compiler for the Cython language.\footnote{See \url{https://cython.org/}.}  The Cython language is a superset of the Python language. It extends Python with explicit type declarations of native C/C++ types.  Cython is a compiled language used to generate CPython extension modules.  The Cython compiler generates C/C++ code from Cython code, which in turn compiles with any C/C++ compiler, and is automatically wrapped in interface code, producing extension modules that can be loaded and used by regular Python code using the \pyLstinline{import} statement, but with significantly less computational overhead at run time. Cython also facilitates wrapping independent C or C++ code into python-importable modules.  With Cython it is possible to call back and forth between Python code, Cython code, and native library code originally developed in Fortran, C, or C++.

% SIP
SIP is a collection of tools that makes it easy to create Python bindings for C and C++ libraries.\footnote{See
 \url{https://pypi.org/project/sip/}.}  It was originally developed in 1998 to create PyQt, the Python bindings for the Qt toolkit, but can be used to create bindings for any C or C++ library. SIP comprises a set of build tools and a module called sip. The build tools process a set of specification files and generates C or C++ code, which is then compiled to create the bindings extension module. Several extension modules may be installed in the same Python package. Extension modules can be built so that they are are independent of the version of Python being used.
% In other words a wheel created from them can be installed with any
% version of Python starting with v3.5.

% PyBindGen

% SWIG
SWIG is a software development tool that connects programs written in C and C++ with a variety of high-level programming languages.\footnote{See \url{http://www.swig.org/}.}  It is different than any of the others tools listed here.  SWIG is used to create bindings to C and C++ code for different types of target languages, and not only Python, including common scripting languages such as Javascript, Perl, PHP, Python, Tcl, or Ruby. The list of supported languages also includes non-scripting languages such as C\#, D, Go language, Java including Android, Lua, OCaml, Octave, Scilab and R. Also several interpreted and compiled Scheme implementations (Guile, MzScheme/Racket) are supported. SWIG is most commonly used to create high-level interpreted or compiled programming environments, user interfaces, and as a tool for testing and prototyping C/C++ software. SWIG is typically used to parse C/C++ interfaces and generate the 'glue code' required for the above target languages to call into the C/C++ code. SWIG can also export its parse tree in the form of XML. SWIG is free software and the code that SWIG generates is compatible with both commercial and non-commercial projects.

% Boost.Python
Boost is a large and sophisticated suite of utility libraries that works with almost every C++ compiler in existence. Boost.Python~\cite{bhsbp-ag-03} is a member of the Boost suit that enables seamless interoperability between C++ and the Python programming language.\footnote{See   \url{https://www.boost.org/doc/libs/1_70_0/libs/python/doc/html/index.html}.} similar to \pyLstinline{cppyy}, Boost.Python focuses at C++; unlike \pyLstinline{cppyy} it uses C++ to specify and build bindings, taking
advantage of the metaprogramming tools and polymorphism in C++. The library includes support for:
\begin{compactitem}
\item References and shared pointers (which are handled extremely gracefully)
\item Globally Registered Type Coercions
\item Automatic Cross-Module Type Conversions
\item Efficient Function Overloading
\item C++ to Python Exception Translation
\item Default Arguments
\item Keyword Arguments
\item Manipulating Python objects in C++
\item Exporting C++ Iterators as Python Iterators
\item Documentation Strings
\end{compactitem}

% Nanobind
Nanobind~\cite{nanobind} is a lightweight header-only library that exposes C++ types in Python and vice versa, mainly to create Python bindings of existing C++ code.\footnote{See \url{https://github.com/wjakob/nanobind}.} It is the evolution of PyBind11.\footnote{See \url{https://github.com/pybind/pybind11}.} Its goal, like the goal of Boost.Python, is to minimize boilerplate code in traditional extension modules by inferring type information using compile-time introspection. Its syntax is also similar to the syntax of Boost.Python. In fact, it is a tiny self-contained version of Boost.Python with everything stripped away that isn't relevant for
binding generation. 
% Without comments, the core header files only require approximately 4K lines of code and depend on Python and the C++ standard library. 
This compact implementation was possible thanks to some of the new C++11 language features (specifically, tuples, lambda functions and variadic templates). Since its creation, these libraries have grown beyond Boost.Python in several ways, leading to simpler binding code in many common situations.

We have concentrated on the three most propitious tools, namely SWIG, Nanobind, and  Boost.Python and conducted several experiments to compare the performance of the bindings generated using the selected tools; see Section~\ref{ssec:comp:exp}.

% -----------------------------------------------------------------------------
\subsection{Experiments}
\label{ssec:comp:exp}
% ----------------------------------------------------------------------------- 
Limited Python bindings for \cgal{} are available as part of an experimental project that uses SWIG; see Section~\ref{ssec:comp:tools}. Naturally, embarking on the existing venture might have saved time in developing new bindings, on the one hand, and in exploiting new bindings by our users on the other. Another advantage of using SWIG, and probably the reason for choosing SWIG in the first place in the aforementioned project, is the support for multiple target languages. Once bindings have been developed for one language, generating bindings for additional languages does not require additional development resources. However, as our experiments show, this profiteering has its cost---the generated bindings are not as time or space efficient as their counterparts generated with other tools. Both Boost.Python and PyBind11 (the origin of Nanobind) are both widely spread and tested, and actively maintained and enhanced. They are developed using advanced features of C++11 and higher versions. These two tools have similar interfaces; that is, the C++ code used to generate the bindings using both tools is similar and in some cases even identical.
As we are very much familiar with developing code that exploits advanced feature of C++, developing bindings using these tools does not present a major hurdle for us. On the other hand, developing bindings using SWIG, requires writing code that describes the bindings in a proprietary language.  Some developers may avoid using Boost.Python, because it requires dealing with a large and complex system. This does not present a drawback for us, because \cgal{} itself depends on several other libraries of Boost; thus, we are familiar with Boost and must deal with it anyway.
The compatibility of Boost with almost every C++ compiler in general has its cost; arcane template tricks and workarounds are necessary to support the oldest and buggiest of compiler specimens.  Now that C++11-compatible compilers are widely available, the heavy machinery required to support the wide range of compilers may take some toll.  However, our main concern was the efficiency of the generated bindings in terms of execution time and memory space.

We have conducted two sets of experiments. In the first set we compared the time it took to access C++ objects from Python code via bindings to the c++ shared pointers of the objects, and in the second set we compared the time it took to call C++ functions from Python code. The first set was further divided into two subsets. In the first subset we compared the time it took to access many objects and in the second subset we compared the time it took to access a single object many times; see Table~\ref{tab:res:refs} for the results.
The second set was further divided into too subsets as well. In the first subset we compared the time it took to call many small functions, and in the second subset we compared the time it took to call few large functions. We also used the second set of experiments to measure the time it took to compute the various tasks in native Python; see Table~\ref{tab:res:calls} for the results. 
All benchmark programs were executed on a machine equipped with an Intel Core~i5 clocked at~3.5GHz with~16GB of RAM.
The results we obtained left nothing in doubt.
% We abandoned using SWIG due do its high overhead and we picked Boost.python for its high efficiency in every category.
We picked Nanobind for its high efficiency in every category.

\begin{table}
  \caption{Time consumption of references measured in seconds.}
  \label{tab:res:refs}
  \centering\begin{tabular}{|l|r|r|r|r|}
    \hline
    & \textbf{Nanobind} & \textbf{pybind11} & \textbf{boost} & \textbf{SWIG}\\
    \hline
    \hline
    Many references to many objects    & \textcolor{blue}{\textbf{0.995}} & 2.632  & 2.306 & 7.874\\
    \hline
    Many references to a single object & \textcolor{blue}{\textbf{1.122}} & 2.300  & 1.478 &  5.095\\
    \hline
  \end{tabular}
\end{table}
\begin{table}
  \caption{Time consumption of function calls measured in seconds.}
  \label{tab:res:calls}
  \centering\begin{tabular}{|l|r|r|r|r|r|}
    \hline
    & \textbf{Nanobind} & \textbf{pybind11} & \textbf{boost} & \textbf{SWIG} & \textbf{Native}\\
    \hline
    \hline
    Many small tasks & \textcolor{blue}{\textbf{6.231}} & 11.532 & 7.545 & 9.112 & 9.264\\
    \hline
    One large task   & \textcolor{blue}{\textbf{0.353}}  & 0.353  & 0.353  & 0.353 & 8.581\\
    \hline
  \end{tabular}
\end{table}

% =============================================================================
\section{Binding Generation}
\label{sec:gen}
% =============================================================================
The most simple form of building an application written in C++ that depends on \cgal{} consists of the stages below. Let \myLstinline{\$APP\_SRC\_DIR} point at the root directory of the application source tree.
\begin{compactenum}
\item Obtain the sources of \cgal{}; let
  \myLstinline{\$CGAL\_SRC\_DIR} point at the root directory of the
  \cgal{} source tree.
\item Create a build directory for \cgal{}; let
  \myLstinline{\$CGAL\_DIR} point at this build directory.
\item Change directory to \myLstinline{\$CGAL_DIR} and run
  '\myLstinline{cmake [options] \$CGAL\_SRC\_DIR}'.
\item Create a build directory for the application; let
  \myLstinline{\$APP\_DIR} point at this build directory.
\item Change directory to \myLstinline{\$APP\_DIR} and run
  '\myLstinline{cmake [options] \$APP\_SRC\_DIR}' followed by
  \myLstinline{make}.
\end{compactenum}
\noindent
In the procedure above \myLstinline{options} refer to optional arguments passed to \myLstinline{cmake}, which govern the generation of the native build environment. When running an application written in Python that depends on \cgal{} instead of building the application, the user must build the bindings. Here, \myLstinline{\$APP\_SRC\_DIR} needs to point at the root directory of the binding source tree, and \myLstinline{\$APP\_DIR} needs to point at the build directory for the bindings. The optional arguments passed to \myLstinline{cmake} in stage~5 mainly consist of variables that determine (i) the object types to be bound and the unique identifier of each such object, and (ii) the name of the libraries that comprise the bindings, in case the default library name is overridden.

% -----------------------------------------------------------------------------
\subsection{Bindings Components}
\label{sec:gen:components}
% -----------------------------------------------------------------------------
Our binding code is divided into modules, which correspond to \cgal{} packages.  Each module has a long (and meaningful) name and a short name; see Table~\ref{tab:moduleName} for the list of relevant modules. Each module is associated with a \cmake{} Boolean flag that determines whether to generate bindings for that particular module. Each module may be associated with zero or more additional flags that specify selections for bindings of entities of the corresponding \cgal{} packages.  For example, the long name of the \iiDArrangementsPackage{} module is \myLstinline{ARRANGEMENT\_ON\_SURFACE\_2}; its short name is \myLstinline{AOS2}; \myLstinline{CGALPY\_ARRANGEMENT\_ON\_SURFACE\_2\_BINDINGS} is the \cmake{} Boolean flag that indicates whether to generate bindings for this module.  This module is associated with the Boolean flag \myLstinline{CGALPY\_AOS2\_POINT\_ LOCATION\_BINDINGS}, which indicates whether to generate bindings for point location queries supported by the \iiDArrangementsPackage{} package. This module is also associated with the \myLstinline{CGALPY_AOS2_GEOMETRY_TRAITS_NAME} string flag, which specify the name of the geometry traits and used to instantiate the \cppLstinline{Arrangement\_2<GeometryTraits_2, Dcel>} class template. The Boolean flags \cppLstinline{CGALPY_AOS2_VERTEX_EXTENDED}, \cppLstinline{CGALPY_AOS2_HALFEDGE_ EXTENDED}, and \cppLstinline{CGALPY_AOS2_FACE_EXTENDED} specify whether to extend the vertex, halfedge, or face records of the DCEL, respectively. See Section~\ref{ssec:modules:aos2} for more details on the \iiDArrangementsPackage{} module.

\begin{table}[!htp]
  \caption{Module names and short names}
  \centering\begin{tikzpicture}
  \matrix (first) [tableModuleName] {
    Module & Name & Short Name\\
    \iiDandiiiDGeometryKernelPackage & \myLstinline{KERNEL} & \myLstinline{KER}\\
    \dDGeometryKernelPackage & \myLstinline{KERNEL\_D} & \myLstinline{KERD}\\
    \iiDArrangementsPackage & \myLstinline{ARRANGEMENT\_ON\_SURFACE\_2} & \myLstinline{AOS2}\\
    \iiDAlphaShapesPackage & \myLstinline{ALPHA\_SHAPE\_2} & \myLstinline{AS2}\\
    \iiiDAlphaShapesPackage & \myLstinline{ALPHA\_SHAPE\_3} & \myLstinline{AS3}\\
    \iiDRegularizedBooleanSetOperationsPackage & \myLstinline{BOOLEAN\_SET\_OPERATIONS\_2} & \myLstinline{BSO2}\\
    Bounding Volumes & \myLstinline{BOUNDING\_VOLUMES} & \myLstinline{BV}\\
    \iiDConvexHullsandExtremePointsPackage & \myLstinline{CONVEX\_HULL\_2} & \myLstinline{CH2}\\
    \iiiDConvexHullsPackage & \myLstinline{CONVEX\_HULL\_3} & \myLstinline{CH3}\\
    \iiDPolygonsPackage & \myLstinline{POLYGON\_2} & \myLstinline{POL2}\\
    \iiDPolygonPartitioningPackage & \myLstinline{POLYGON\_PARTITIONING} & \myLstinline{PP}\\
    \iiDMinkowskiSumsPackage & \myLstinline{MINKOWSKI\_SUM\_2} & \myLstinline{MS2}\\
    \dDSpatialSearchingPackage & \myLstinline{SPATIAL\_SEARCHING} & \myLstinline{SS}\\
    \iiDTriangulationsPackage & \myLstinline{TRIANGULATION\_2} & \myLstinline{TRI2}\\
    \iiiDTriangulationsPackage & \myLstinline{TRIANGULATION\_3} & \myLstinline{TRI3}\\
    \surfaceMeshPackage & \myLstinline{SURFACE\_MESH} & \myLstinline{SM}\\
    \iiiDPolyhedralSurfacesPackage & \myLstinline{POLYHEDRON\_3} & \myLstinline{POL3}\\
    \PolygonMeshProcessing & \myLstinline{POLYGON_MESH_PROCESSING} & \myLstinline{PMP}\\
    };
  \end{tikzpicture}
  \label{tab:moduleName}
\end{table}

The exposed names of the bound entities of different modules are gathered in distinct Python namespaces to prevent name conflicts. The namespace is the short name of the module in lower-case except for the first letter (which is in upper-case); for example, the namespaces of the bound types and free functions of the \iiDandiiiDGeometryKernelPackage{} and the \iiDRegularizedBooleanSetOperationsPackage{} packages are \pyLstinline{Ker} and \pyLstinline{Bso2}, respectively.  Each of these two namespaces has the attribute \myLstinline{intersection} (i.e., \myLstinline{Ker.intersection()} and \myLstinline{Bso2.intersection()}). We introduce as many (Python) namespaces as modules, and nest every exposed name of a bound entity in the appropriate (Python) namespace according to the module of the entity. We provide implementation details in the next section.

The nesting of constructs in C++ is reflected in Python code. For example, assume that the binding module is called \myLstinline{CGALPY} and it includes bindings for an instance of the class template \cppLstinline{Arrangement_2<GeometryTraits, Dcel>}, where the \cppLstinline{GeometryTraits} template parameter is substituted with a traits class that handles segments, and the \cppLstinline{Dcel} template parameter is substituted with the default DCEL; see Section~\ref{ssec:modules:aos2}. The following code excerpt constructs the segment $(0,0),(1,0)$.

\begin{lstlisting}[style=pythonStyle]
>>> import CGALPY
>>> Aos2 = CGALPY.Aos2        # define the module namespace
>>> Arr = Aos2.Arrangement_2  # define the arrangement type
>>> Point = Arr.Point_2       # define the point type
>>> Segment = Arr.Curve_2     # define the curve type
>>> seg = Segment(Point(0, 0), Point(1, 0))
\end{lstlisting}

The code that binds all functions and classes of different modules is nested in distinct C++ namespaces (even if the C++ \cgal{} module does not have a namespace). It prevents name conflicts between bound entities in different modules. Within a (C++) namespace we first set the Python namespace to match the name of a module. Then, we bind all functions and classes of \cgal{} packages associated with the module. The following code except shows how it is done for the \iiDandiiiDGeometryKernelPackage{}
module:

\begin{lstlisting}[style=cppNumberedStyle]
#ifdef CGALPY_KERNEL_BINDINGS
  auto ker_m = m.def_submodule("Ker");
  export_kernel(ker_m);
#ifdef CGALPY_KERNEL_INTERSECTION_BINDINGS
  export_intersections(ker_m);
#endif
#endif
\end{lstlisting}

For every module there exists a function called \cppLstinline{export_<name>}, where \cppLstinline{<name>} is the module name. This function contains the binding code for the essential functions and classes of the module. The \iiDandiiiDGeometryKernelPackage{} module is associated with the \cmake{} \myLstinline{KERNEL_INTERSECTION_BINDINGS} flag (see Table~\ref{tab:ker}), which determines whether to generate bindings for intersections; hence, the compile-time conditional calls to the function \cppLstinline{export_kernel_intersections()}, which contains the binding code for intersections. This code presents a certain challenge described in the Section~\ref{sec:code:intersectionDetection}.

% -----------------------------------------------------------------------------
\subsection{Binding Library Name}
\label{sec:gen:libraryName}
% -----------------------------------------------------------------------------
By default, the base name of the generated library is \myLstinline{CGALPY}. However, the user can override the name with a generated string that maps to the set of bound types. This is imperative when more then one instance of the same template must be bound. Each execution of \myLstinline{cmake} followed by make generates a single library. The \cmake{} flag \myLstinline{CGALPY\_FIXED\_LIBRARY\_NAME} determines whether the base name of the generated library is \myLstinline{CGALPY} or not. If not, it has the prefix \myLstinline{CGALPY_} followed by as many as substrings as modules, the bindings of which are enabled, separated by an underscore (\myLstinline{_}). Each such substring starts with the short name of the module in lower case followed by strings that map to the selections for bindings within the module. Each such string is a single word that starts with a capital letter. For example, the name \myLstinline{GALPY_kernelEpecInt_Aos2SegPl} of a generated library, names a library that contains bindings for (i) the Exact-Predicate-Exact-Construction (EPEC) \iiDandiiiDGeometryKernelPackage{} module and intersections, and (ii) the \iiDArrangementsPackage{} module, where the \myLstinline{Arrangement\_2<>} class template is instantiated with a traits class that handles segments and the default DCEL (see Section~\ref{ssec:modules:aos2}), and point location queries.

The generated library is dynamically linked---it must be so. However, the library itself can be compiled of either static or dynamic (dependent) libraries. If you intend to generate and use just a single library that contains the bindings, you have the freedom to choose between generating a library compiled of static libraries or dynamic libraries. However, if you intend to generate several libraries and use them all in a single Python module, it is recommended using binding libraries compiled of dynamic libraries. The attributes of a kernel type, such as \myLstinline{Kernel::Point_2}, in different bindings libraries compiled of dynamic libraries refer to the same object. Otherwise, objects of the same bound type cannot be interchanged. (Also, different generated libraries might be compiled of conflicting static libraries.) The \cmake{} flag \myLstinline{CGALPY\_USE\_SHARED\_LIBS} indicates whether the generated library is compiled of static or dynamic libraries; it is true by default.

The content of the library and its name are governed by flags provided to \myLstinline{cmake}. All flags have the prefix \myLstinline{CGALPY_}. In Tables~\ref{tab:args}, \ref{tab:sel}, \ref{tab:ker}, \ref{tab:kerd}, \ref{tab:aos}, \ref{tab:tri2}, \ref{tab:as2}, \ref{tab:tri3}, \ref{tab:as3}, and \ref{tab:ss} this
prefix is omitted.

%%%%%%%%
\begin{table}[!htp]
  \caption{General Arguments}
  \centering\tikzset{%
    tableGA/.style={table,nodes={minimum height=1.6em},
    row 4/.style={nodes={minimum height=2.8em}},
    column 1/.style={nodes={text width=11em}},
    column 2/.style={nodes={text width=4.8em}},
    column 3/.style={nodes={text width=3.8em}},
    column 4/.style={nodes={text width=28em}}}}
  \centering\begin{tikzpicture}
    \matrix (first) [tableGA] {
Name & Type & Default & Description\\
\myLstinline{USE\_SHARED\_LIBS}    & \myLstinline{Boolean} & \myLstinline{true} & Determines whether to compile shared libraries\\
\myLstinline{BUILD\_SHARED\_LIBS}  & \myLstinline{Boolean} & \myLstinline{true} & Determines whether to generate shared libraries\\
\myLstinline{FIXED\_LIBRARY\_NAME} & \myLstinline{Boolean} & \myLstinline{true} & Determines whether the library name is fixed \myLstinline{cgalpy.so} or set based on other selections\\
    };
  \end{tikzpicture}
  \label{tab:args}
\end{table}

%%%%%%%%
\begin{table}[!htp]
  \caption{Binding Selection}
  \centering\begin{tikzpicture}
    \matrix (first) [tableBindingSelection] {
Name & Type & Default & Description\\
\cmakeLstinline{KERNEL\_BINDINGS}                      & Boolean & \myLstinline{true}  & Determines whether to generate bindings for 2D and 3D Kernel types\\
\cmakeLstinline{KERNEL\_D\_BINDINGS}                   & Boolean & \myLstinline{false} & Determines whether to generate bindings for dD Kernel types\\
\cmakeLstinline{ARRANGEMENT\_ON\_SURFACE\_2\_BINDINGS} & Boolean & \myLstinline{false} & Determines whether to generate bindings for 2D arrangement instances\\
\cmakeLstinline{ALPHA\_SHAPE\_2\_BINDINGS}             & Boolean & \myLstinline{false} & Determines whether to generate bindings for 2D Alpha shape instances\\
\cmakeLstinline{ALPHA\_SHAPE\_3\_BINDINGS}             & Boolean & \myLstinline{false} & Determines whether to generate bindings for 3D Alpha shape instances\\
\cmakeLstinline{BOOLEAN\_SET\_OPERATIONS\_2\_BINDINGS} & Boolean & \myLstinline{false} & Determines whether to generate bindings for 2D Boolean set operation instances\\
\cmakeLstinline{BOUNDING\_VOLUMES\_BINDINGS}           & Boolean & \myLstinline{false} & Determines whether to generate bindings for bounding volume instances\\
\cmakeLstinline{CONVEX\_HULL\_2\_BINDINGS}             & Boolean & \myLstinline{false} & Determines whether to generate bindings for 2D convex hull instances\\
\cmakeLstinline{CONVEX\_HULL\_3\_BINDINGS}             & Boolean & \myLstinline{false} & Determines whether to generate bindings for 3D convex hull instances\\
\cmakeLstinline{POLYGON\_2\_BINDINGS}                  & Boolean & \myLstinline{false} & Determines whether to generate bindings for 2D polygon instances\\
\cmakeLstinline{POLYGON\_PARTITIONING\_BINDINGS}       & Boolean & \myLstinline{false} & Determines whether to generate bindings for 2D polygon partitioning instances\\
\cmakeLstinline{MINKOWSKI\_SUM\_2\_BINDINGS}           & Boolean & \myLstinline{false} & Determines whether to generate bindings for 2D Minkowski sum instances\\
\cmakeLstinline{SPATIAL\_SEARCHING\_BINDINGS}          & Boolean & \myLstinline{false} & Determines whether to generate bindings for spatial searching instances\\
\cmakeLstinline{TRIANGULATION\_2\_BINDINGS}            & Boolean & \myLstinline{false} & Determines whether to generate bindings for 2D triangulation instances\\
\cmakeLstinline{TRIANGULATION\_3\_BINDINGS}            & Boolean & \myLstinline{false} & Determines whether to generate bindings for 3D triangulation instances\\
\cmakeLstinline{SURFACE_MESH_BINDINGS}                 & Boolean & \myLstinline{false} & Determines whether to generate bindings for surface mesh instances\\
\cmakeLstinline{POLYHEDRON_3_BINDINGS}                 & Boolean & \myLstinline{false} & Determines whether to generate bindings for 3D polyhedron instances\\
\cmakeLstinline{POLYGON_MESH_PROCESSING_BINDINGS}      & Boolean & \myLstinline{false} & Determines whether to generate bindings for polygon mesh processing instances\\
  };
 \end{tikzpicture}
 \label{tab:sel}
\end{table}

% =============================================================================
\section{Binding Modules}
\label{sec:modules}
% =============================================================================
We describe the bindings generated by our system for various \cgal{}
packages and exemplify the use of the generated bindings.
When developing code in Python that uses the bindings, the statement
that imports the binding library, that is,
\begin{lstlisting}[style=pythonStyle]
>>> import CGALPY
\end{lstlisting}
(assuming the default binding library name is retained) must precede
any statement that use the binding. This statement is omitted in all
examples hereafter.

% -----------------------------------------------------------------------------
\subsection{Two- and Three-Dimensional Kernel Bindings}
\label{ssec:modules:kernel}
% -----------------------------------------------------------------------------
The \iiDandiiiDGeometryKernelPackage{}
package~\cite{cgal:bfghhkps-lgk23-22} of \cgal{} consists of constant-size non-modifiable geometric primitive objects and operations on these objects. The objects are sets of points in $d$-dimensional affine Euclidean space, where $d = 2,3$.  Each point is uniquely represented either by Cartesian coordinates or by homogeneous coordinates. An object type can be defined either precatively as a member of a kernel type or imperatively as a global class-template parameterized by a kernel type, defined in the C++ \cppLstinline{CGAL} namespace. For example, assume that \cppLstinline{Kernel} is a kernel type; the type that represents a two-dimensional point of this kernel is either \cppLstinline{Kernel::Point_2} or \cppLstinline{CGAL::Point_2<Kernel>}.  The generated bindings better reflects the latter types; that is, defining a two-dimensional point in Python amounts to the code below.\footnote{When writing code in C++ the precative style is advantageous, as it enables the extension of kernel object types, which is irrelevant when generating bindings.}
\begin{lstlisting}[style=pythonStyle,basicstyle=\normalsize]
>>> Ker = CGALPY.Ker      # define the module namespace
>>> Point_2 = Ker.Point_2 # define the point type in the module namespace
>>> p = Point_2(0, 0);
\end{lstlisting}
Table~\ref{tab:ker} lists the \cmake{} flags associated with the \iiDandiiiDGeometryKernelPackage{} module.  The kernel type is an instance of a chain of C++ class templates. For convenience, \cgal{} provides the following predefined types of generally useful kernels:
\begin{compactenum}
\item \cppLstinline{Exact_predicates_inexact_constructions_kernel}---provides exact geometric predicates, but geometric constructions may be inexact due to round-off errors.
\item
  \cppLstinline{Exact_predicates_exact_constructions_kernel}---provides exact geometric constructions, in addition to exact geometric predicates.
\item
  \cppLstinline{Exact_predicates_exact_constructions_kernel_with_sqrt}---same as \cppLstinline{Exact_predicates_exact _constructions_kernel}, but the number type is a model of the concept that requires operations that perform square roots, namely \cppLstinline{FieldWithSqrt}.\footnote{See
    \url{https://doc.cgal.org/latest/Algebraic_foundations/classFieldWithSqrt.html}.}
\item
  \cppLstinline{Exact_predicates_exact_constructions_kernel_with_kth_root}---same as
  \cppLstinline{Exact_predicates_ exact_constructions_kernel}, but the number type is a model of the concept that requires operations that perform $k$-th roots, namely \cppLstinline{FieldWithKthRoot}.\footnote{See
    \url{https://doc.cgal.org/latest/Algebraic_foundations/classFieldWithKthRoot.html}}
\item \cppLstinline{Exact_predicates_exact_constructions_kernel_with_root_of}---same as
\cppLstinline{Exact_predicates_ exact_constructions_kernel}, but the number type is a model of the concept that requires operations that computes the root of univariate polynomial, namely \cppLstinline{FieldWithRootOf}.\footnote{See \url{https://doc.cgal.org/latest/Algebraic_foundations/classFieldWithRootOf.html}.}
\end{compactenum}

%%%%%%%% Kernel Module Flags
\begin{table}[!htp]
  \caption{\iiDandiiiDGeometryKernelPackage{} module flags}
  \centering\begin{tikzpicture}
  \matrix (first) [tableModuleFlags,
    column 1/.style={nodes={text width=17em}},
    column 2/.style={nodes={text width=3.5em}},
    column 3/.style={nodes={text width=4em}},
    column 4/.style={nodes={text width=23.5em}},
    row 3/.style={nodes={minimum height=2.8em}}
  ] {
    Name & Type & Default & Description\\
    \cmakeLstinline{KERNEL_NAME} & String & \myLstinline{epic} &
      The kernel type used\\
    \cmakeLstinline{KERNEL_INTERSECTION_BINDINGS} & Boolean &
      \myLstinline{true} &
      Determines whether to generate bindings for intersection functions\\
  };
 \end{tikzpicture}
 \label{tab:ker}
\end{table}

All the predefined types are of Cartesian kernels. The \cmake{} flag \cmakeLstinline{CGALPY_KERNEL_NAME} specifies which kernel type should be used for the generated bindings.  Currently, only three predefined types and two specific types are supported; see Table~\ref{tab:ker:name}.

%%%%%%%% Kernel Name Options
\begin{table}[!htp]
  \caption{Kernel name options.}
  \centering\begin{tikzpicture}
    \matrix (first) [tableName,
      column 1/.style={nodes={text width=8em}},
      column 2/.style={nodes={text width=34em}}
    ] {
    \cmakeLstinline{KERNEL_NAME} & Predefined Type\\
    \myLstinline{epic} &
      \cppLstinline{Exact_predicates_inexact_constructions_kernel}\\
    \myLstinline{epec} &
      \cppLstinline{Exact_predicates_exact_constructions_kernel}\\
    \myLstinline{epecws} &
      \cppLstinline{Exact_predicates_exact_constructions_kernel_with_sqrt}\\
   };
 \end{tikzpicture}
   \centering\begin{tikzpicture}
    \matrix (first) [tableName,
      row 2/.append style={nodes={minimum height=2.8em}},
      row 3/.append style={nodes={minimum height=2.8em}},
      column 1/.style={nodes={text width=19em}},
      column 2/.style={nodes={text width=23em}}
    ] {
    \cmakeLstinline{KERNEL_NAME} & Predefined Type\\
    \myLstinline{filteredSimpleCartesianDouble} &
      \setlength{\tabcolsep}{0pt}\renewcommand{\arraystretch}{0}
      \begin{tabular}{l}
        \cppLstinline{NT = double}\\[5pt]
        \cppLstinline{Filtered_kernel<Simple_cartesian<NT>>}
      \end{tabular}\\
    \myLstinline{filteredSimpleCartesianLazyGmpq} &
      \setlength{\tabcolsep}{0pt}\renewcommand{\arraystretch}{0}
      \begin{tabular}{l}
        \cppLstinline{NT = Lazy_exact_nt<Gmpq>}\\[5pt]
        \cppLstinline{Filtered_kernel<Simple_cartesian<NT>>}
      \end{tabular}\\
   };
 \end{tikzpicture}
 \label{tab:ker:name}
\end{table}

The kernel type determines the underlying number type used to represent coefficients and coordinates of kernel objects and for evaluating mathematical expressions that involve these coefficients and coordinates. The Python attribute \pyLstinline{FT}, nested in the Python namespace \pyLstinline{Ker}, exposes the C++ underlying field umber type when it is not a primitive data type; that is, when the selected kernel is nor \cppLstinline{Exact_predicates_inexact_constructions_kernel} neither \cppLstinline{Filtered_kernel<Simple_cartesian<double>>}. (In both cases the underlying number type is \cppLstinline{double}.) Similar to the Python code above, the code below defines a two-dimensional point; here, the coordinates are explicitly converted to the underlying
number type.
\begin{lstlisting}[style=pythonStyle]
>>> p = Point_2(Ker.FT(0), Ker.FT(0))
\end{lstlisting}

The code excerpt shown in Listing~\ref{lst:cmake:epecFixed}, in the \cmake{} language, sets the \cmake{} flags for our first example. When \myLstinline{cmake} is applied with these settings followed by the native build commands (e.g., \bashLstinline{make} on Linux platforms) a library called \myLstinline{CGALPY} is generated. This library consists of the bindings necessary to run the Python example shown in Listing~\ref{lst:py:kerIntersection}, which (i) determines whether two segments intersects, and (ii) computes their intersection point, using the generated binding. Observe that bindings of the \iiDandiiiDGeometryKernelPackage{} module are generated by default, whereas bindings for all other modules must be specifically requested.

\begin{lstfloat}
  \def\filename{epec_fixed_release.cmake}%
  \lstinputlisting[style=cmakeFramedStyle,basicstyle=\normalsize,
                   label={lst:cmake:epecFixed},
                   caption={\cmake{} flag settings used to generate bindings for the exact-predicates and exact-constructions kernel.}]
                  {\cmakeDir\filename}
\end{lstfloat}

\begin{lstfloat}
  \def\filename{ker_intersection.py}%
  \lstinputlisting[style=pythonFramedStyle,basicstyle=\normalsize,
                   firstline=15,
                   label={lst:py:kerIntersection},
                   caption={Computing the intersection between two line segments.}]
                  {\pythonDir\filename}
\end{lstfloat}

% -----------------------------------------------------------------------------
\subsection{$d$-Dimensional Kernel Bindings}
\label{ssec:modules:kerneld}
% -----------------------------------------------------------------------------
\cgal{} includes a separate package that consists of constant-size non-modifiable geometric primitive objects in arbitrary dimensions, and operations on these objects, called \dDGeometryKernelPackage{}~\cite{cgal:s-gkd-22}. Similar to the two- and three-dimensional kernels, the objects of the $d$-dimensional kernels are sets of points in some $d$-dimensional affine Euclidean space, where the dimension $d$ is either static across a kernel type or dynamic; see below for more details. Each point is uniquely represented either by Cartesian coordinates or by homogeneous coordinates. For convenience, \cgal{} provides the following
predefined types of generally useful kernels:
\begin{compactenum}
\item \cppLstinline{Epick_d<DimensionTag>}---provides exact geometric predicates, but geometric constructions may be inexact due to round-off errors.
\item \cppLstinline{Epeck_d<DimensionTag>}---provides exact geometric constructions, in addition to exact geometric predicates.
\end{compactenum}
Table~\ref{tab:kerd} lists the \cmake{} flags associated with the \dDGeometryKernelPackage{} module.

%%%%%%%% D Kernel Module Flags
\begin{table}[!htp]
  \caption{\dDGeometryKernelPackage{} module flags}
  \centering\begin{tikzpicture}
  \matrix (first) [tableModuleFlags,
    column 3/.style={nodes={text width=4.5em}},
    column 4/.style={nodes={text width=20.8em}}
  ] {
    Name & Type & Default & Description\\
    \cmakeLstinline{KERNEL_D_NAME} & String & \myLstinline{epicd} &
      The kernel type used\\
     \cmakeLstinline{KERNEL_D_DIMENSION_TAG} & String & dynamic &
      Determines whether the dimension is dynamic\\
    \cmakeLstinline{KERNEL_D_DIMENSION} & Integer & \myLstinline{2} &
      The dimension of the ambient Euclidean space\\
  };
 \end{tikzpicture}
 \label{tab:kerd}
\end{table}

The \cmake{} flag \cmakeLstinline{CGALPY_KERNEL_D_NAME} specifies which kernel type should be used for the generated bindings; see Table~\ref{tab:kerd:name}.

%%%%%%%% D Kernel Name Options
\begin{table}[!htp]
  \caption{$d$-dimensional Kernel name options.}
  \centering\begin{tikzpicture}
  \matrix (first) [tableName, column
    1/.style={nodes={text width=19em}},
    column 2/.style={nodes={text width=30em}}
  ] { \cmakeLstinline{KERNEL_D_NAME} & Predefined Type\\
    \cmakeLstinline{epicd} & \cppLstinline{Epick_d<DimensionTag>}\\
    \cmakeLstinline{epecd} & \cppLstinline{Epeck_d<DimensionTag>}\\
    \cmakeLstinline{cartesiandDouble} & \cppLstinline{Cartesian_d<double>}\\
    \cmakeLstinline{cartesiandLazyGmpq} &
      \cppLstinline{Cartesian_d<Lazy_exact_nt<Gmpq>>}\\
  };
 \end{tikzpicture}
 \label{tab:kerd:name}
\end{table}

When either \cppLstinline{Epick_d<DimensionTag>} or \cppLstinline{Epeck_d<DimensionTag>} are instantiated the template parameter must be substituted with a type that represents the
dimension of the ambient Euclidean space. It may be either \cppLstinline{Dimension_tag<d>} where $d$ is an integer or \cppLstinline{Dynamic_dimension_tag}. In the latter case, the dimension of the space is specified for each point when it is constructed, so it does not need to be known at compile-time of the bindings. The \cmake{} \cmakeLstinline{CGALPY_KERNEL_D_DIMENSION_TAG} flag
specifies whether the dimension is static or dynamic. If it is static, the dimension is extracted from the \cmake{} \cmakeLstinline{CGALPY_KERNEL_D_DIMENSION} \cmake{} flag; ; see
Table~\ref{tab:kerd:dim}.

%%%%%%%% D Kernel Dimension Tag Options
\begin{table}[!htp]
  \caption{$d$-dimensional Kernel dimension tag options.}
  \centering\begin{tikzpicture}
  \matrix (first) [tableName, column
    1/.style={nodes={text width=19em}},
    column 2/.style={nodes={text width=30em}}
  ] { \cmakeLstinline{KERNEL_D_DIMENSION_TAG} & Predefined Type\\
    \cmakeLstinline{static} & \cppLstinline{Dimension_tag<d>}\\
    \cmakeLstinline{dynamic} & \cppLstinline{Dynamic_dimension_tag}\\
  };
 \end{tikzpicture}
 \label{tab:kerd:dim}
\end{table}

The kernel type determines the underlying number type. It is possible to have different underlying number types for the \iiDandiiiDGeometryKernelPackage{} and the \dDGeometryKernelPackage{} models. However, when the number types differ, expensive conversions
might be necessary to combine operations from both kernels, or it may not be possible at all using binding code developed thus far.

The code excerpt in the \cmake{} language shown in Listing~\ref{lst:cmake:cdlg}, sets the \cmake{} flags for our second example. When \myLstinline{cmake} is applied with these settings
followed by the native build commands, a library called \myLstinline{CGALPY_kerdCdlgDynamic} is generated. This library consists of the bindings necessary to run the Python example shown in
Listing~\ref{lst:py:kerdDoIntersect}, which determines whether two segments in four dimensions intersect using the generated binding.

\begin{lstfloat}
  \def\filename{cdlg_release.cmake}%
  \lstinputlisting[style=cmakeFramedStyle,basicstyle=\normalsize,
                   label={lst:cmake:cdlg},
                   caption={\cmake{} flag settings used to generate bindings for the $d$-dimensional kernel.}]
                  {\cmakeDir\filename}
\end{lstfloat}

\begin{lstfloat}
  \def\filename{kerd_do_intersect.py}%
  \lstinputlisting[style=pythonFramedStyle,basicstyle=\normalsize,
                   firstline=15,
                   label={lst:py:kerdDoIntersect},
                   caption={Determining whether two segments in 4D itersect.}]
                  {\pythonDir\filename}
\end{lstfloat}

% -----------------------------------------------------------------------------
\subsection{2D Arrangement Bindings}
\label{ssec:modules:aos2}
% -----------------------------------------------------------------------------
The \cgal{} \emph{arrangements} packages constitute a large component of the \cgal{} library. This component is particularly intricate, partly due to the interplay between combinatorial algorithms and algebra~\cite{fhw-caass-12}. Arrangements are space subdivisions induced by curves and surfaces, which have been intensively studied in discrete and computational geometry~\cite{hs-a-18}, and have applications in various domains, from robotics~\cite{my-hs-atarr-95} and assembly planning~\cite{hlw-gfapm-00} through Geographic
Information Systems (GIS)~\cite{DBLP:conf/gis/KreveldSW04} to protein structure determination~\cite{DBLP:journals/jcb/MartinYBZD11}, to mention just a few uses. The arrangements packages of \cgal{} have been developed since the early days of \cgal, first for planar arrangements and maps~\cite{DBLP:journals/jea/FlatoHHNE00,   DBLP:journals/comgeo/WeinFZH07}, Boolean operations, and Minkowski sums~\cite{bfhhm-epsph-15}. Then, envelopes of surfaces in three-dimensions have been added~\cite{m-rgece-06}. Finally, a major effort has been undertaken to supports two-dimensional arrangements on (not necessarily planar) surfaces~\cite{bfhks-apsca-10,bfhmw-apsgf-10}.

Given a surface $S$ in \Rthree{} and a set $\calC$ of curves embedded in this surface, the curves subdivide $S$ into cells of dimension~$0$ (vertices),~$1$ (edges), and~$2$ (faces). This subdivision is the arrangement $\calA(\calC)$ induced by $\calC$ on $S$~\cite{fhw-caass-12}. Arrangements embedded in curved surfaces in \Rthree{} are generalizations of arrangements embedded in the plane. The \iiDArrangementsPackage{} package~\cite{cgal:wfzh-a2-22}
can be used to construct, maintain, alter, and display 2D arrangements embedded in ruled curved surfaces, such as, spheres, ellipses, tori, cones, paraboloids, and the plane. It also supports queries on such arrangements, such as point location and vertical ray shooting. One of the main
components of the \iiDArrangementsPackage{} package is the \cppLstinline{Arrangement_2<Traits, Dcel>} class template. An instance of this template is used to represent an arrangement embedded
in the plane.  Table~\ref{tab:aos} lists the \cmake{} flags associated with the \iiDArrangementsPackage{} module.
% Currently, only instances of the
% \cppLstinline{Arrangement_2<Traits,Dcel>} class template are supported.
A description of the two template parameters of this class template follows.

%%%%%%%% Arrangement Module Flags
\begin{table}[!htp]
 \caption{\iiDArrangementsPackage{} module flags}
  \centering\begin{tikzpicture}
  \matrix (first) [tableModuleFlags,
    column 1/.style={nodes={text width=17.4em}},
    column 2/.style={nodes={text width=4.6em}},
    column 3/.style={nodes={text width=4.6em}},
    column 4/.style={nodes={text width=21.5em}},
    row 6/.style={nodes={minimum height=2.8em}}
  ] {
    Name & Type & Default & Description\\
    \cmakeLstinline{AOS2_GEOMETRY_TRAITS_NAME} & \cmakeLstinline{String} &       \myLstinline{segment} & The basic geometry traits\\
    \myLstinline{AOS2_EXTEND_VERTEX} & \cmakeLstinline{Boolean} & \myLstinline{false} & Determines whether to extend the vertex type\\
    \myLstinline{AOS2_EXTEND_HALFEDGE} & \cmakeLstinline{Boolean} & \myLstinline{false} & Determines whether to extend the halfedge type\\
    \myLstinline{AOS2_EXTEND_FACE} & \cmakeLstinline{Boolean} & \myLstinline{false} & Determines whether to extend the face type\\
    \myLstinline{AOS2_POINT_LOCATION_BINDINGS} & \cmakeLstinline{Boolean} &       \myLstinline{true} & Determines whether to generate bindings for point location and vertical ray shooting queries\\
  };
 \end{tikzpicture}
 \label{tab:aos}
\end{table}

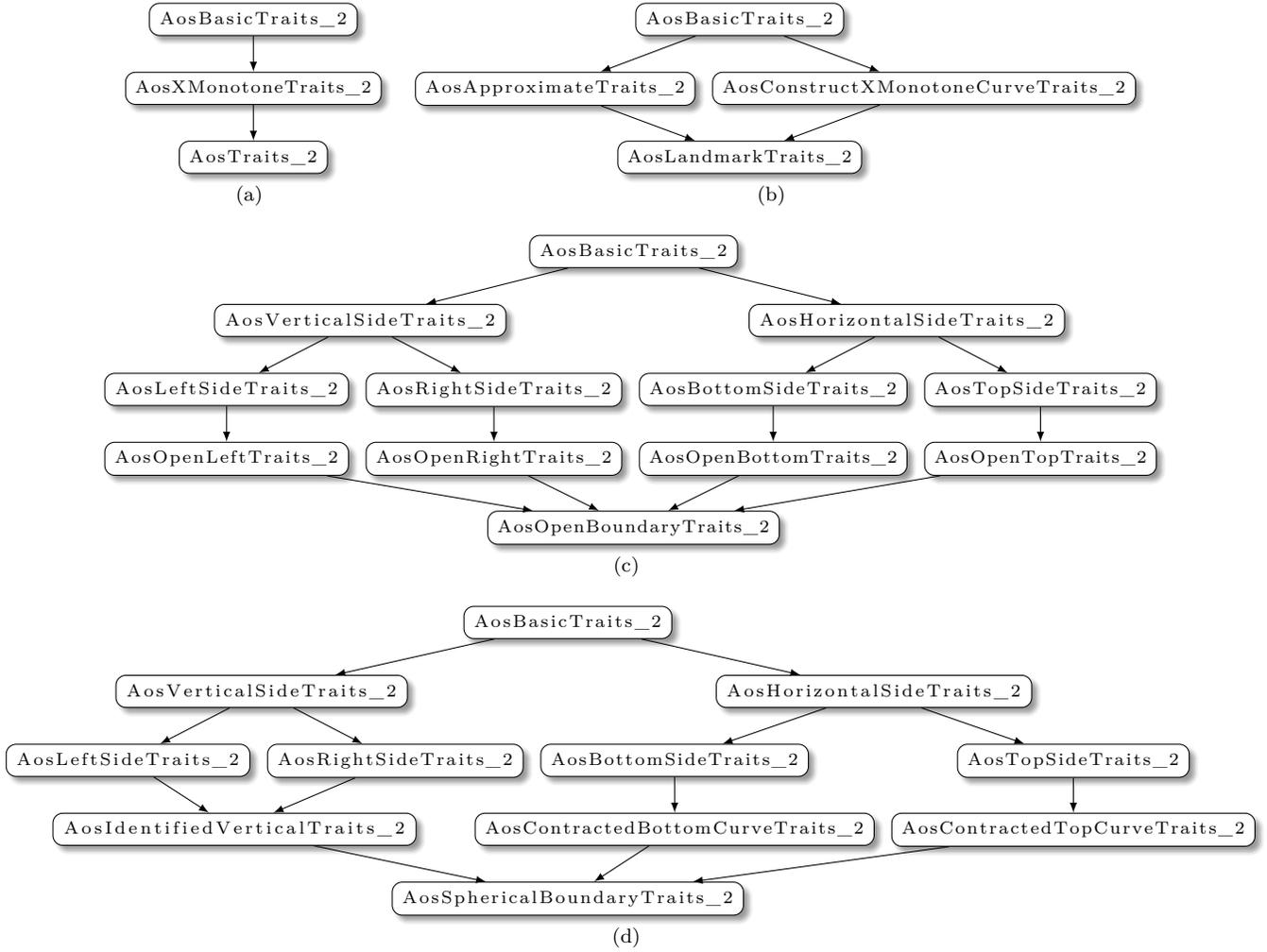
\begin{figure}[!htp]
  \def\name#1{\scriptsize\cppLstinline{#1}}
  \tikzset{concept/.style={rectangle split,rectangle split parts=1,draw,
    fill=white,blur shadow,rounded corners,align=center}}
  \captionsetup[subfigure]{justification=centering}
  \centering%
  \subfloat[]{\label{fig:atc:central}
    \begin{forest}
      [\name{AosBasicTraits_2},for tree={concept,edge={-latex}}
        % forked edges
        [\name{AosXMonotoneTraits_2}
          [\name{AosTraits_2}]
        ]
      ]
    \end{forest}}\quad
  \subfloat[]{\label{fig:atc:landmark}
    \begin{forest}
      [\name{AosBasicTraits_2},name=abt,for tree={concept,edge={-latex}}
        % forked edges
        [\name{AosApproximateTraits_2}
          [,phantom]
          [\name{AosLandmarkTraits_2},name=alt,before drawing tree={x=0}]
        ]
        [\name{AosConstructXMonotoneCurveTraits_2},name=acxmt]
      ]
      \draw[-latex] (acxmt) to (alt);
      \end{forest}}\\
  \subfloat[]{\label{fig:atc:openBoundary}
    \begin{forest}
      [\name{AosBasicTraits_2},name=abt,before drawing tree={x=0pt},for tree={concept,edge={-latex}}
        % forked edges
        [\name{AosVerticalSideTraits_2}
          [\name{AosLeftSideTraits_2}
            [\name{AosOpenLeftTraits_2},name=olt
              [,phantom]
              [\name{AosOpenBoundaryTraits_2},name=oyt,before drawing tree={x=0pt}]
            ]
          ]
          [\name{AosRightSideTraits_2}
            [\name{AosOpenRightTraits_2},name=ort]
          ]
        ]
        [\name{AosHorizontalSideTraits_2}
          [\name{AosBottomSideTraits_2}
            [\name{AosOpenBottomTraits_2},name=obt]
          ]
          [\name{AosTopSideTraits_2}
            [\name{AosOpenTopTraits_2},name=ott]
          ]
        ]
      ]
      \draw[-latex] (ort) to (oyt);
      \draw[-latex] (obt) to (oyt);
      \draw[-latex] (ott) to (oyt);
      \end{forest}}\\
  \subfloat[]{\label{fig:atc:sphericalBoundary}
    \begin{forest}
      [\name{AosBasicTraits_2},name=abt,before drawing tree={x=0pt},for tree={concept,edge={-latex}}
        % forked edges
        [\name{AosVerticalSideTraits_2}
          [\name{AosLeftSideTraits_2}
            [,phantom]
            [\name{AosIdentifiedVerticalTraits_2},name=ivt
              [,phantom]
              [\name{AosSphericalBoundaryTraits_2},name=sbt,before drawing tree={x=0pt}]
            ]
          ]
          [\name{AosRightSideTraits_2},name=rst]
        ]
        [\name{AosHorizontalSideTraits_2}
          [\name{AosBottomSideTraits_2}
            [\name{AosContractedBottomCurveTraits_2},name=cbt]
          ]
          [\name{AosTopSideTraits_2}
            [\name{AosContractedTopCurveTraits_2},name=ctt]
          ]
        ]
      ]
      \draw[-latex] (cbt) to (sbt);
      \draw[-latex] (ctt) to (sbt);
      \draw[-latex] (rst) to (ivt);
      \end{forest}}
  \caption[]{%
    \subref{fig:atc:central}~The central cluster.%
    \subref{fig:atc:landmark}~The landmark cluster.%
    \subref{fig:atc:openBoundary}~The open boundary cluster.%
    \subref{fig:atc:sphericalBoundary}~The spherical boundary cluster.}
  \label{fig:atc}
\end{figure}

\begin{itemize}
  \item The \cppLstinline{Traits} template-parameter determines the family of curves that induce the arrangement. The parameter should be substituted with a model of the basic arrangement traits concept or one or more concepts that refine the basic concept. A model of the basic traits concept defines the types of x-monotone curves and two-dimensional points and supports basic geometric predicates on them. A rather large directed acyclic graph is required to capture the entire hierarchy of the geometry traits-class concepts; therefore, we typically use subgraphs to describe the refinement relations among closely related concepts, and refer to these subgraph as clusters. Figure~\ref{fig:atc} depicts four clusters. The list of supported traits class templates follows. For each class template we describe the family of curves it handles.
  \begin{compactenum}
  \item \cppLstinline{Arr_non_caching_segment_basic_traits_2<>}---handles segments.
  \item \cppLstinline{Arr_segment_traits_2<>}---handles segments, where each segment is represented a by its supporting line in addition to its two endpoints.
  \item \cppLstinline{Arr_linear_traits_2<>}---handles linear curves, i.e., segments, rays, and lines.
  \item \cppLstinline{Arr_polyline_traits_2<>}---handles polylines.
  \item \cppLstinline{Arr_circle_segment_traits_2<>}---handles segments and circular arcs.
  \item \cppLstinline{Arr_conic_traits_2<>}---handle conic arcs.
  \item \cppLstinline{Arr_rational_function_traits_2<>}---handle rational functions.
  \item \cppLstinline{Arr_Bezier_curve_traits_2<>}---handles B\'ezier curves of arbitrary degrees.
  \item \cppLstinline{Arr_algebraic_segment_traits_2<>}---handles algebraic curves of arbitrary degrees.
  \item \cppLstinline{Arr_polycurve_traits_2<>}---handle polycurves, which are
    piecewise curves that are not necessarily linear.
  \end{compactenum}

  The \cmake{} flag \cmakeLstinline{CGALPY_AOS2_GEOMETRY_TRAITS_NAME} specifies which geometry traits should be used for the generated bindings; see Table~\ref{tab:aos:gt}. Observe that instances of the class templates \cppLstinline{General_polygon_set_2<>} and \cppLstinline{Polygon_set_2<>} (see Section~\ref{ssec:modules:bso2}) employ 2D arrangement types. Bindings for instances of \cppLstinline{Polygon_set_2<>} and   \cppLstinline{General_polygon_set_2<>} are enabled as part of the bindings for the \iiDRegularizedBooleanSetOperationsPackage{} package. When bindings for the \iiDRegularizedBooleanSetOperationsPackage{} package are enabled, bindings of the \iiDArrangementsPackage{} package must be explicitly enabled as well. The traits type that substitutes the traits parameter determines the type of curves that bound the polygons or generalized polygons, and the type must be explicitly indicated too (unless the segment type is selected, which is the default). The traits type must also model the \cppLstinline{GeneralPolygonSetTraits_2} concept; see Figure~\ref{fig:atc:gps} for the relevant traits-concept cluster. The final traits type that is used for the bindings is automatically extended to satisfy this requirement. The details of this extension is given in Section~\ref{sec:extenders:arrTraits} of the appendix.
\item The \cppLstinline{Dcel} template-parameter should be substituted with a type that models the \cppLstinline{ArrangementDcel} concept, which is used to represent the topological layout of the arrangement.\footnote{See     \url{https://doc.cgal.org/latest/Arrangement_on_surface_2/classArrangementDcel.html}.} This layout is, in particular, represented by a doubly-connected edge list data-structure (DCEL for short), which consists of containers of vertices, edges and faces and maintains the incidence   relations among these objects. We substitute this type with an instance of the template \cppLstinline{CGAL::Arr_dcel_base<V, H, F>}, where \cppLstinline{V}, \cppLstinline{H}, and
  \cppLstinline{F} are models of the concepts\linebreak
  \cppLstinline{ArrangementDcelVertex}, \cppLstinline{ArrangementDcelHalfedge}, and
  \cppLstinline{ArrangementDcelFace}, respectively; by default they are substituted with
  \cppLstinline{Arr_vertex_base<Traits::Point_2>},   \cppLstinline{Arr_halfedge_base<Traits::X_monotone_curve_2>}, and
  \cppLstinline{Arr_face_base}, respectively. In many applications it is necessary to extend the types of the DCEL main features. This is governed by three \cmake{} Boolean flags as follows. If any one of the \cmake{} variables \cmakeLstinline{CGALPY_AOS2_VERTEX_EXTENDED},
  \cmakeLstinline{CGALPY_AOS2_HALFEDGE_EXTENDED}, or   \cmakeLstinline{CGALPY_AOS2_FACE_EXTENDED} is set to true, the corresponding template parameter, \cppLstinline{V}, \cppLstinline{H}, or \cppLstinline{F}, is substituted with instances of \cppLstinline{Arr_extended_vertex<Vb, VertexData>},
  \cppLstinline{Arr_extended_halfedge<Hb, HalfedgeData>}, or
  \cppLstinline{Arr_extended_face<Fb, FaceData>}, respectively, where \cppLstinline{Vb}, \cppLstinline{Hb}, and \cppLstinline{Fb} are the basic types above. It is impossible to define a custom C++ type from Python code. Therefore, when the bindings are generated each one of the
  \cppLstinline{VertexData}, \cppLstinline{HalfedgeData}, and \cppLstinline{FaceData} template parameters must be substituted with a C++ type known at the time the bindings were   implemented. Flexibility is nevertheless retained by substituting every one of these parameters with the generic Python object \cppLstinline{py::object} when the respective cell is extended.\footnote{For more information on Python objects, see, e.g., \url{https://docs.python.org/3/library/functions.html?highlight=object\#object}.} This object provides a general interface to Python objects.
  If bindings for the \iiDRegularizedBooleanSetOperationsPackage{} package is enabled, the template parameters \cppLstinline{H} and \cppLstinline{F} above are substituted with types that also model the concepts \cppLstinline{GeneralPolygonSetDcelHalfedge}, and \cppLstinline{GeneralPolygonSetDcelFace}, respectively; see Section~\ref{ssec:modules:bso2}. The final types that are used for the bindings are automatically extended to respect this requirement. The details of this extension is given in Section~\ref{sec:extenders:arrDcel} of the appendix.
\end{itemize}

%%%%%%%% Arrangement Geometry Traits
\begin{table}[!htp]
  \caption{2D arrangement geometry traits options}
  \centering\begin{tikzpicture}
   \matrix (first) [tableName,
     column 1/.style={nodes={text width=18em}},
     column 2/.style={nodes={text width=31em}}
   ] {
     \cmakeLstinline{AOS2_GEOMETRY_TRAITS_NAME} & Type\\
     \myLstinline{nonCachingSegment} &
       \cppLstinline{Arr_non_caching_segment_basic_traits_2<Kernel>}\\
     \myLstinline{segment} & \cppLstinline{Arr_segment_traits_2<Kernel>}\\
     \myLstinline{linear} & \cppLstinline{Arr_linear_traits_2<Kernel>}\\
     \myLstinline{conic} &
       \cppLstinline{Arr_conic_traits_2<RatKernel, AlgKernel, NtTraits>}\\
     \myLstinline{circleSegment} &
       \cppLstinline{Arr_circle_segment_traits_2<Kernel>}\\
     \myLstinline{algebraic} &
       \cppLstinline{Arr_algebraic_segment_traits_2<Coefficient>}\\
   };
  \end{tikzpicture}
  \label{tab:aos:gt}
\end{table}

The code excerpt shown in Listing~\ref{lst:cmake:aos2EpecFaceExtended} in the \cmake{} language sets the \cmake{} flags for our next example shown in Listing~\ref{lst:py:aos2Fex}. This example constructs an arrangement with two faces. The arrangement is induced by line segments and its face type is extended. The properties of the bounded face and the unbounded face are initialized with Python integer objects '0' and '1', respectively.

\begin{lstfloat}
  \def\filename{aos2_epec_face_extended_release.cmake}%
  \lstinputlisting[style=cmakeFramedStyle, basicstyle=\normalsize,%
                   label={lst:cmake:aos2EpecFaceExtended},%
                   caption={\cmake{} flag settings used to generate bindings for 2D arrangements with their faces extended.}]%
                  {\cmakeDir\filename}
\end{lstfloat}

\begin{lstfloat}
  \def\filename{aos2_fex.py}%
  \lstinputlisting[style=pythonFramedStyle,basicstyle=\normalsize,
                   firstline=14,
                   label={lst:py:aos2Fex},
                   caption={Constructing an arrangement with its faces extended and initializing their data.}]
                  {\pythonDir\filename}
\end{lstfloat}

% -----------------------------------------------------------------------------
\subsection{2D Regularized Boolean Set Operation Bindings}
\label{ssec:modules:bso2}
% -----------------------------------------------------------------------------
The \cgal{} package \iiDRegularizedBooleanSetOperationsPackage{} consists of the implementation of regularized Boolean set-operations, intersection predicates, and point containment predicates on point sets bounded by weakly $x$-monotone curves in two-dimensional Euclidean space~\cite{cgal:fwzh-rbso2-22}. The \iiDRegularizedBooleanSetOperationsPackage{} module is not associated with \cmake{} flags (besides the flag \cmakeLstinline{CGALPY_BOOLEAN_SET_OPERATIONS_2_BINDINGS}, which indicates whether to generate bindings for this module). The Boolean set operations supported by this package depend on the
\iiDArrangementsPackage{} package. If the operations are applied on (linear) polygons they also depend on the \iiDPolygonsPackage{} package. Recall, that both further depend on the
\iiDandiiiDGeometryKernelPackage{} package. Therefore, bindings for these packages must be explicitly enabled as well when bindings for the \iiDRegularizedBooleanSetOperationsPackage{} package are enabled.

The code excerpt shown in Listing~\ref{lst:cmake:bso2EpecCs} in the \cmake{} language sets the \cmake{} flags for our next example. Applying these settings followed by the native build commands generates a library, the basename of which is \myLstinline{CGALPY_kerEpec_ aos2Cs_bso2_pol2}. It supports bindings for the types below and operations on these types, but nothing else.
\begin{compactitem}
\item kernel types,
\item line segment and circular arc,
\item arrangement in the plane induced by curves of the above types,
\item generalized polygon and generalized polygon with holes bounded by curves of the above types, and
\item Boolean operations on generalized polygons of the above types.
\end{compactitem}

\begin{lstfloat}
  \def\filename{bso2_epec_cs_release.cmake}%
  \lstinputlisting[style=cmakeFramedStyle, basicstyle=\normalsize,%
                   label={lst:cmake:bso2EpecCs},%
                   caption={\cmake{} flag settings used to generate bindings for 2D regularized Boolean operations on generalized polygons bounded by line segments and circular arcs.}]%
                  {\cmakeDir\filename}
\end{lstfloat}

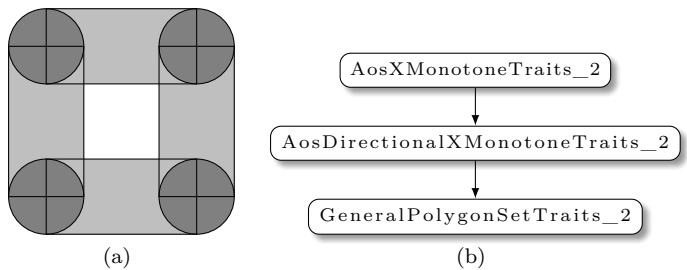
\begin{figure}[!htp]
  \def\name#1{\scriptsize\cppLstinline{#1}}
  \tikzset{concept/.style={rectangle split,rectangle split parts=1,draw,
      fill=white,blur shadow,rounded corners,align=center}}
  \captionsetup[subfigure]{justification=centering}
  \centering%
  \subfloat[]{\label{fig:csp}
    \begin{tikzpicture}
    \fill[lightgray] (0.5,0) rectangle (2.5,1);
    \fill[lightgray] (0.5,2) rectangle (2.5,3);
    \fill[lightgray] (0,0.5) rectangle (1,2.5);
    \fill[lightgray] (2,0.5) rectangle (3,2.5);
    \filldraw[fill=gray,draw=black](0.5,0.5) circle (0.5);
    \filldraw[fill=gray,draw=black](2.5,0.5) circle (0.5);
    \filldraw[fill=gray,draw=black](0.5,2.5) circle (0.5);
    \filldraw[fill=gray,draw=black](2.5,2.5) circle (0.5);
    \draw[black] (0.5,0) rectangle (2.5,1);
    \draw[black] (0.5,2) rectangle (2.5,3);
    \draw[black] (0,0.5) rectangle (1,2.5);
    \draw[black] (2,0.5) rectangle (3,2.5);
    \end{tikzpicture}}\quad
  \subfloat[]{\label{fig:atc:gps}
    \begin{forest}
      [\name{AosXMonotoneTraits_2},for tree={concept,edge={-latex}}
        % forked edges
        [\name{AosDirectionalXMonotoneTraits_2}
          [\name{GeneralPolygonSetTraits_2}]
        ]
      ]
    \end{forest}}
  \caption[]{%
    \subref{fig:csp}~A generalized polygon with holes bounded by circular
      arcs and line segments.
    \subref{fig:atc:gps}~The general point-set cluster of concepts.}
  \label{fig:bso2}
\end{figure}

This library can be used to execute the Python example shown in Listing~\ref{lst:py:circleSegmentPolygons}; the example constructs a general polygon-set that represents the point set depicted in Figure~\ref{fig:csp}. It is the result of the union of four disjoint circles and four rectangles. Each circle is represented as a generalized polygon bounded by two x-monotone circular arcs. The union is computed incrementally, resulting with a single generalized polygon with a single hole. Note that as the four circles are disjoint, their
union is computed with the \cppLstinline{insert()} function, while the union with the rectangles is computed with the \cppLstinline{join()} function.

\begin{lstfloat}
  \def\filename{Boolean_set_operations_2/circle_segment.py}%
  \lstinputlisting[style=pythonFramedStyle, basicstyle=\normalsize,%
                   firstline=15,lastline=55,%
                   label={lst:py:circleSegmentPolygons},%
                   caption={Constructing a general polygon-set that represents the point set depicted in Figure~\ref{fig:csp}.}]%
                  {\pythonDir\filename}
\end{lstfloat}

An arrangement data structure is internally used to represent the point set maintained by a general polygon-set object; it is possible to obtain it and apply further operations on it, as demonstrated by the code excerpt shown in Listing~\ref{lst:py:circleSegmentPolygonsArr}.

\begin{lstfloat}
  \def\filename{Boolean_set_operations_2/circle_segment.py}%
  \lstinputlisting[style=pythonFramedStyle, basicstyle=\normalsize,%
                   firstline=56,%
                   label={lst:py:circleSegmentPolygonsArr},%
                   caption={Extracting the arrangement from a point set.}]%
                  {\pythonDir\filename}
\end{lstfloat}

The bindings for the \iiDArrangementsPackage{} package includes bindings for the geometry traits and the DCEL suitable for Boolean operations. In particular, the traits must model the concept \cppLstinline{GeneralPolygonSetTraits_2}; see Figure~\ref{fig:atc:gps}.

% -----------------------------------------------------------------------------
\subsection{2D Minkowski Sums}
\label{ssec:modules:ms2}
% -----------------------------------------------------------------------------
Given two sets of points $\calA, \calB \in \Rd$ their Minkowski sum, denoted by $\calA \oplus \calB$, is their point-wise sum, namely the set $\{a+b|\, a \in \calA,b \in \calB\}$. The \cgal{} package \iiDMinkowskiSumsPackage{} contains functions that compute the planar Minkowski sum of two polygons and the planar Minkowski sum of a simple polygon and a disc---an operation also referred to as offsetting or dilating a polygon. The package also supports inner offsetting a polygon (also referred to as insetting), which is equivalent to the complement of the offset of (i) a disk and (ii) the complement of a polygon~\cite{cgal:w-rms2-22}. The \iiDMinkowskiSumsPackage{} module is not associated with any \cmake{} flags (besides the flag \lstinline[language={Cmake}]{CGALPY_MINKOWSKI_SUM_2_BINDINGS}, which indicates whether to generate bindings for this module). Similar to the case of generating bindings of Boolean operations on (linear) polygons supported by the \iiDRegularizedBooleanSetOperationsPackage{} package, the operations supported by this package depend on the \iiDandiiiDGeometryKernelPackage{}, \iiDArrangementsPackage{}, and
\iiDPolygonsPackage{} packages. Therefore, bindings for these packages must be explicitly enabled as well when bindings for the \iiDMinkowskiSumsPackage{} package are enabled.  The result of the inset and offset operations is a generalized polygon bounded by line
segments and circular arc. Binding for a type that represents such polygons, that is, an instance of \cppLstinline{CGAL::General_polygon_with_holes_2<>} is generated as well.

The function \cppLstinline{CGAL::minkowski_sum()} is extremely overloaded. It can be used to compute the Minkowski sum of two polygons either applying the \emph{convex-decomposition} approach or the \emph{reduced-convolution} approach. When applying the \emph{convex-decomposition} approach we first decompose each summand
into convex sub-polygons. The function template \cppLstinline{CGAL::minkowski_sum_by_full_convolution_2()} applies a third approach, namely \emph{full convolution}. For more information on the various approaches refer to the manual.

There are two overloaded function templates \cppLstinline{CGAL::minkowski_sum()} that apply the \emph{reduced-convolution} approach; both accepts the two input polygons as input; one also accepts a specific geometry traits as input (while the other constructs and uses a default traits object). Each type of the summands must represent either a simple polygon or a polygon with holes. The signature of the former is shown in Listing~\ref{lst:cpp:minkSum}, where each of \cppLstinline{PolygonType1} and \cppLstinline{PolygonType2} can be substituted either with an instance of \cppLstinline{Polygon_2<Kernel, Container>} or with and instance of \cppLstinline{Polygon_with_holes_2<Kernel, Container>}.
Thus, given a specific kernel and container types, we get eight overloaded instances that apply the \emph{reduced-convolution} approach in total. (The Container type determines the representation of the polygon's extreme points in memory.)

\begin{lstfloat}
  \begin{lstlisting}[style=cppFramedStyle,basicstyle=\normalsize,
                     label={lst:cpp:minkSum},
                     caption={The signature of one of the \cppLstinline{CGAL::minkowski_sum_2} overloaded function templates.}]
template <typename Kernel, typename Container>
Polygon_with_holes_2<Kernel, Container>
minkowski_sum_2(const PolygonType1<Kernel, Container>& p,
                const PolygonType2<Kernel, Container>& q)
\end{lstlisting}
\end{lstfloat}

There are two sets of function templates \cppLstinline{CGAL::minkowski_sum()} that apply the \emph{convex-decomposition} approach; all functions accepts two polygons as input; functions in one set also accept a specific geometry traits as input. As with the functions that apply the \emph{reduced-convolution} approach, each type of the summands must be substituted either with a type that represents a simple polygon or a type that represents a polygon with holes. Each set consists of two function templates; one is parameterized with the type that represents a single decomposition strategy that should be applied to both summands and another one that is parameterized with two types that
represent two decomposition strategies that should be applied to the two summands, respectively. The package provides four types of decomposition strategies; however, only two can be applied to a polygon with holes. Listing~\ref{lst:minkSumConvexDecomposition} shows the signatures of the two function templates that do not accept a traits parameter.
We get 10~overloaded instances of functions that apply the \emph{convex-decomposition} approach, do not accept a traits parameter, and are parameterized with a single decomposition strategy. We get 36~overloaded instances of functions that apply the
\emph{convex-decomposition} approach, do not accept a traits parameter, and are parameterized with two decomposition strategies. Thus, given a specific kernel and container types, we get $2 \times (10 + 36) = 92$ overloaded instances of functions that apply the \emph{convex-decomposition} approach.

The \cppLstinline{approximated_inset_2(P, r, eps, oi)} function template accepts a polygon $P$, an inset radius $r$, (a floating-point number\index{number!floating-point}) $\epsilon > 0$, and an output iterator\index{iterator!output} \cppLstinline{oi}; dereferencing the iterator must yield an instance of the class template
\cppLstinline{CGAL::Gps_circle_segment_traits_2<Kernel>::Polygon_2}. It constructs an approximation of the inset of $P$ by the radius $r$, where the approximation error is bounded by $\epsilon$. The function returns the polygons that approximate the inset polygon through the output iterator \cppLstinline{oi}.

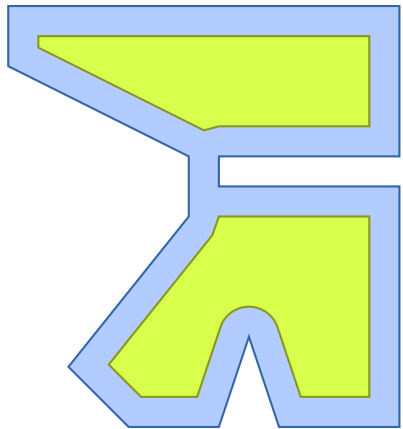
\begin{SCfigure}[][!t]
 \begin{tikzpicture}[scale=0.4]
    \coordinate (c1) at (8,3);
    \coordinate (p11) at (8.94868,3.31623);
    \coordinate (p12) at (7.05132,3.31623);
    \coordinate (c2) at (6,7);
    \coordinate (p21) at (6.78087,6.3753);
    \coordinate (p22) at (7,7);      %\draw (6,9) circle (3pt);
    \filldraw [color=polyADarkColor,fill=polyALightColor,thick]
      (4, 0)--(7,0)--(8,3)--(9,0)--(13,0)--(13,8)--(7,8)--(7,9)--
      (13,9)--(13,14)--(0,14)--(0,12)--(6,9)--(6,7)--(2,2)--cycle;
    \filldraw [color=polyCDarkColor,fill=polyCLightColor,thick]
      let \p1 = ($(p11) - (c1)$),
        \p2 = ($(p12) - (c1)$),
        \n0 = {veclen(\x1,\y1)}, % radius
        \n1 = {atan2(\y1,\x1)},  % initial angle
        \n2 = {atan2(\y2,\x2)},  % Final angle
        \p3 = ($(p21) - (c2)$),
        \p4 = ($(p22) - (c2)$),
        \n3 = {veclen(\x3,\y4)}, % radius
        \n4 = {atan2(\y3,\x4)},  % initial angle
        \n5 = {atan2(\y3,\x4)}   % Final angle
      in
        (7,7)--(12,7)--(12,1)--(9.72076,1)--(9.29057,2.29057)--(p11)arc(\n1:\n2:\n0)--
        (6.3932,1.34189)--(6.27924,1)--(4.41421,1)--(3.34,2.07422)--(3.53148,2.31357)--
        (p21)arc(\n4:\n5:\n3)--cycle;
    \filldraw [color=polyCDarkColor,fill=polyCLightColor,thick]
      (1,13)--(12,13)--(12,10)--(7,10)--
      (6.5,9.86603)--
      (6.44721,9.89443)--(4.78885,10.7236)--(1,12.618)--(1,13)--cycle;
      %\draw (7,9) circle (3pt);
      %\draw (6,9) circle (3pt);
  \end{tikzpicture}
  \caption{%
    The inset (yellow) of a polygon (blue) with a tight corridor consists of two generalized polygons bounded by line segments and circular arcs.}
  \label{fig:ms2:inset}
\end{SCfigure}

The code excerpt shown in Listing~\ref{lst:py:approxInset} demonstrates the construction of an approximated inner offset; see Figure~\ref{fig:ms2:inset}.

\begin{lstfloat}
  \def\filename{approx_inset.py}%
  \lstinputlisting[style=pythonFramedStyle, basicstyle=\normalsize,%
                   firstline=35,%
                   label={lst:py:approxInset},%
                   caption={Computing the approximated inner offset.}]%
                  {\pythonDir\filename}
\end{lstfloat}

% -----------------------------------------------------------------------------
\subsection{2D Triangulation Bindings}
\label{ssec:modules:tri2}
% -----------------------------------------------------------------------------
Triangulation is perhaps the most common term in the lexicon of computational geometry. Triangulations are ubiquitous geometric data structures, which are used in numerous areas, such as, GIS, robotics, geometric modeling and meshing to name a few; see, e.g.,~\cite{by-ag-98} for a survey on triangulations. The triangulation packages of \cgal{} are integral parts of the library and have constantly improved and enhanced
since the early days of \cgal{}. In two-dimensions \cgal{} offers basic, Delaunay and regular triangulations, as well as constrained triangulations and constrained Delaunay triangulations. In three dimensions \cgal{} offers basic, Delaunay, and regular triangulations. \cgal{} also offers periodic triangulations both in the plane and in space~\cite{ct-cpt-09}.

A triangulation of a set of points $\calP$ in \Rtwo is a partition of the convex hull of $\calP$ into triangles whose vertices are the points of $\calP$. Together with the unbounded face having the convex hull boundary as its frontier, the triangulation forms a partition of \Rtwo. Any two facets (2-face) are either disjoint or share a common
edge (1-face) or vertex (0-face). A triangulation can be described as a simplicial complex. The binding module \iiDTriangulationsPackage{} consists of bindings of types provided by the \cgal{} packages \iiDTriangulationsPackage{} and \iiDPeriodicTriangulationsPackage{}. Table~\ref{tab:tri2} lists the \cmake{} flags associated with the \iiDTriangulationsPackage{} module.

%%%%%%%% 2D Triangulation Module Flags
\begin{table}[!htp]
  \caption{\iiDTriangulationsPackage{} module flags.}
  \centering\begin{tikzpicture}
  \matrix (first) [tableModuleFlags,
    column 1/.style={nodes={text width=16.0em}},
    column 3/.style={nodes={text width=5.7em}},
    column 4/.style={nodes={text width=21.6em}},
    row 6/.append style={nodes={minimum height=2.8em}}
  ] {
    Name & Type & Default & Description\\
    \cmakeLstinline{TRI2_NAME} & String & \myLstinline{plain} &
      The 2D triangulation type\\
    \cmakeLstinline{TRI2_VERTEX_WITH_INFO}      & Boolean & \myLstinline{false} & Determines whether the vertex type is extended\\
    \cmakeLstinline{TRI2_FACE_WITH_INFO}        & Boolean & \myLstinline{false} & Determines whether the face type is extended\\
    \cmakeLstinline{TRI2_INTERSECTION_TAG_NAME} & String  & \myLstinline{ncirc} & The intersection tag\\
    \cmakeLstinline{TRI2_HIERARCHY}               & Boolean & \myLstinline{false} & Determines whether to generate the binding for a hierarchy triangulation\\
  };
  \end{tikzpicture}
  \label{tab:tri2}
\end{table}

The \iiDTriangulationsPackage{} package~\cite{cgal:y-t2-22} provides several class templates, instances of which can be used to represent a variety of 2D triangulations. In particular, the package provides the following class templates:
\begin{compactenum}
\item \cppLstinline{Triangulation_2<Traits, Tds>},
\item \cppLstinline{Regular_triangulation_2<Traits, Tds>},
\item \cppLstinline{Delaunay_triangulation_2<Traits, Tds>},
\item \cppLstinline{Constrained_triangulation_2<Traits, Tds, Itag>},
\item \cppLstinline{Constrained_Delaunay_triangulation_2<Traits, Tds, Itag>}, and
\item \cppLstinline{Triangulation_hierarchy_2<Triangulation_2>}
\end{compactenum}
Instances of the \cppLstinline{Delaunay_triangulation_2<>} and \cppLstinline{Regular_triangulation_2<>} class templates can be used to represent Delaunay and regular triangulation, respectively. In a regular triangulation points have an associated weight, and some points can be hidden and do not result in vertices of the
triangulation. The class template \cppLstinline{Triangulation_hierarchy_2} enables fast point location queries. When instantiated its template parameter must be substituted
with an instance of any other triangulation class template.

The \iiDPeriodicTriangulationsPackage{} package~\cite{cgal:k-pt2-13-22} supports triangulations of sets of points in the two-dimensional flat torus~\cite{ct-cpt-09}. This
package provides the class templates
\begin{compactenum}
\item \cppLstinline{Periodic_2_triangulation_2<Traits, Tds>},
\item \cppLstinline{Periodic_2_Delaunay_triangulation_2<Traits, Tds>}, and
\item \cppLstinline{Periodic_2_triangulation_hierarchy_2<PeriodicTriangulation>}
\end{compactenum}

The \cmake{} flag \cmakeLstinline{CGALPY_TRI2_NAME} specifies the particular type of triangulation, bindings for which should be generated; see Table~\ref{tab:tri2:names}.

\begin{table}[!htp]
  \caption{2D triangulation name options}
  \centering\begin{tikzpicture}
  \matrix (first) [tableName,
    column 1/.style={nodes={text width=11.5em}},
    column 2/.style={nodes={text width=37.5em}}
  ] {
    \cmakeLstinline{TRI2_NAME} & Type\\
    \myLstinline{plain} & \cppLstinline{Triangulation_2<Traits, Tds>}\\
    \myLstinline{regular} & \cppLstinline{Regular_triangulation_2<Traits, Tds>}\\
    \myLstinline{delaunay} & \cppLstinline{Delaunay_triangulation_2<Traits, Tds>}\\
    \myLstinline{constrained} & \cppLstinline{Constrained_triangulation_2<Traits, Tds, Itag>}\\
    \myLstinline{constrainedDelaunay} & \cppLstinline{Constrained_Delaunay_triangulation_2<Traits, Tds, Itag>}\\
    \myLstinline{periodicPlain} & \cppLstinline{Periodic_2_triangulation_2<Traits, Tds>}\\
    \myLstinline{periodicDelaunay} & \cppLstinline{Periodic_2_Delaunay_triangulation_2<Traits, Tds>}\\
  };
  \end{tikzpicture}
  \label{tab:tri2:names}
\end{table}

\begin{figure}[!htp]
  \def\name#1{\scriptsize\cppLstinline{#1}}
  \tikzset{%
    concept/.style={rectangle,rounded corners,align=center,
      draw,fill=white,blur shadow,% drop shadow%
    }
  }
  \forestset{
    my tree style/.style={
      for tree={concept,
        parent anchor=south,
        child anchor=north,
        edge={->,>=stealth,black,thick},
        edge path={% |-| style branches from parent to child
          \noexpand\path [draw, \forestoption{edge}] (!u.parent anchor)--(.child anchor)\forestoption{edge label};
        },
        if n children=3{% if a node has 3 children
          for children={% align the middle child with its parent
            if n=2{calign with current}{}
          }
        }{},
      }
    }
  }
  \centering%
  \begin{forest}
    my tree style
    [ \name{TriTr2}
      [ \name{RegularTriTr2}
        [ \name{WeightedAlphaShapeTr2} ]
      ]
      [ \name{Pr2TriTr2},name=p ]
      [ \name{DelaunayTriTr2},name=d
        [ \name{Pr2DelaunayTriTr2},name=pd
	  [ \name{Pr2AlphaShapeTr2} ]
        ]
        [ \name{AlphaShapeTr2} ]
      ]
      [ \name{ConstrainedTriTr2}
        [ \name{ConstrainedDelaunayTriTr2},name=cd ]
      ]
    ]
    \draw[->,>=stealth] (d.parent anchor)--(cd.child anchor);
    \draw[->,>=stealth] (p.parent anchor)--(pd.child anchor);
  \end{forest}
  \caption{The 2D triangulation traits concept hierarchy.
    \cppLstinline{Triangulation}, \cppLstinline{Periodic_2} and
    \cppLstinline{Traits_2} are abbreviated as \cppLstinline{Tri},
    \cppLstinline{Pr2}, and \cppLstinline{Tr2},
    respectively.}
  \label{tab:tri2:concept-hierarchy}
\end{figure}
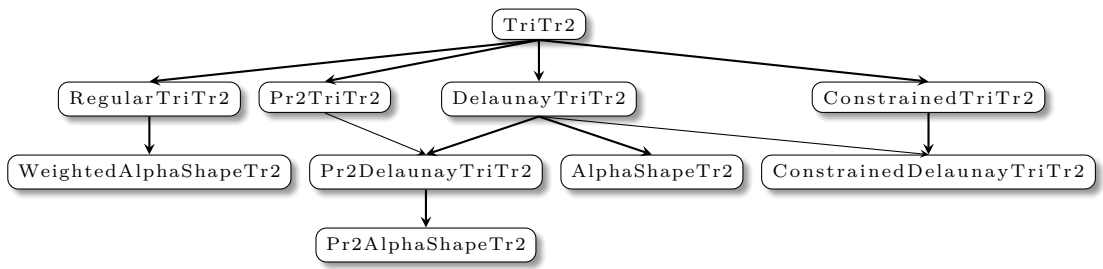

%% \cppLstinline{TriangulationTraits_2}
%% \cppLstinline{RegularTriangulationTraits_2}
%% \cppLstinline{DelaunayTriangulationTraits_2}
%% \cppLstinline{ConstrainedTriangulationTraits_2}
%% \cppLstinline{ConstrainedDelaunayTriangulationTraits_2}
%% \cppLstinline{AlphaShapeTraits_2}
%% \cppLstinline{WeightedAlphaShapeTraits_2}
%% \cppLstinline{Periodic_2TriangulationTraits_2}
%% \cppLstinline{Periodic_2DelaunayTriangulationTraits_2}
%% \cppLstinline{Periodic_3AlphaShapeTraits_2}
When any template above is instantiated the template parameter \cppLstinline{Traits} must be substituted with a model of a suitable geometric traits concept; this model is referred to as the geometric traits class; it provides the type of points to use as well as elementary operations on points of the indicated type. The type of traits used for the generated bindings is determined based on the selection of the triangulation type as explained below. It is conveniently defined in C++ as \cppLstinline{Dt::Geom_traits}, where \cppLstinline{Dt} is a triangulation instance. This type is exposed as a Python attribute with the same name under \pyLstinline{Triangulation_2} (which in turn is nested under the Python namespace \pyLstinline{Tri2}). Figure~\ref{tab:tri2:concept-hierarchy} depicts the 2D triangulation traits concept hierarchy. Any kernel instance is a model of any non-periodic traits concept; thus, when one of the templates
\begin{compactenum}
\item \cppLstinline{Triangulation_2<Traits, Tds>},
\item \cppLstinline{Regular_triangulation_2<Traits, Tds>},
\item \cppLstinline{Delaunay_triangulation_2<Traits, Tds>},
\item \cppLstinline{Constrained_triangulation_2<Traits, Tds, Itag>}, or
\item \cppLstinline{Constrained_Delaunay_triangulation_2<Traits, Tds, Itag>}
\end{compactenum}
is instantiated, the \cppLstinline{Traits} parameter is substituted with the \cppLstinline{Kernel} type (the selected kernel; see Section~\ref{ssec:modules:kernel}). When one of the templates
\begin{compactenum}
\item \cppLstinline{Periodic_2_triangulation_2<Traits, Tds>} or
\item \cppLstinline{Periodic_2_Delaunay_triangulation_2<Traits, Tds>}
\end{compactenum}
is instantiated, the \cppLstinline{Traits} parameter is substituted with \cppLstinline{Periodic_2_triangulation_traits_2<Kernel>} or \cppLstinline{Periodic_2_Delaunay_triangulation_traits_2<Kernel>}, respectively.
Observe that when a Delaunay triangulation instance is used to define an alpha shape type (see Section~\ref{ssec:modules:as2}), the traits parameter must be substituted with a traits class that models the \cppLstinline{AlphaShapeTraits_2} concept. Similarly, when a regular triangulation instance is used to define a fixed alpha shape type (see Section~\ref{ssec:modules:as2}), the traits parameter must be substituted with a traits class that models the \cppLstinline{WeightedAlphaShapeTraits_2} concept. Also in these
cases the selected kernel serves as the traits.

The \cmake{} Boolean flag \cmakeLstinline{CGALPY_TRI2_HIERARCHY} indicates whether the type of triangulation, bindings for which should be generated, is an instance of one of the triangulation hierarchy templates, namely,
\begin{compactitem}
\item \cppLstinline{Triangulation_hierarchy_2<Triangulation_2>} and
\item \cppLstinline{Periodic_triangulation_hierarchy_2<PeriodicTriangulation_ 2>}.
\end{compactitem}
The template parameter in both cases is substituted with the triangulation selected via the \cmake{} flag \cmakeLstinline{CGALPY_TRI2_ NAME}, which also determines whether to use the non-periodic or periodic version above.

The type that substitutes the \cppLstinline{Tds} parameter models the concept \cppLstinline{TriangulationDataStructure_2}. An object of this type stores the combinatorial structure of the triangulation; it is an instance of the class template
\cppLstinline{Triangulation_data_structure_2<V, F>}. The types that substitute the \cppLstinline{V} and \cppLstinline{F} template parameters when the template
\cppLstinline{Triangulation_data_structure_2<V, F>} is instantiated represent the type of the vertex and the type of the face of the triangulation, respectively; they must model the concepts \cppLstinline{TriangulationDSVertexBase_2} and \cppLstinline{TriangulationDSFaceBase_2}, respectively. If the binding is generated for a periodic triangulation, these parameters must be substituted with types that also model the concepts \cppLstinline{Periodic_2TriangulationDSVertexBase_ 2} and
\cppLstinline{Periodic_2TriangulationDSFaceBase_2}, respectively. If the triangulation is used to define an alpha shape type, these parameters must be substituted with types that also model the concepts \cppLstinline{AlphaShapeVertex_2} and \cppLstinline{AlphaShapeFace_2}, respectively; see Section~\ref{ssec:modules:as2}. Finally, if the binding is generated for a hierarchy triangulation, e.g., an instance of the template parameter \cppLstinline{Triangulation_hierarchy_2<Triangulation_2>},
the \cppLstinline{V} template parameter must be substituted with a model of the concept
\cppLstinline{TriangulationHierarchyVertexBase_2}. (Observe that there is no special requirements on the type that substitutes the \cppLstinline{F} parameter in this case.) Similar to the 2D arrangement data structure, it is possible to extend the vertex and
face types of the triangulation. This is governed by two \cmake{} Boolean flags as follows. If any one of the \cmake{} variables \cmakeLstinline{CGALPY_TRI2_VERTEX_WITH_INFO} or
\cmakeLstinline{CGALPY_TRI2_FACE_WITH_INFO} is set to true, the corresponding template parameter, \cppLstinline{V} or \cppLstinline{F}, is substituted with instances of
\cppLstinline{CGAL::Triangulation_vertex_base_with_info_2<py::object, Traits, Vb>} or \cppLstinline{CGAL::Triangulation_face_base_with_info_2<py::object, Traits, Vb>},
respectively, where \cppLstinline{Vb} and \cppLstinline{Fb} are types that model the concepts above. The final vertex type is selected accordingly; it is explained in details in Section~\ref{sec:extenders:iitri} of the appendix.

The final vertex type is conveniently defined in C++ as \cppLstinline{Dt::Vertex}, where \cppLstinline{Dt} is the triangulation instance. This type is exposed as a Python attribute with the same name under \pyLstinline{Triangulation_2}.
The final face type is conveniently defined in C++ as \cppLstinline{Dt::Face}, where \cppLstinline{Dt} is the triangulation instance. This type is also exposed as a Python attribute with the same name under \pyLstinline{Triangulation_2}.

% -----------------------------------------------------------------------------
\subsection{2D Alpha Shape Bindings}
\label{ssec:modules:as2}
% -----------------------------------------------------------------------------
The alpha shape (a.k.a.\ alpha complex) of a set of points is one of
several notions of a shape formed by the set. Given a set of points
sampled in a 2D body, an alpha shape is demarcated by a frontier,
which is a linear approximation of the original boundary of the
body. A two-dimensional alpha shape object maintains an underlying
triangulation of a set of input points in the plane. There are two
distinguished versions of alpha shapes as follows. Basic alpha shapes
are based on the Delaunay triangulation and weighted alpha shapes are
based on its generalization, the regular triangulation, where the
euclidean distance is replaced by the power to weighted points.  The
package \iiDAlphaShapesPackage{}~\cite{cgal:d-as2-22} provides the
class template \cppLstinline{Alpha_shape_2<Dt,
  ExactAlphaComparisonTag>}. In a 2D alpha shape object represented by
an instance of this class template each $k$-simplex of the underlying
triangulation is associated with an interval that specifies for which
values of $\alpha$ the $k$-simplex belongs to the alpha shape.
Table~\ref{tab:as2} lists the \cmake{} flags associated with the
\iiDAlphaShapesPackage{} module.

%%%%%%%% 2D Alpha Shape Module flags
\begin{table}[!htp]
  \caption{\iiDAlphaShapesPackage{} module flags}
  \centering\begin{tikzpicture}
  \matrix (first) [tableModuleFlags,
    column 1/.style={nodes={text width=16em}},
    column 3/.style={nodes={text width=5.7em}},
    column 4/.style={nodes={text width=21.6em}}
  ] {
    Name & Type &  Default & Description\\
    \myLstinline{AS2_EXACT_COMPARISON} & Boolean & \myLstinline{false} &
      Determines whether to apply exact comparisons\\
  };
 \end{tikzpicture}
 \label{tab:as2}
\end{table}

% -----------------------------------------------------------------------------
\subsection{3D Triangulation Bindings}
\label{ssec:modules:tri3}
% -----------------------------------------------------------------------------
A triangulation of a set of points $\calP$ in \Rthree is a partition
of the convex hull of $\calP$ into tetrahedra whose vertices are the
points of $\calP$. Similar to the two-dimensional triangulation,
together with the unbounded cell having the convex hull boundary as
its frontier, the triangulation forms a partition of \Rthree. Any two
cells (3-face) are either disjoint or share a common facet (2-face),
edge (1-face) or vertex (0-face).  The binding module
\iiiDTriangulationsPackage{} consists of bindings of types provided by
the \cgal{} packages \iiiDTriangulationsPackage{} and
\iiiDPeriodicTriangulationsPackage{}. Table~\ref{tab:tri3} lists the
\cmake{} flags associated with the \iiiDTriangulationsPackage{}
module.

%%%%%%%% 3D Triangulation Module Flags
\begin{table}[!htp]
  \caption{\iiiDTriangulationsPackage{} module flags.}
  \centering\begin{tikzpicture}
  \matrix (first) [tableModuleFlags,
    row 5/.append style={nodes={minimum height=2.8em}}
  ] {
    Name & Type & Default & Description\\
    \cmakeLstinline{TRI3_NAME}                   & String & \myLstinline{plain}    & The 3D triangulation type\\
    \cmakeLstinline{TRI3_CONCURRENCY_NAME}      & String & \myLstinline{sequential} & The concurrency method\\
    \cmakeLstinline{TRI3_LOCATION_POLICY_NAME} & String & \myLstinline{compact}    & The location policy\\
    \cmakeLstinline{TRI3_HIERARCHY}             & Boolean & \myLstinline{false} & Determines whether to generate the binding for a hierarchy triangulation\\
  };
  \end{tikzpicture}
  \label{tab:tri3}
\end{table}

The \iiiDTriangulationsPackage{} package~\cite{cgal:pt-t3-22}
provides several class templates, instances of which can be used to
represent a variety 3D triangulations. In particular, the package
provides the following class templates:
\begin{compactenum}
\item \cppLstinline{Triangulation_3<Traits, Tds, Slds>},
\item \cppLstinline{Regular_triangulation_3<Traits, Tds, Slds>}, and
\item \cppLstinline{Delaunay_triangulation_3<Traits, Tds, LocationPolicy, Slds>}
\item \cppLstinline{Triangulation_hierarchy_3<Triangulation_3>}
\end{compactenum}
Instance of the \cppLstinline{Delaunay_triangulation_3<>} and
\cppLstinline{Regular_triangulation_3<>} class templates can be used
to represent Delaunay and regular triangulation, respectively. In a
regular triangulation points have an associated weight, and some
points can be hidden and do not result in vertices in the
triangulation. The template \cppLstinline{Triangulation_hierarchy_3} enables fast
point location queries. When instantiated its template parameter must
be substituted with an instance of any other triangulation class template.

The \iiiDPeriodicTriangulationsPackage{}
package~\cite{cgal:ct-pt3-22} supports triangulations of sets of
points in the three-dimensional flat torus~\cite{ct-cpt-09}. This
package provides the class templates
\begin{compactenum}
\item \cppLstinline{Periodic_3_triangulation_3<Traits, Tds>},
\item \cppLstinline{Periodic_3_regular_triangulation_3<Traits, Tds>}, and
\item \cppLstinline{Periodic_3_Delaunay_triangulation_3<Traits, Tds>}
\item \cppLstinline{Periodic_3_triangulation_hierarchy_3<PeriodicTriangulation>}
\end{compactenum}

The \cmake{} flag \myLstinline{CGALPY_TRI3_NAME} specifies the particular
types of triangulation, bindings for which should be generated; see
Table~\ref{tab:tri3:names}.

\begin{table}[!htp]
  \caption{3D triangulation name options}
  \centering\begin{tikzpicture}
  \matrix (first) [tableName,
    column 1/.style={nodes={text width=10em}},
    column 2/.style={nodes={text width=39em}}
  ] {
    \myLstinline{TRI3_NAME} & Type\\
    \myLstinline{plain} & \cppLstinline{Triangulation_3<Traits, Tds, Slds>}\\
    \myLstinline{regular} & \cppLstinline{Regular_triangulation_3<Traits, Tds, Slds>}\\
    \myLstinline{delaunay} & \cppLstinline{Delaunay_triangulation_3<Traits, Tds, LocationPolicy, Slds>}\\
    \myLstinline{periodicPlain} & \cppLstinline{Periodic_3_triangulation_3<Traits, Tds>}\\
    \myLstinline{periodicRegular} & \cppLstinline{Periodic_3_regular_triangulation_3<Traits, Tds>}\\
    \myLstinline{periodicDelaunay} & \cppLstinline{Periodic_3_Delaunay_triangulation_3<Traits, Tds>}\\
  };
  \end{tikzpicture}
  \label{tab:tri3:names}
\end{table}

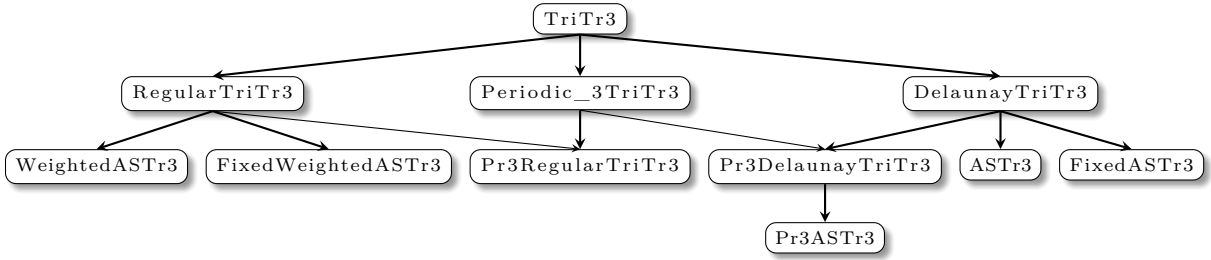
\begin{figure}[!htp]
  \def\name#1{\scriptsize\cppLstinline{#1}}
  \tikzset{%
    concept/.style={rectangle,rounded corners,align=center,
      draw,fill=white,blur shadow,% drop shadow%
    }
  }
  \forestset{
    my tree style/.style={
      for tree={concept,
        parent anchor=south,
        child anchor=north,
        edge={->,>=stealth,black,thick},
        edge path={% |-| style branches from parent to child
          \noexpand\path [draw, \forestoption{edge}] (!u.parent anchor)--(.child anchor)\forestoption{edge label};
        },
        if n children=3{% if a node has 3 children
          for children={% align the middle child with its parent
            if n=2{calign with current}{}
          }
        }{},
      }
    }
  }
  \centering
  \begin{forest}
    my tree style
    [ \name{TriTr3}
      [ \name{RegularTriTr3},name=r
        [ \name{WeightedASTr3} ]
        [ \name{FixedWeightedASTr3} ]
      ]
      [ \name{Periodic_3TriTr3},name=p
        [ \name{Pr3RegularTriTr3},name=pr ]
      ]
      [ \name{DelaunayTriTr3},name=d
        [ \name{Pr3DelaunayTriTr3},name=pd
	  [ \name{Pr3ASTr3} ]
        ]
        [ \name{ASTr3} ]
        [ \name{FixedASTr3} ]
      ]
    ]
    \draw[->,>=stealth] (r.parent anchor)--(pr.child anchor);
    \draw[->,>=stealth] (p.parent anchor)--(pd.child anchor);
  \end{forest}
  \caption{The 3D triangulation traits concept hierarchy.
    \cppLstinline{Triangulation}, \cppLstinline{AlphaShape},
    \cppLstinline{Traits_3} and \cppLstinline{Periodic_3} are
    abbreviated as \cppLstinline{Tri}, \cppLstinline{AS},
    \cppLstinline{Pr3}, and \cppLstinline{Tr3}, respectively.}
  \label{fig:tri3:concept-hierarchy}
\end{figure}

%% \cppLstinline{TriangulationTraits_3}
%% \cppLstinline{RegularTriangulationTraits_3}
%% \cppLstinline{DelaunayTriangulationTraits_3}
%% \cppLstinline{AlphaShapeTraits_3}
%% \cppLstinline{FixedAlphaShapeTraits_3}
%% \cppLstinline{WeightedAlphaShapeTraits_3}
%% \cppLstinline{FixedWeightedAlphaShapeTraits_3}
%% \cppLstinline{Periodic_3TriangulationTraits_3}
%% \cppLstinline{Periodic_3RegularTriangulationTraits_3}
%% \cppLstinline{Periodic_3DelaunayTriangulationTraits_3}
%% \cppLstinline{Periodic_3AlphaShapeTraits_3}
When any template above is instantiated the template parameter
\cppLstinline{Traits} must be substituted with a model of a suitable
geometric traits concept; this model is referred to as the geometric
traits class; it provides the type of points to use as well as
elementary operations on points of the indicated types. The type of
traits used for the generated bindings is determined based on the
selection of the triangulation type as explained below. It is
conveniently defined in C++ as \cppLstinline{Dt::Geom_traits}, where
\cppLstinline{Dt} is the triangulation instance. This type is exposed
as a Python attribute with the same name under
\pyLstinline{Triangulation_3} (which in turn is nested under the
Python namespace \pyLstinline{Tri3}).
Figure~\ref{fig:tri3:concept-hierarchy} depicts the 3D triangulation
traits concept hierarchy. Any kernel instance is a model of any
non-periodic traits concept; thus, when one of the templates
\begin{compactenum}
\item \cppLstinline{Triangulation_3<Traits, Tds>},
\item \cppLstinline{Regular_triangulation_3<Traits, Tds>}, abd
\item \cppLstinline{Delaunay_triangulation_3<Traits, Tds>}
\end{compactenum}
is instantiated, the \cppLstinline{Traits} parameter is substituted
with the \cppLstinline{Kernel} type (the selected kernel; see
Section~\ref{ssec:modules:kernel}). When one of the templates
\begin{compactenum}
\item \cppLstinline{Periodic_3_triangulation_3<Traits, Tds>},
\item \cppLstinline{Periodic_3_regular_triangulation_3<Traits, Tds>}, or
\item \cppLstinline{Periodic_3_Delaunay_triangulation_3<Traits, Tds>}
\end{compactenum}
is instantiated, the \cppLstinline{Traits} parameter is substituted
with \cppLstinline{Periodic_3_triangulation_traits_3<Kernel>},
\cppLstinline{Periodic_3_regular_triangulation_traits_3<Kernel>}, or
\cppLstinline{Periodic_3_Delaunay_triangulation_ traits_3<Kernel>},
respectively.
Observe that when a Delaunay triangulation instance is used to define
either a plain or a fixed alpha shape type (see
Section~\ref{ssec:modules:as3}), the traits parameter must be
substituted with a traits class that models either the
\cppLstinline{AlphaShapeTraits_3} or the
\cppLstinline{FixedAlphaShapeTraits_3} concept,
respectively. Similarly, when a regular triangulation instance is used
to define either a plain or a fixed alpha shape type (see
Section~\ref{ssec:modules:as3}, the traits parameter must be
substituted with a traits class that models either the
\cppLstinline{WeightedAlphaShapeTraits_3} or the
\cppLstinline{FixedWeightedAlphaShapeTraits_3} concept,
respectively. Also in these cases the selected kernel serves as the
traits.

The \cmake{} Boolean flag \cmakeLstinline{CGALPY_TRI3_HIERARCHY} indicates
whether the type of triangulation, bindings for which should be
generated, is an instance of one of the triangulation hierarchy
templates, namely,
\begin{compactitem}
\item \cppLstinline{Triangulation_hierarchy_3<Triangulation_2>} and
\item \cppLstinline{Periodic_triangulation_hierarchy_3<PeriodicTriangulation_3>}.
\end{compactitem}
The template parameter in both cases is substituted with the
triangulation selected via the \cmake{} flag
\cmakeLstinline{CGALPY_TRI3_ NAME}, which also determines whether to use the
non-periodic or periodic version above.

The type that substitutes the \cppLstinline{Tds} parameter models the
concept \cppLstinline{TriangulationDataStructure_3}. An object of this
type stores the combinatorial structure of the triangulation; it is an
instance of the class template
\cppLstinline{Triangulation_data_structure_3<V, C, ConcurrencyTag>}.
The types that substitute the \cppLstinline{V} and \cppLstinline{C}
template parameters when the
\cppLstinline{Triangulation_data_structure_3<V, C>} template
is instantiated represent the type of the vertex and the type of the
cell of the triangulation, respectively; they must model the
\cppLstinline{TriangulationDS VertexBase_3} and
\cppLstinline{TriangulationDSCellBase_3} concepts, respectively. If the
binding is generated for a periodic triangulation, these parameters
must be substituted with types that also model the concepts
\cppLstinline{Periodic_3 TriangulationDSVertexBase_3} and
\cppLstinline{Periodic_3TriangulationDSCellBase_3}, respectively.
If the triangulation is used to define a plain alpha shape type, these
parameters must be substituted with types that model the concepts
\cppLstinline{AlphaShapeVertex_3} and
\cppLstinline{AlphaShapeCell_3}, respectively; see
Section~\ref{ssec:modules:as3}. If the triangulation is used to define
a fixed alpha shape type, these parameters must be substituted with
types that model the concepts \cppLstinline{FixedAlphaShapeVertex_3}
and \cppLstinline{FixedAlphaShapeCell_3}, respectively. Finally, if
the binding is generated for a hierarchy triangulation, e.g., an
instance of the template parameter
\cppLstinline{Triangulation_hierarchy_3<Triangulation_3>}, the
\cppLstinline{V} template parameter must be substituted with a model
of the concept \cppLstinline{Triangulation HierarchyVertexBase_3}.
(Observe that there is no special requirements on the type that
substitutes the \cppLstinline{C} parameter in this case.)
Similar to the 2D arrangement and 2D triangulation data structures,
it is possible to extend the vertex and cell types of the
triangulation. This is governed by two \cmake{} Boolean flags as
follows. If any one of the \cmake{} variables
\cmakeLstinline{CGALPY_TRI3_VERTEX_WITH_INFO} or
\cmakeLstinline{CGALPY_TRI3_CELL_WITH_INFO} is set to true, the
corresponding template parameter, \cppLstinline{V} or \cppLstinline{C},
is substituted with instances of
\cppLstinline{CGAL::Triangulation_vertex_base_ with_info_3<py::object, Traits, Vb>}
or
\cppLstinline{CGAL::Triangulation_cell_base_with_info_3<py::object, Traits, Cb>},
respectively, where
\cppLstinline{Vb} and \cppLstinline{Cb} are types that model the
concepts above. The final vertex type is selected accordingly; it is
explained in details in Section~\ref{sec:extenders:iiitri} of the appendix.

The final vertex type is conveniently defined in C++ as
\cppLstinline{Dt::Vertex}, where \cppLstinline{Dt} is the
triangulation instance. This type is exposed as a Python attribute
with the same name under \pyLstinline{Triangulation_3}.
The final face type is conveniently defined in C++ as
\cppLstinline{Dt::Face}, where \cppLstinline{Dt} is the triangulation
instance. This type is also exposed as a Python attribute with the
same name under \pyLstinline{Triangulation_3}.

The template parameter \cppLstinline{ConcurrencyTag} is substituted
with either \cppLstinline{Sequential_tag} or
\cppLstinline{Parallel_tag} when the template
\cppLstinline{Triangulation_data_structure_3<V, C, ConcurrencyTag} is
instantiated. It enables the use of a concurrent
container to store vertices and cells. The \cmake{} flag
\cmakeLstinline{CGALPY_TRI3_CONCURRENCY_NAME} determines the selection; see
Table~\ref{tab:tri3:concurrency}.

\begin{table}[!htp]
  \caption{3D triangulation concurrency options}
  \centering\begin{tikzpicture}
  \matrix (first) [tableName,
    column 1/.style={nodes={text width=18em}},
    column 2/.style={nodes={text width=31em}}
  ] {
    \cmakeLstinline{TRI3_CONCURRENCY_NAME} & Type\\
    \myLstinline{sequential} & \cppLstinline{Sequential_tag}\\
    \myLstinline{parallel} & \cppLstinline{Parallel_tag}\\
  };
  \end{tikzpicture}
  \label{tab:tri3:concurrency}
\end{table}

The template parameter \cppLstinline{LocationPolicy} is substituted
with either \cppLstinline{Fast_location} or
\cppLstinline{Compact_location} when the template
\cppLstinline{Delaunay_triangulation_3<Traits, Tds, LocationPolicy>}
is instantiated. It enables a faster point location at the account of
memory space. This is useful when performing point locations or random
point insertions in large data sets.  The \cmake{} flag
\cmakeLstinline{CGALPY_TRI3_LOCATION_POLICY_NAME} determines the
selection; see Table~\ref{tab:tri3:location}.

\begin{table}[!htp]
  \caption{3D triangulation location policy options}
  \centering\begin{tikzpicture} \matrix (first) [tableName, column
    1/.style={nodes={text width=18em}}, column 2/.style={nodes={text
        width=31em}} ] {
    \cmakeLstinline{TRI3_LOCATION_POLICY_NAME} & Type\\ \myLstinline{fast} &
    \cppLstinline{Fast_location}\\ \myLstinline{compact} &
    \cppLstinline{Compact_location}\\ };
  \end{tikzpicture}
  \label{tab:tri3:location}
\end{table}

The code excerpt in the \cmake{} language shown in
Listing~\ref{lst:cmake:tri2} sets the \cmake{} flags for our next
example in Python shown in Listing~\ref{lst:py:tri3}. The example
constructs a three-dimensional Delaunay triangulation from six points
and verifies that the triangulation is valid.

\begin{lstfloat}
  \def\filename{tri3_del_epic_release.cmake}%
  \lstinputlisting[style=cmakeFramedStyle,basicstyle=\normalsize,
                   label={lst:cmake:tri2},
                   caption={\cmake{} flag settings used to generate bindings for 3D triangulations.}]
                  {\cmakeDir\filename}
\end{lstfloat}

\begin{lstfloat}
  \def\filename{tri3_del.py}%
  \lstinputlisting[style=pythonFramedStyle,basicstyle=\normalsize,
                   firstline=15,
                   label={lst:py:tri3},
                   caption={Constructing a 3D~triangulations of six vertices.}]
                  {\pythonDir\filename}
\end{lstfloat}

% -----------------------------------------------------------------------------
\subsection{3D Alpha Shape Bindings}
\label{ssec:modules:as3}
% -----------------------------------------------------------------------------
Given a set of points sampled in a 3D body, an alpha shape
(a.k.a.\ alpha complex) is demarcated by a frontier, which is a linear
approximation of the original boundary of the body. Similar to the 2D
alpha shape object, A 3D alpha shape object maintains an underlying 3D
triangulation of a set of input points. Basic alpha shapes are based
on the Delaunay triangulation and weighted alpha shapes are based on
regular triangulation, where the euclidean distance is replaced by the
power to weighted points. The package
\iiiDAlphaShapesPackage{}~\cite{cgal:dy-as3-22} provides the class
templates \cppLstinline{Alpha_shape_3<Dt, ExactAlphaComparisonTag>}
and \cppLstinline{Fixed_alpha_shape_3<Dt>}. Instances of both
templates can be used to represent a large variety of alpha shapes for
a given set of points. In a plain alpha shape each $k$-face of this
triangulation is associated with an interval specifying for which
values of $\alpha$ the face belongs to the alpha complex. In a fixed
alpha shape each $k$-face is associated with a classification that
specifies its status in the alpha complex, alpha being
fixed. Table~\ref{tab:as3} lists the \cmake{} flags associated with
the \iiiDAlphaShapesPackage{} module.

%%%%%%%% 3D Alpha Shape flags
\begin{table}[!htp]
  \caption{\iiiDAlphaShapesPackage{} module flags}
  \centering\begin{tikzpicture}
  \matrix (first) [tableModuleFlags,
    column 1/.style={nodes={text width=16em}},
    column 3/.style={nodes={text width=5.7em}},
    column 4/.style={nodes={text width=21.6em}}
  ] {
Name & Type &  Default & Description\\
\myLstinline{AS3_NAME}              & String  & \myLstinline{plain} & The 3D Alpha shape type\\
\myLstinline{AS3_EXACT_COMPARISON} & Boolean & \myLstinline{false} & Determines whether to apply exact comparisons\\
  };
 \end{tikzpicture}
 \label{tab:as3}
\end{table}

The \cmake{} flag \myLstinline{AS3_NAME} specifies which alpha shape
should be used for the generated bindings; see
Table~\ref{tab:as3:types}

%%%%%%%% 3D Alpha Shapes Module Types
\begin{table}[!htp]
 \caption{\iiiDAlphaShapesPackage{} types}
  \centering\begin{tikzpicture}
  \matrix (first) [tableName,
    column 1/.style={nodes={text width=12em}},
    column 2/.style={nodes={text width=37em}}
  ] {
     \myLstinline{AS3_NAME} & Type\\
     \myLstinline{plain} & \cppLstinline{Alpha_shape_3<Tri, Ec>}\\
     \myLstinline{fixed} & \cppLstinline{Fixed_alpha_shape_3<Tri>}\\
  };
 \end{tikzpicture}
 \label{tab:as3:types}
\end{table}

The template parameter \cppLstinline{Dt} must be substituted with an
instance of one of the following class templates that can be used to
represent a triangulation:
\begin{compactenum}
\item \cppLstinline{Delaunay_triangulation_3},
\item \cppLstinline{Regular_triangulation_3},
\item \cppLstinline{Periodic_3_Delaunay_triangulation_3}, or
\item \cppLstinline{Periodic_3_regular_triangulation_3};
\end{compactenum}
see Section~\ref{ssec:modules:tri3}.  Note that
\cppLstinline{Dt::Geom_traits} must be model of a suitable alpha-shape
traits concept, and \cppLstinline{Dt::Vertex} and
\cppLstinline{Dt::Face} must be models suitable alpha-shape vertex and
cell concepts, respectively; see Section~\ref{ssec:modules:tri3}.

The following code excerpt in the \cmake{} language shown in
Listing~\ref{lst:cmake:as3} sets the \cmake{} flags for our next
Python example shown in Listing~\ref{lst:py:as3}. The examples
constructs an alpha shape object.

\begin{lstfloat}
  \def\filename{as3_del_compact_parallel_epic_release.cmake}%
  \lstinputlisting[style=cmakeFramedStyle,basicstyle=\normalsize,
                   label={lst:cmake:as3},
                   caption={\cmake{} flag settings used to generate bindings for 3D alpha shapes.}]
                  {\cmakeDir\filename}
\end{lstfloat}

\begin{lstfloat}
  \def\filename{as3_del_compact_epic.py}%
  \lstinputlisting[style=pythonFramedStyle,basicstyle=\normalsize,
                   firstline=15,
                   label={lst:py:as3},
                   caption={Constructing a 3D alpha shape object.}]
                  {\pythonDir\filename}
\end{lstfloat}

% -----------------------------------------------------------------------------
\subsection{Spatial Searching Bindings}
\label{ssec:modules:ss}
% -----------------------------------------------------------------------------
The \dDSpatialSearchingPackage{} package~\cite{cgal:tf-ssd-22}
implements exact and approximate distance browsing by providing
implementations of algorithms supporting
\begin{compactenum}
\item both nearest and furthest neighbor searching,
\item both exact and approximate searching,
\item (approximate) range searching,
\item (approximate) $k$-nearest and $k$-furthest neighbor searching,
\item (approximate) incremental nearest and incremental furthest neighbor searching, and
\item query items representing points and spatial objects.
\end{compactenum}
In these searching problems a set $\calP$ of data points in
$d$-dimensional space is given. The points in $\calP$ are preprocessed
into a tree data structure, so that given any query item $q$ the
points of $\calP$ can be browsed efficiently. The approximate
\dDSpatialSearchingPackage{} package is designed for data sets that
are small enough to store the search structure in main memory (in
contrast to approaches from databases that assume that the data reside
in secondary storage).

%%%%%%%%
\begin{table}[!htp]
  \caption{\dDSpatialSearchingPackage{} module flags}
  \centering\begin{tikzpicture}
  \matrix (first) [tableModuleFlags,
    column 3/.style={nodes={text width=3.5em}},
    column 4/.style={nodes={text width=21.8em}}
  ] {
    Name & Type &  Default & Description\\
    \cmakeLstinline{SPATIAL_SEARCHING_DIMENSION} & Integer & \cppLstinline{2} &
      The dimension of spatial searching related classes\\
  };
 \end{tikzpicture}
 \label{tab:ss}
\end{table}

The code excerpt in the \cmake{} language shown in
Listing~\ref{lst:cmake:ss}, sets the \cmake{} flags for our last
example shown in Listing~\ref{lst:py:ss}. The example constructs a
\kdtree{} of 2D points and applies various queries on the tree.
Observe that the \kdtree{} dimension (and the dimension of related
data structures) is set via the \cmake{} flag
\myLstinline{CGALPY_SPATIAL_SEARCHING_DIMENSION} during compile time.
The code in the listings assumes that the dimension is~2. In order to
assert the dimension the bindings were compiled with (see Line~10 in
the listing), we have introduced and exposed a free function called
\myLstinline{get_spatial_searching_dimension()} that returns the
dimension as an \pyLstinline{int}.

\begin{lstfloat}
  \def\filename{ss_2_cdlg_release.cmake}%
  \lstinputlisting[style=cmakeFramedStyle,basicstyle=\normalsize,
                   label={lst:cmake:ss},
                   caption={\cmake{} flag settings used to generate bindings for spatial searching types.}]
                  {\cmakeDir\filename}
\end{lstfloat}

\begin{lstfloat}
  \def\filename{kd_tree.py}
  \lstinputlisting[style=pythonFramedStyle,basicstyle=\normalsize,
                   label={lst:py:ss},
                   firstline=17,
                   lastline=64,
                   caption={Applying various queries on a \kdtree{} of 2D points.}]
                  {\pythonDir\filename}
\end{lstfloat}

%% 2. kernel-d is used by spatial searching

% =============================================================================
\section{Binding code}
\label{sec:code}
% =============================================================================
Nanobind allows users to expose C++ classes and functions to Python using nothing more than a C++ compiler. It exploits metaprogramming techniques to implement a rich set of features and high-level user-friendly interface. Nanobind extracts as much as information as possible from the source code to be wrapped. This approach is referred to as \emph{user guided wrapping}. Additional information that cannot be automatically deduced is explicitly supplied by the user. The interface specification is written in the same full-featured C++ language as the code being exposed, which has the potential of generating efficient binding. In this section we describe some of the techniques we use to ensure that.

% -----------------------------------------------------------------------------
\subsection{Return value policy}
\label{sec:code:rvp}
% -----------------------------------------------------------------------------
Python and C++ use fundamentally different ways of managing the memory and lifetime of objects. This can lead to issues when creating bindings for functions that return a non-trivial type. Just by looking at the type, it is unclear whether Python should take charge of the returned value and eventually free its resources, or whether this is handled by the C++ code. When writing code in C++, it is usually considered a good practice to use smart pointers, which exactly describe ownership semantics. Still, even good C++ interfaces use raw references and pointers sometimes. In some cases, in order to assure proper memory management of a return value from a function, explicitly specifying a return value policy is needed. The policy may depend on whether the returned object is a newly created object, a reference to some already existing (internal) object, or a \cppLstinline{const} reference to some existing object. Additionally, in some cases we need to tie the lifetime of the result to the lifetime of the arguments, or the other way around.

For some functions of kernel objects the kind of return value depends on the kernel type; thus, the return value policy needs to be set accordingly. For example, when \cppLstinline{Exact\_predicates\_exact\_constructions\_kernel} is the kernel type, the return type of some functions that return a coordinate or a coefficient is a \cppLstinline{const} reference, while when the kernel type is \cppLstinline{Exact\_predicates\_inexact\_constructions\_kernel} the same functions return an object by value. This is why for these functions a \myLstinline{Kernel_return_value_policy} type is being passed as the return value policy. The type is defined differently based on the kernel being exposed.

\subsection{Intersection Detection}
\label{sec:code:intersectionDetection}
% -----------------------------------------------------------------------------
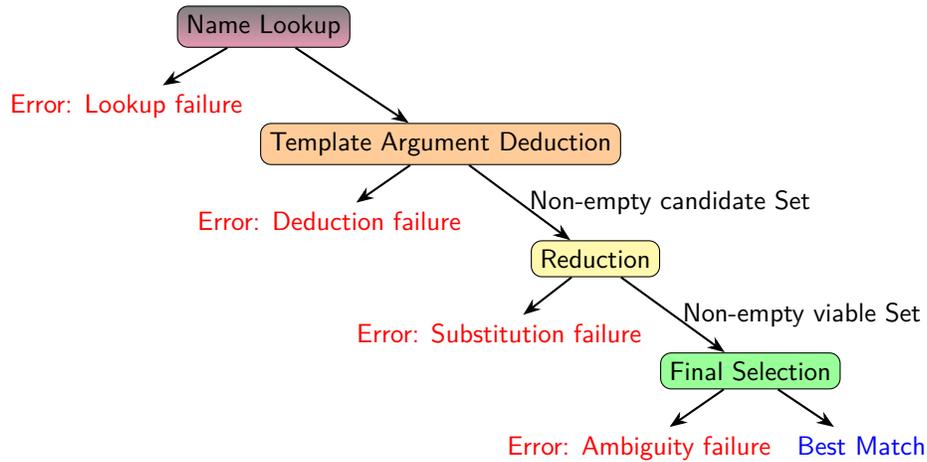
\begin{figure}[!tp]
  \centerline{\begin{tikzpicture}[>=stealth,align=center,every node/.style={font=\sffamily}]
    \tikzstyle{state}=[text centered,node distance=0.5cm,%
      rectangle,rounded corners,fill=white,draw=black,%
      minimum width=1.0cm,minimum height=0.45cm]
    % All state styles
    \tikzstyle{lookup}=[state,bottom color=purple!40]
    \tikzstyle{deduction}=[state,fill=orange!40]
    \tikzstyle{reduction}=[state,fill=yellow!40]
    \tikzstyle{selection}=[state,fill=green!40]
    % All states
    \node (lookup)[lookup]{Name Lookup};
    \node (lookupFailure)[below left=0.5 and -1 of lookup]
      {\textcolor{red}{Error: Lookup failure}};
    \node (deduction)[deduction,below right=1 and -1.2 of lookup]
      {Template Argument Deduction};
    \node (deductionFailure)[below left=0.5 and -2.8 of deduction]
      {\textcolor{red}{Error: Deduction failure}};
    \node (reduction)[reduction,below right=1 and -1.2 of deduction] {Reduction};
    \node (reductionFailure)[below left=0.5 and -1.6 of reduction]
      {\textcolor{red}{Error: Substitution failure}};
    \node (selection)[selection,below right=1 and 0.0 of reduction]
      {Final Selection};
    \node (success)[below right=0.5 and -0.7 of selection]{\textcolor{blue}{Best Match}};
    \node (slectionFailure)[below left=0.5 and -1.6 of selection]
      {\textcolor{red}{Error: Ambiguity failure}};
    % Edges
    \draw [thick,->,>=Stealth] (lookup) -- (deduction);
    \draw [thick,->,>=Stealth] (lookup) -- (lookupFailure);
    \draw [thick,->,>=Stealth] (deduction) -- node[anchor=west]{Non-empty candidate Set}(reduction);
    \draw [thick,->,>=Stealth] (deduction) -- (deductionFailure);
    \draw [thick,->,>=Stealth] (reduction) -- node[anchor=west]{Non-empty viable Set} (selection);
    \draw [thick,->,>=Stealth] (reduction) -- (reductionFailure);
    \draw [thick,->,>=Stealth] (selection) -- (success);
    \draw [thick,->,>=Stealth] (selection) -- (slectionFailure);
  \end{tikzpicture}}
  \caption{Overload resolution steps}
  \label{fig:ors}
\end{figure}

The function template \cppLstinline{CGAL::do_intersect(const T1& o1, const T2& o2)} determines whether two geometric objects intersect. The function is overloaded with several implementations that handle different combinations of types of arguments provided by the \iiDandiiiDGeometryKernelPackage{} package (see Section~\ref{ssec:modules:kernel}). However, not every combination of two types is implemented in this package. For example, the intersection of a line segment and a circle in the general case is a point with algebraic coordinates, and thus not supported by the package.\footnote{For a complete list of valid combinations of \cppLstinline{do_intersect()} argument types refer to the reference manual at   \url{https://doc.cgal.org/latest/Kernel_23/group__do__intersect__linear__grp.html}.} We use \emph{substitution failure is not an error} (SFINAE) to generate bindings for all supported combinations while avoiding getting compilation errors for unsupported combinations. SFINAE refers to a situation in C++ where an invalid substitution of template parameters should not be treated as an error. In particular, it may occur during a process called \emph{overload resolution}. In order to compile a function call, the compiler creates a set of candidate functions the names of which match the call and that can be accessed by the caller. Then, it reduces the candidate set to a set of viable functions that includes all function instances the parameters of which match the call arguments. Finally, the compiler selects the best match among the viable functions according to the C++ standard;\footnote{For the full specification of the C++ standard see \url{https://isocpp.org/std/the-standard}.} a flowchart that describes the process is shown in Figure~\ref{fig:ors}. When the substitution of an explicitly specified or deduced type for the template parameter fails, the specialization is discarded from the overload set instead of causing a compilation error.\footnote{For more information on SFINAE see, e.g., \url{https://en.cppreference.com/w/cpp/language/sfinae}.}

Our entry point is the function \cppLstinline{export_intersections_2(py::module_& m)}; see Line~8 in Listing~\ref{lst:exportIntersection}. It accepts a wrapper for Python extension modules, which is used to wrap the \cppLstinline{CGAL::do_intersect()} functions (Line 7 in Listing~\ref{lst:do_intersect}). It also serves as the entry point for the generation of bindings for the function \cppLstinline{CGAL::intersection(const T1& o1, const T2& o2)}; see Section~\ref{sec:code:intersectionComputation}. Let $\calT_2$ and $\calT_3$ denote the complete sets of types of two-dimensional and three-dimensional objects, respectively, supported by the \iiDandiiiDGeometryKernelPackage{} package; see Listing~\ref{lst:exportIntersection} Lines~9--12. Our goal is the automatic instantiation of the class template \cppLstinline{Wrapper} (Listing~\ref{lst:exportIntersection} Line 1) for every pair (\cppLstinline{T1}, \cppLstinline{T2}) in the Cartesian products $\calT_2 \times \calT_2$ and $\calT_3 \times \calT_3$, where the template parameter pack \cppLstinline{... Types} is substituted with the pair \cppLstinline{T1}, \cppLstinline{T2}. In particular, we seek for the evaluation of the expression \cppLstinline{Wrapper<T1, T2>::operator()(m)} for each instance, which in turn, leads to the evaluation of the expression \cppLstinline{bind_do_intersect<T1, T2>(m, true)} for every such pair. The expansion of the constant expression in Line~13 of Listing~\ref{lst:exportIntersection} as part of the compilation process results in the evaluation of the expression \cppLstinline{Wrapper<T1, T2>::operator()(m)} for each (\cppLstinline{T1}, \cppLstinline{T2}) $\in \calT_2 \times \calT_2$. Similarly, the expansion of the constant expression in Line~14 of Listing~\ref{lst:exportIntersection} results in the evaluation of the expression \cppLstinline{Wrapper<T1, T2>::operator()(m)} for each (\cppLstinline{T1}, \cppLstinline{T2}) $\in \calT_3 \times \calT_3$.

\begin{lstfloat}
  \begin{lstlisting}[style=cppFramedStyle,basicstyle=\normalsize,%
                     label={lst:exportIntersection},%
                     caption={Entry points for the code that automatically generates
                     binding for \cppLstinline{CGAL::do_intersect()} and
                     \cppLstinline{CGAL::intersection()} with arguments
                     that represent planar geometric objects.}]
template <typename Arg, typename ... Types> struct Wrapper {
  void operator()(Arg& arg) {
    bind_do_intersect<Types...>(arg, true);
    bind_intersection<Types...>(arg, true);
  }
};

void export_intersections_2(py::module_& m) {
  CGALPY::Type_list<Iso_rectangle_2, Line_2, Ray_2, Segment_2, Triangle_2,
                    Point_2, Circle_2> type_list_2;
  CGALPY::Type_list<Iso_cuboid_3, Line_3, Ray_3, Segment_3, Tetrahedron_3,
                    Triangle_3, Point_3, Sphere_3> type_list_3;
  cartesian_product<Wrapper>(m, type_list_2, type_list_2);
  cartesian_product<Wrapper>(m, type_list_3, type_list_3);
}
  \end{lstlisting}
\end{lstfloat}

The function template \cppLstinline{cartesian_product} (see Listing~\ref{lst:cartesianProduct} Line~17) can be used to evaluate expressions for each member of the Cartesian product of a given sequence of sets of types.  The code in Listing~\ref{lst:cartesianProduct} was presented to us, so we merely provide the code. When the template function is instantiated the \cppLstinline{Wrap} template parameter must be substituted with a class template that has an \cppLstinline{operator()} member function. The template class that substitutes the \cppLstinline{Wrap} template parameter is instantiated for every member of the Cartesian product of the sets of types that are used to define the types of the \cppLstinline{type_lists} argument pack. Observe that the type of each object in this pack is an instance of the class template \cppLstinline{Type_list}, which in turn is defined by substituting a template parameter pack with a sequence of types. The \cppLstinline{arg} argument is propagated all the way and passed to the \cppLstinline{operator()} member function of the wrapper instantiated object. Finally, the expression \cppLstinline{Wrap<T1, T2>::operator()(arg)} is evaluated for every member of the Cartesian product.
The function \cppLstinline{cartesian_product} is implemented with the help of the (compile-time) recursive function template \cppLstinline{cartesian_product_recursive()}, which exploits a unary right fold expression; see Listing~\ref{lst:cartesianProduct} Line 13.\footnote{For more information about fold expressions see \url{https://en.cppreference.com/w/cpp/language/fold}.} Let $\mathrm{E}$ denote an expression that contains a variadic template pack $\mathrm{\calT}$. Consider a block of code that contains a unary right fold expression $(\mathrm{E}\,op\,...)$, and assume that the template pack $\mathrm{\calT}$ is substituted with the sequence of types $\mathrm{T}_1,\mathrm{T}_2,\ldots,\mathrm{T}_n$, when the code block is instantiated. The unary right fold expression $(\mathrm{E}\,op\,...)$ expands to $(\mathrm{E}_1\,op\,(\ldots\,op\,(\mathrm{E}_{n-1}\,op\,\mathrm{E}_n)))$, where $\mathrm{E}_i$ is the expression $\mathrm{E}$ with the template pack $\mathrm{\calT}$ substituted with $\mathrm{T}_i$. In our case (lines 13-14 in the listing) (i) $op$ is the (binary) comma operator,\footnote{For more information about the comma operator see \url{https://en.cppreference.com/w/cpp/language/operator_other\#Built-in_function_call_operator}.} (ii) $\mathrm{E}$ is the recursive call to \cppLstinline{cartesian_product_recursive()}, and (iii) the variadic template pack is \cppLstinline{... Types}. Let \cppLstinline{type_lists_1}, \cppLstinline{type_lists_2},\ldots,\cppLstinline{type_lists_m} denote the sequence of objects that comprise the \cppLstinline{type_lists} argument pack. In all evaluations of the recursive expression at level $i$ of the recursion tree, that represents the iterated recurrence, the template pack \cppLstinline{... Types} is substituted with the sequence of types that are used to define the type of the \cppLstinline{type_lists_i}.

\begin{lstfloat}
  \begin{lstlisting}[style=cppFramedStyle,basicstyle=\normalsize,%
                     label={lst:cartesianProduct},%
                     caption={Automatic expansion of expressions with types that
                     are members of the Cartesian product of sets of types.}]
template <typename... Types> class Type_list {};

template <typename... Types> class Call_args {};

template <template <typename ...> class Wrapper, typename Arg, typename... Types>
void cartesian_product_recursive(Arg& arg, Call_args<Types...>)
{ Wrapper<Arg, Types...>()(arg); }

template <template <typename ...> class Wrapper, typename Arg,
          typename... CallTypes, typename... Types, typename ... TypeLists>
void cartesian_product_recursive(Arg& arg, Call_args<CallTypes...>,
                                 Type_list<Types...>, TypeLists... type_lists) {
  (cartesian_product_recursive<Wrapper>(arg, Call_args<CallTypes..., Types>(),
                                        type_lists...), ...);
}

template <template <typename ...> class Wrapper, typename Arg, typename... TypeLists>
void cartesian_product(Arg& arg, TypeLists... type_lists) {
  cartesian_product_recursive<Wrapper>(arg, Call_args<>(), type_lists...);
}
  \end{lstlisting}
\end{lstfloat}

Consider two types \cppLstinline{Type1} and \cppLstinline{Type2}, and assume that the \cppLstinline{CGAL::do_intersect(const Type1& o1, const Type2& o2)} overloaded version is not implemented but all other three combinations in the Cartesian product of \{\cppLstinline{Type1}, \cppLstinline{Type2}\}$\times$\{\cppLstinline{Type1}, \cppLstinline{Type2}\} are, and refer to Listing~\ref{lst:do_intersect}. The \cppLstinline{bind_do_intersect()} function template is overloaded with two implementations (Line~5 and~7). As a consequence, assuming that the code compiles, one overload of \cppLstinline{bind_do_intersect()} is selected for every member of the Cartesian product of \{\cppLstinline{Type1}, \cppLstinline{Type2}\}$\times$\{\cppLstinline{Type1}, \cppLstinline{Type2}\}. The primary implementation of \cppLstinline{bind_do_intersect()} (Line~1) uses \emph{variadic arguments} and serves as a fall-through when the evaluation of the second implementation fails during the resolution process.\footnote{For more information on \emph{variadic arguments} see \url{https://en.cppreference.com/w/cpp/language/variadic_arguments}.} The type of the second parameter of the second implementation is defined as the return type of \cppLstinline{CGAL::do_intersect<Kernel>(T1(), T2())}. When \cppLstinline{T1} and \cppLstinline{T2} are substituted with \cppLstinline{Type1} and \cppLstinline{Type2}, respectively, the evaluation fails, and the blank implementation is selected, as this is the only candidate left. In all other three cases the evaluation succeeds and the second implementation is selected, as functions that accept variadic arguments have a lower rank for the purpose of overload resolution.
% The return type of \cppLstinline{CGAL::do_intersect<Kernel>(T1(), T2())} is resolved at compile time.
Observe that in the call to \cppLstinline{bind_do_intersect()} (Line~2) we pass a \cppLstinline{bool} argument of value \cppLstinline{true}. This value can be \cppLstinline{false} just as well, but its type, that is \cppLstinline{bool}, must match the return type of \cppLstinline{CGAL::do_intersect<Kernel>(T1(), T2())} when defined. The result is not only compact code that is easy to maintain, but also code that supports bindings for every potential \cppLstinline{do_intersect()} overload that might be implemented in the future.

\begin{lstfloat}
  \begin{lstlisting}[style=cppFramedStyle,basicstyle=\normalsize,%
                     label={lst:do_intersect},%
                     caption={Automatic generation of binding  for \cppLstinline{CGAL::do_intersect()} exploiting SFINAE.}]
template <typename, typename> void bind_do_intersect(py::module_&, ...) {}

template <typename T1, typename T2>
void bind_do_intersect(py::module_& m,
                       decltype(CGAL::do_intersect<Kernel>(T1(), T2()))) {
  using Do_intersect = bool(*)(const T1&, const T2&);
  m.def("do_intersect", static_cast<Do_intersect>(&CGAL::do_intersect<Kernel>));
}
  \end{lstlisting}
\end{lstfloat}

% function \cppLstinline{bind_do_intersect()} calls
%\cppLstinline{bind_do_intersect_1T<T1>()} twice; in the first call
%\cppLstinline{T1} is substituted with \cppLstinline{Type1} (Line~15)
%and in the second call \cppLstinline{T1} it is substituted with
%\cppLstinline{Type2} (Line~16).  The function template
%\cppLstinline{bind_do_intersect_1T()} calls a third function, called
%\cppLstinline{bind_do_intersect_pair{}, twice; first it calls
%cppLstinline{bind_do_intersect_pair<T1, Type1>(true)} (Line~10); then,
% it calls \cppLstinline{bind_do_intersect_pair<T1, Type2>(true)} (Line~11).

% -----------------------------------------------------------------------------
\subsection{Intersection Computation}
\label{sec:code:intersectionComputation}
% -----------------------------------------------------------------------------
The situation presented by the overloaded function template \cppLstinline{CGAL::intersection(const T1& o1, const T2& o2)} that computes the intersection of two geometric objects is a bit more complicated. The various implementations are also provided by the \iiDandiiiDGeometryKernelPackage{} package. Similar to the \cppLstinline{CGAL::do_intersect()} function templates, different implementations handle different combination of types of arguments and not all combinations are supported. The intersection of two geometric objects can be either empty, a single object, or several points. Unlike the \cppLstinline{CGAL::do_intersect()} function templates, the return types of which are all simple \cppLstinline{bool}, the return type of each \cppLstinline{CGAL::intersection()} function template is a polymorphic type that either implicitly converts to \cppLstinline{false} if the intersection is empty or represents the intersection and depends on the types of the input arguments.

The mechanism introduced in the previous section results in the evaluation of the expression \cppLstinline{bind_intersection<T1, T2>(m, true)} for every pair (\cppLstinline{T1}, \cppLstinline{T2}) in the Cartesian products $\calT_2 \times \calT_2 \cup \calT_3 \times \calT_3$; see Listing~\ref{lst:exportIntersection} Line 4. See Line~1 and Line~3 in Listing~\ref{lst:intersection} for the primary and specialized implementations, respectively, of the function template \cppLstinline{bind_intersection()}. Given a combination of two specific types, \cppLstinline{T1} and \cppLstinline{T2}, we still evaluate the return value of the function template \cppLstinline{CGAL::intersection(const T1& o1, const T2& o2)} as a mean to determine whether the function is defined, but we do not try to match this type to a type of an argument. Instead, we introduce an unnamed template parameter with a default value that evaluates to the return type of \cppLstinline{CGAL::intersection(const T1& o1, const T2& o2)}; see Line~4 in the listing. Observe that the outcome of the evaluation is irrelevant. The only thing that matters is whether the evaluation succeeds---it succeeds only if the function template \cppLstinline{CGAL::intersection(const T1& o1, const T2& o2)} is defined. If it is defined, the specialized implementation is selected by the overload-resolution process, and the function \cppLstinline{cgalpy_intersection(const T1& t1, const T2& t2)}, which actually computes the intersection, is exposed; see Listing~\ref{lst:intersectionImpl}.

\begin{lstfloat}
  \begin{lstlisting}[style=cppFramedStyle,basicstyle=\normalsize,%
                     label={lst:intersection},%
                     caption={Automatic generation of binding for \cppLstinline{CGAL::intersection()} exploiting SFINAE}.]
template <typename, typename> void bind_intersection(py::module_& m, ...) {}

template <typename T1, typename T2,
          typename = decltype(CGAL::intersection<Kernel>(T1(), T2()))>
void bind_intersection(py::module_& m, bool)
{ m.def("intersection", &cgalpy_intersection<T1, T2>); }
  \end{lstlisting}
\end{lstfloat}

The exposed Python function that computes the intersection returns \pyLstinline{None} if the intersection is empty; see Line~17 in Listing~\ref{lst:intersectionImpl}. If the intersection consists of a single geometric object, the function returns a single Python object the type of which is the exposed \cgal{} C++ type of the geometric object; see Line~4. Finally, if several points comprise the intersection, the return type is a list of Python objects (see Line~10), where the type of each element in the list is the exposed type of \cppLstinline{Kernel::Point_2}. For more information of functions that return collections of object see Section~\ref{sec:code:output}.

\begin{lstfloat}
  \begin{lstlisting}[style=cppFramedStyle,basicstyle=\normalsize,%
                     label={lst:intersectionImpl},%
                     caption={Implementation of the binding for \cppLstinline{CGAL::intersection()}.}]
class Intersection_visitor : public boost::static_visitor<py::object> {
public:
  template<typename T>
  py::object operator()(T& operand) const { return py::cast(operand); }

  // Handle vector of points
  py::object operator()(std::vector<Point_2>& operand) const {
    py::list lst;
    for (const auto& p : operand) lst.append(p);
    return lst;
  }
};

template <typename T1, typename T2>
py::object cgalpy_intersection(const T1& t1, const T2& t2) {
  auto result = CGAL::intersection<Kernel>(t1, t2);
  if (! result) return py::object();    // no intersection
  return boost::apply_visitor(Intersection_visitor(), *result);
}
  \end{lstlisting}
\end{lstfloat}

% -----------------------------------------------------------------------------
\subsection{Minkowski Sum Construction}
\label{sec:code:minkSumComputation}
% -----------------------------------------------------------------------------
The situation presented by the function template \cppLstinline{CGAL::minkowski_sum_2()} that applies the \emph{convex-decomposition approach} (see Section~\ref{ssec:modules:ms2}) is even more complicated. The function template shown in Line~1 of Listing~\ref{lst:minkSumConvexDecomposition} has 10 instances of valid combinations of argument types. There is another set, of the same cardinality, of function templates that also accept a fourth traits parameter. The number of instances of the function template shown in Line~8 is 36, and also here there is another set, of the same cardinality, of function templates that accept an additional traits parameter. Consider the function template shown in Line~1. We need to generate bindings for instances of this function only for valid combinations of \cppLstinline{PolygonType1}, \cppLstinline{PolygonType2}, and \cppLstinline{PolygonConvexDecomposition_2} types. Hereafter and in the code listing we use the respective short names \cppLstinline{PT1}, \cppLstinline{PT2}, and \cppLstinline{PP} instead. A valid combination exists only if \cppLstinline{PP} is a type of a polygon-partition unary function that can be applied to a polygon of type \cppLstinline{PT1} and to a polygon of type \cppLstinline{PT2}. Some supported polygon-partition functions cannot be applied to polygon with holes, hence invalid combinations. In particular, if \cppLstinline{pp} is a polygon-partition function, the calls \cppLstinline{pp(pgn1, it)} and \cppLstinline{pp(pgn2, it)} must be valid, where \cppLstinline{pgn1} and \cppLstinline{pgn2} are input polygons of types \cppLstinline{PT1} and \cppLstinline{PT2}, respectively, and \cppLstinline{it} is an output iterator of a container of the resulting polygons. Dereferencing the iterator must yield a type that can represent a simple polygon. If \cppLstinline{PT1} is identical to \cppLstinline{Polygon_2}, then \cppLstinline{std::list<PT1>::iterator} can serve as an output iterator; otherwise, \cppLstinline{PT1} must represent a polygon with holes, and in this case \cppLstinline{std::list<PT1::Polygon_2>::iterator} can serve as an output iterator; the same holds for \cppLstinline{PT2}. We use SFINAE yet again to make this distinction, \footnote{This usage is similar to the example shown in \url{https://en.wikipedia.org/wiki/Substitution_failure_is_not_an_error\#C++11_simplification}.} and we use SFINAE one more time to generate binding for \cppLstinline{CGAL::minkowski_sum_2()} only for valid combinations of argument types.

\begin{lstfloat}
  \begin{lstlisting}[style=cppframedStyle,basicstyle=\normalsize,%
                     label={lst:minkSumConvexDecomposition},%
                     caption={Signatures of function templates that compute the 2D Minkwoski sum using the convex-decomposition approach.}]
template <typename Kernel, typename Container,
          typename PolygonConvexDecomposition_2>
Polygon_with_holes_2<Kernel, Container>
minkowski_sum_2(const PolygonType1<Kernel, Container>& p,
                const PolygonType2<Kernel, Container>& q,
                const PolygonConvexDecomposition_2& decomp)

template <typename Kernel, typename Container,
          typename PolygonConvexDecompositionP_2,
          typename PolygonConvexDecompositionQ_2>
Polygon_with_holes_2<Kernel, Container>
minkowski_sum_2(const PolygonType1<Kernel, Container>& p,
                const PolygonType2<Kernel, Container>& q,
                const PolygonConvexDecompositionP_2& decomp_p,
                const PolygonConvexDecompositionQ_2& decomp_q)
  \end{lstlisting}
\end{lstfloat}

Our entry point is the function \cppLstinline{void export_minkowski_sum_2()}; see Line~11 in Listing~\ref{lst:minkowskiSumExpansion}. We need to capture valid combinations of three types and valid combinations of four types. Recall that the code in Listing~\ref{lst:do_intersect} captures valid combinations of two types. Following the method presented in the previous section, we introduce two wrapper template classes with \cppLstinline{operator()} members, which call \cppLstinline{bind_mink_sum_one_strategy()} and \cppLstinline{bind_mink_sum_two_strategies()}, respectively; see Line~3 and Line~8 in Listing~\ref{lst:minkowskiSumExpansion}.

\begin{lstfloat}
  \begin{lstlisting}[style=cppframedStyle,basicstyle=\normalsize,%
                     label={lst:minkowskiSumExpansion},%
                     caption={Automatic expansion of expressions that generate
                              bindings for Minkowski-sum computation.}]
template <typename Arg, typename ... Types> struct Wrapper_one_strategy {
  void operator()(Arg& arg)
  { bind_mink_sum_decomp_one_strategy<Types...>(arg, true); }
};

template <typename Arg, typename ... Types> struct Wrapper_two_strategies {
  void operator()(Arg& arg)
  { bind_mink_sum_decomp_two_strategies<Types...>(arg, true); }
};

void export_minkowski_sum_2(py::module_& m) {
  using Pnp = ms2::Polygon_nop_decomposition_2;
  using Pvd = pp2::Polygon_vertical_decomposition_2;
  using Ptd = pp2::Polygon_triangulation_decomposition_2;
  using Ssabd = pp2::Small_side_angle_bisector_decomposition_2;

  CGALPY::Type_list<Polygon_2, Polygon_with_holes_2> polygon_types;
  CGALPY::Type_list<Pnp, Pvd, Ptd, Ssabd> strategy_types;

  CGALPY::cartesian_product<Wrapper_one_strategy>(m, polygon_types, polygon_types,
                                                  strategy_types);
  CGALPY::cartesian_product<Wrapper_two_strategies>(m, polygon_types, polygon_types,
                                                    strategy_types, strategy_types);
}
  \end{lstlisting}
\end{lstfloat}

First, we introduce a class template called \cppLstinline{target}; see Line~3 in Listing~\ref{lst:minkSumResolution}. It is parameterized with a named parameter \cppLstinline{T} and an unnamed parameter that defaults to \cppLstinline{void}. It delegates the type \cppLstinline{std::list<T>::iterator}. We also introduce a specialization (Line 6) that delegates the type \cppLstinline{std::list<T::Polygon_2>::iterator}. The second parameter is substituted with a type provided by a utility template called \cppLstinline{void_t<>} (Line~1). If the type \cppLstinline{T::Polygon_2} is undefined, the evaluation of \cppLstinline{void_t<T::Polygon_2>} fails, and in turn the \cppLstinline{Target<>} specialization is discarded from the overloaded resolution set. Otherwise, \cppLstinline{void_t<T::Polygon_2>} successfully evaluates (to \cppLstinline{void}). In this case the \cppLstinline{Target<>} specialization remains a viable option, and it is selected over the primary implementation, since specializations are ranked higher. Observe that the type delegated by the \cppLstinline{void_t<>} template and the type of the unnamed template parameter of the primary implementation of \cppLstinline{target<>} must match for the ordering to apply. These types are arbitrarily chosen to be \cppLstinline{void}.

We need to capture valid combinations of three types for Minkowski-sum computation using a single convex decomposition strategy and valid combinations of four types for Minkowski-sum computation using two convex decomposition strategies. Recall that the code in the listings of the previous sections captures valid combinations of two types. Following the method presented in the previous sections, we introduce a wrapper template class with an \cppLstinline{operator()} member, which calls \cppLstinline{bind_mink_sum()}; see Lines~10,13 in Listing~\ref{lst:minkSumResolution} for the primary and specialized implementations, respectively. The polygon-partition function returns an output iterator that points to the element, which is next to the last element of the container. Passing an argument, the type of which matches the type of the returned value of the polygon-partition function, to the function \cppLstinline{bind_mink_sum()} is complicated, as this type is not a simple type (e.g., \cppLstinline{bool}) anymore. Instead, we introduce two unnamed template parameters with default values that evaluate to the return types of the polygon-partition function when applied to a polygon of type \cppLstinline{PT1} or \cppLstinline{PT2}, respectively; see Lines~14--15. Observe that the outcome of an evaluation is irrelevant. The only thing that matters is whether the evaluation succeeds---it succeeds only if the matching function is defined.

\begin{lstfloat}
  \begin{lstlisting}[style=cppFramedStyle,basicstyle=\normalsize,
                     label=lst:minkSumResolution,
                     caption={Automatic generation of binding for \cppLstinline{CGAL::minkowski_sum_2}            exploiting SFINAE.}]
template <typename... Ts> using void_t = void;

template <typename T, typename = void> struct target
{ using type = typename std::list<T>::iterator; };

template <typename T>
struct target<T, void_t<typename T::Polygon_2>>
{ using type = typename std::list<typename T::Polygon_2>::iterator; };

template <typename T1, typename T2, typename T3>
void bind_mink_sum(py::module_&, ...) {}

template <typename PT1, typename PT2, typename PP,
          typename = decltype(PP()(PT1(), typename target<PT1>::type())),
          typename = decltype(PP()(PT2(), typename target<PT2>::type()))>
void bind_mink_sum(py::module_& m, bool) {
  m::def<Polygon_with_holes_2(const PT1&, const PT2&, const PP&)>
    ("minkowski_sum_2", &CGAL::minkowski_sum_2<Kernel, Point_2_container, PP>);
}
  \end{lstlisting}
\end{lstfloat}

%% templated with some supported type
%% {T1}. \myLstinline{bind_do_intersect_1T()}, templated with
%% \myLstinline{T1}, calls \myLstinline{bind_do_intersect_2T} templated
%% with \myLstinline{T1} and some supported type \myLstinline{T2}, and a
%% \myLstinline{bool} argument.  The declaration of
%% \myLstinline{bind_do_intersect_2T}, templated with both
%% \myLstinline{T1} and \myLstinline{T2} is only valid when templated
%% with a combination of types for which an overloaded implementation of
%% \myLstinline{CGAL::do_intersect} exists. Otherwise, due to SFINAE, the
%% blank implementation template of \myLstinline{bind_do_intersect_2T}
%% expecting (...) as an argument will be invoked (it would normally not
%% be invoked as variadic parameters have the lowest rank for the purpose
%% of overload resolution).

\noindent%
We exploit SFINAE to generate bindings of other types as described in the following sections.

% -----------------------------------------------------------------------------
\subsection{Arrangement Extension}
\label{sec:code:ae}
% -----------------------------------------------------------------------------
In many applications it is necessary to extend the type that represents certain \cgal{} data structures with some types that represent new properties. In \cgal{} it is more convenient, and thus more common, to extend the types that represent the topological features of the data structures (rather than the types that represent the geometric features), e.g., the DCEL of the arrangement data structure; see Section~\ref{ssec:modules:aos2}. When developing code in pure C++, there is no restriction on the property types used to extend the data structure. However, when the data structure interface is wrapped with Python bindings, the property type must be known when the bindings are compiled. Nevertheless, we support extensions with generic Python objects. If a certain feature must be extended with a certain property, the type of which is defined in the C++ code, it is always possible to use an external property map, where the properties are indexed by a Python object, such as an integer or a string. When generating the bindings the user must set the flag that enables the extension; see Table~\ref{tab:aos}. Attaching an index to and extracting an index from a feature is done via the \cppLstinline{set_data()} and \cppLstinline{data()} exposed functions. Each cell type, namely, \cppLstinline{Vertex}, \cppLstinline{Halfedge}, or \cppLstinline{Face}, can be extended independently, which implies that there are eight plausible combinations that the user can select from to form the final DCEL type of the arrangement.

Extending the arrangement features is useful for many applications, but essential for applications that compute the overlay of arrangements as discussed in the next Section.

% Add a discussion about extending the triangulation data structure in 2D and 3D
% Change the title to be more general to reflect other extensions

% -----------------------------------------------------------------------------
\subsection{Arrangement Overlay Traits}
\label{sec:code:overlay}
% -----------------------------------------------------------------------------
The map overlay of two arrangements $\calA_1$ and $\calA_2$, conveniently referred to as the \textcolor{red}{red} and \textcolor{blue}{blue} arrangements, is a third arrangement $\calA$, such that there is a cell $c$ in $\calA$ iff there are cells $c_1$ and $c_2$ in $\calA_1$ and $\calA_2$, respectively, and $c$ is a maximal connected component of $c_1 \cap c_2$.
% Arrangements are subdivisions of some ambient space (see Section~\ref{ssec:modules:aos2})
Computing the overlay of two arrangements is useful for many applications, including Boolean operations; see Section~\ref{ssec:modules:bso2}. Indeed, the \iiDArrangementsPackage{} package provides function templates that computes the overlay of two arrangements, namely overloaded \cppLstinline{CGAL::overlay()}. In particular, the call \cppLstinline{CGAL::overlay(arr_r, arr_b, arr)} computes the overlay of the arrangements \cppLstinline{arr_r} and \cppLstinline{arr_b} and stores the result in the arrangement \cppLstinline{arr}. All three types must be instances of the \cppLstinline{Arrangement_2<Traits, Dcel>} class template; see Section~\ref{ssec:modules:aos2}. When the function is used in pure C++ code, the geometry-traits classes that substitute the \cppLstinline{Traits} template parameters of the input arrangements must be convertible to the geometry-traits class of the resulting arrangement. We only wrap instances of the overlay function with Python bindings, where all three arrangements use the exact same geometry-traits classes and the same DCEL structures, and thus the types of the three arrangements of the exposed \cppLstinline{CGAL::overlay()} function are identical.

The \cppLstinline{CGAL::overlay()} function template is overloaded with a variant that accepts four arguments. The last argument, referred to as the overlay traits, enables the update of cells of the output arrangement, and in particular their properties. Using the overlay traits is typically accompanied with the extension of one or more types of the arrangement cells that hold the properties. The overlay traits must model the \cppLstinline{OverlayTraits} concept, which requires the provision of ten functions that handle all possible overlapping cases as listed below. Let $v_r$, $e_r$, and $f_r$ denote input \textcolor{red}{red} cells, i.e.,~a vertex, an edge, and a face, respectively, $v_b$, $e_b$, and $f_b$ denote input \textcolor{blue}{blue} cells, and $v$, $e$, and $f$ denote output
cells.
\begin{compactenum}
\item A new vertex $v$ is induced by coinciding vertices $v_r$ and $v_b$.
\item A new vertex $v$ is induced by a vertex $v_r$ that lies on an edge $e_b$.
\item An analogous case of a vertex $v_b$ that lies on an edge $e_r$.
\item A new vertex $v$ is induced by a vertex $v_r$ that is contained in a face $f_b$.
\item An analogous case of a vertex $v_b$ contained in a face $f_r$.
\item A new vertex $v$ is induced by the intersection of two edges $e_r$ and $e_b$.
\item A new edge $e$ is induced by the (possibly partial) overlap of two edges $e_r$ and $e_b$.
\item A new edge $e$ is induced by the an edge $e_r$ that is contained in a face $f_b$.
\item An analogous case of an edge $e_b$ contained in a face $f_r$.
\item A new face $f$ is induced by the overlap of two faces $f_r$ and $f_b$.
\end{compactenum}

Evidently a custom C++ overlay traits cannot be defined in Python; more precisely, the traits type must be known when the bindings are compiled. We introduce and expose with Python bindings two models of the \cppLstinline{OverlayTraits} concept. The most general between the two, called \cppLstinline{Arr_overlay_traits}, defines ten functions that correspond to the list above, each accepting three arguments, namely, two objects that represent input cells, and one object that represents an output cell. By default these functions do nothing. A user (that is, a Python programmer) can override any subset of these functions. One constructor of this model accepts all the ten functions at once. Another constructor accepts a single function, which corresponds to Function (10) in the list above.

\begin{figure}[!htp]
  \captionsetup[subfigure]{justification=centering}
  \centering%
  \subfloat[]{\label{fig:faceExt}
    \begin{tikzpicture}
      \fill[polyALightColor] (0,0) rectangle (2,2);
      \fill[polyBLightColor,opacity=0.5] (1,1) rectangle (3,3);
      \draw[thick,polyADarkColor] (0,0) rectangle (2,2);
      \draw[thick,polyBDarkColor] (1,1) rectangle (3,3);
      \node at (2.5,0.5) {[0]};
      \node at (0.5,0.5) {[1]};
      \node at (2.5,2.5) {[1]};
      \node at (1.5,1.5) {[2]};
      \node[point] at (0,0) {};
      \node[point] at (2,0) {};
      \node[point] at (0,2) {};
      \node[point] at (2,2) {};
      \node[point] at (1,1) {};
      \node[point] at (3,1) {};
      \node[point] at (1,3) {};
      \node[point] at (3,3) {};
      \node[point] at (2,1) {};
      \node[point] at (1,2) {};
    \end{tikzpicture}}\quad\quad
  \subfloat[]{\label{fig:vertexExt}
    \begin{tikzpicture}[scale=0.2]
      \draw[thick,polyADarkColor] (3,4) circle (5);
      \draw[thick,polyADarkColor] (-3,-4) circle (5);
      \draw[thick,polyBDarkColor] (4,-3) circle (5);
      \draw[thick,polyBDarkColor] (-4,3) circle (5);
      \node[point=red!50!white,label={[label distance=-1pt]180:4}] at (-8, -4) {}; % 4
      \node[point=green,label={[label distance=-1pt]180:6}] at (-7, -1) {};        % 6
      \node[point=red!50!white,label={[label distance=-1pt]180:2}] at (-9, 3) {};  % 2
      \node[point=red!50!white,label={[label distance=-1pt]180:4}] at (-2, 4) {};  % 4
      \node[point=green,label={[label distance=-1pt]90:6}] at (-1, 7) {};          % 6
      \node[point=red!50!white,label={[label distance=-1pt]180:2}] at (-1, -3) {}; % 2
      \node[point=blue,label={[label distance=-1pt]0:12}] at (0, 0) {};            % 12
      \node[point=green,label={[label distance=-1pt]-90:6}] at (1, -7) {};         % 6
      \node[point=red!50!white,label={[label distance=-1pt]0:2}] at (1, 3) {};     % 2
      \node[point=red!50!white,label={[label distance=-1pt]0:4}] at (2, -4) {};    % 4
      \node[point=green,label={[label distance=-1pt]0:6}] at (7, 1) {};            % 6
      \node[point=red!50!white,label={[label distance=-1pt]0:4}] at (8, 4) {};     % 4
      \node[point=red!50!white,label={[label distance=-1pt]0:2}] at (9, -3) {};    % 2
    \end{tikzpicture}}
  \caption[]{%
    \subref{fig:faceExt} The arrangement faces are extended with integers that indicate the number of overlapping bounded faces, shown in brackets.
    \subref{fig:vertexExt} The arrangement vertices are extended with integers that indicate the weighted degree of the vertices.}
  \label{fig:extend}
\end{figure}
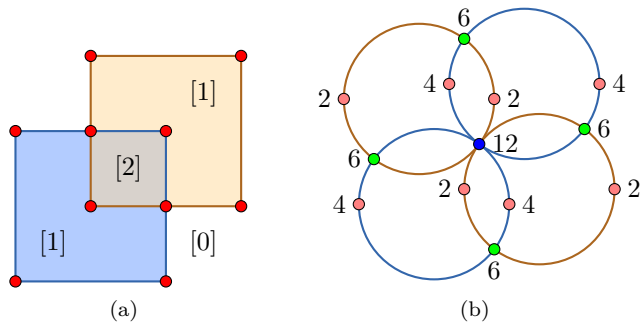

In the following we assume that the face type of the arrangement
% \cppLstinline{Arrangement_2} instance
is extended. The code excerpt in Listing~\ref{lst:twoFaceExtend} constructs two arrangements with two faces each. For each arrangement the properties of the unbounded face and the bounded face are initialized with Python integer objects~$0$ and~$1$, respectively.
The code excerpt in Listing~\ref{lst:overlayTraits} computes the overlay of the two arrangements constructed by the code in Listing~\ref{lst:twoFaceExtend} using the overlay traits. The property of each face of the resulting arrangement is updated as part of the overlay computation to indicate the number of overlapping bounded faces; see Figure~\ref{fig:faceExt}.
Observe that while the computation of the overlay is carried out by the compiled C++ code, the summation is carried out by a Python \pyLstinline{lambda} function, which accepts as the third argument the face to be updated. By default, arguments are copied into new Python objects. We pass the new face by reference, overriding the default, to enforce persistent updates.

\begin{lstfloat}
  \def\filename{overlay.py}%
  \lstinputlisting[style=pythonFramedStyle,basicstyle=\normalsize,
                   firstline=14,lastline=37,
                   label={lst:twoFaceExtend},
                   caption={Constructing two arrangements with their faces extended and initializing their data.}]
                  {\pythonDir\filename}
\end{lstfloat}

\begin{lstfloat}
  \begin{lstlisting}[style=pythonFramedStyle,basicstyle=\normalsize,
                     label={lst:overlayTraits},
                     caption={Computing the map overlay of two arrangements using the general overlay traits.}]
traits = Aos2.Arr_overlay_traits(lambda f1,f2,f: f.set_data(f1.data()+f2.data()))
Aos2.overlay(arr1, arr2, result, traits)
  \end{lstlisting}
\end{lstfloat}

In many cases, such as in the example above, the property of the new cell depends solely on the properties of the overlapping two cells. To this end we introduce (and expose) a second model of the \cppLstinline{OverlayTraits} concept called \cppLstinline{Arr_overlay_function_traits}. This model also defines ten functions that correspond to the list above. However, here each function accepts the two Python objects that extend the overlapping cells as input arguments and returns the Python object that extends the resulting cell. The code excerpt in Listing~\ref{lst:overlayFunctionTraits} has the same effect the code in Listing~\ref{lst:overlayTraits} has, but is more compact.

\begin{lstfloat}
  \begin{lstlisting}[style=pythonFramedStyle,basicstyle=\normalsize,
                     label={lst:overlayFunctionTraits},
                     caption={Computing the map overlay of two arrangements using the functional overlay traits.}]
traits = Aos2.Arr_overlay_function_traits(lambda x, y: x+y)
os2.overlay(arr1, arr2, result, traits)
  \end{lstlisting}
\end{lstfloat}

Each one of the ten functions supported by the \cppLstinline{Arr_overlay_function_traits} model returns a Python object that is passed back to the compiled C++ code. Then, the object is stored with the new cell. Thus, there is no need to pass arguments by reference (even though it could be more efficient in certain cases); however, the implementation of this model presents a new coding challenge---if a certain cell is not extended, attempting to access its non-existing extension would cause a compilation error.
Assume that only the vertex type of the \cppLstinline{Arrangement_2} instance is extended. The Python code excerpt in Listing~\ref{lst::overlayCircleSegment} computes the overlay of two arrangements and updates the property of each vertex of the resulting arrangement to indicate the weighted degree of the vertex, where blue incident edges weigh twice as much as red incident edges; see Figure~\ref{fig:vertexExt}. We set all the six functions that update output vertices in the \cppLstinline{Arr_overlay_function_traits} object using dedicated setters. The name of a setter matches the pattern \lstinline[language={C++},mathescape=true,columns=fixed]{set_$c_rc_b$_$c$()}, where each of $c_r$, $c_b$, and $c$ can be substituted with \cppLstinline{v}, \cppLstinline{e}, or \cppLstinline{f}. $c_r$ and $c_b$ determine the input red and blue cell types, respectively, and $c$ determines the output cell type.

\begin{lstfloat}
  \begin{lstlisting}[style=pythonFramedStyle,basicstyle=\normalsize,
                     label={lst::overlayCircleSegment},
                     caption={Constructing two arrangements with their vertices extended and computing their map overlay using the overlay function traits.}]
Aos2.insert(arr1,[Curve(Circle(Point(3,4),25)),Curve(Circle(Point(-3,-4),25))])
Aos2.insert(arr2, [Curve(Circle(Point(3,4),25)), Curve(Circle(Point(3,4),25))])
for arr in [arr1, arr2]:
  for v in arr.vertices():
    v.set_data(v.degree())
traits = Aos2.Arr_overlay_function_traits()
traits.set_vv_v(lambda x, y: 2*x+y)
traits.set_ve_v(lambda x, y: 2*x+2)
traits.set_vf_v(lambda x, y: 2*x)
traits.set_ev_v(lambda x, y: 4+y)
traits.set_fv_v(lambda x, y: y)
traits.set_ee_v(lambda x, y: 6)
Aos2.overlay(arr1, arr2, result, traits)
  \end{lstlisting}
\end{lstfloat}

Consider a generic function called \cppLstinline{apply()} that accepts two objects that represent input cells, namely \cppLstinline{r} and \cppLstinline{b}, one object that represents an output cell, namely \cppLstinline{o}, and a function called \cppLstinline{f}. The objective of \cppLstinline{apply()} is to apply \cppLstinline{f} to the properties of the cells \cppLstinline{r} and \cppLstinline{b} and store the return value as the property of the cell \cppLstinline{o}. Recall that a property of a cell \cppLstinline{c} is obtained and set via the calls \cppLstinline{c.data()} and \cppLstinline{c.set_data()}, respectively. If the type of the output cell is not extended, however, \cppLstinline{apply()} should become idle. Similarly, if the type of an input cell is not extended, the \pyLstinline{None} Python object (represented by \cppLstinline{py::object}) should be passed to \cppLstinline{f} instead. We use SFINAE yet again, twice, to address the above. First, we introduce an overload of \cppLstinline{apply()} that is idle and serves as a fall-through; see Line~9 in Listing~\ref{lst:data}. When a call to \cppLstinline{apply()} is made, the idle overload is selected by the overload resolution process if the output cell does not have a member called \cppLstinline{set_data()}. If the output cell does have this member the other implementation is selected and calls \cppLstinline{data(r)} and \cppLstinline{data(b)} to obtain the properties of the \textcolor{red}{red} and \textcolor{blue}{blue} input cells, respectively. Second, we introduce an overload of \cppLstinline{data()} that serves as a fall-through (Line~2). It does nothing but return the \pyLstinline{None} Python object. It is selected if the corresponding cell does not have a member called \cppLstinline{data()}. If the cell does have this member the other implementation is selected; it returns the property of the cell.

\begin{lstfloat}
  \begin{lstlisting}[style=cppFramedStyle,basicstyle=\normalsize,
                     label={lst:data},
                     caption={Automatic generation of bindings for the overlay function traits exploiting SFINAE.}]
// Fall-through; A::data() does not exist
template <typename A> py::object data(...) { return py::none(); }

// A::data() exists
template <typename A, typename = decltype(std::declval<A>().data())>
const py::object& data(const A* a) { return a->data(); }

// Fall-through; O::set_data() does not exist
template <typename R, typename B, typename O, typename F> void apply(...) {}

// O::set_data() does exist
template <typename R, typename B, typename O, typename F,
          typename =
            decltype(std::declval<O>().set_data(std::declval<typename O::Data>()))>
void apply(const R* r, const B* b, O* o, F f) {
  o->set_data(f(data<R>(r), data<B>(b)));
}
  \end{lstlisting}
\end{lstfloat}

% -----------------------------------------------------------------------------
\subsection{Miscellaneous}
\label{sec:code:misc}
% -----------------------------------------------------------------------------
In this section we list additional techniques used by the binding code.

% -----------------------------------------------------------------------------
\subsubsection{Handle, iterators, and Circulators}
\label{sec:code:misc:iters}
% -----------------------------------------------------------------------------
The \cppLstinline{Handle} concept, provided by the \HandlesandCirculatorsPackage{} package~\cite{cgal:dksy-hc-22}, describes types that are akin to pointers to objects. A handle provides the dereference operator (\cppLstinline{operator*()}) and the member access operator (\cppLstinline{->()}). Since Python does not have any kind of pointers, both a handle to an object in \cgal{} and the object itself are converted to the same Python object. For example, both types \cppLstinline{Arrangement_2::Vertex} and \cppLstinline{Arrangement_2::Vertex_handle} are converted to the Python type \pyLstinline{Vertex}, which is an attribute of the Python type \pyLstinline{Arrangement_2}.

The \cppLstinline{Iterator} concept and its refinements describe types that can be used to identify and traverse the elements of container that contains sequential data.\footnote{For more information about STL iterators, see, e.g., \url{https://en.cppreference.com/w/cpp/iterator}.} An iterator provides the increment or decrement operators. In \cgal{} quite often an iterator is convertible to a handle, for example, \cppLstinline{Arrangement_2::Vertex_iterator} is convertible to \cppLstinline{Arrangement_2::Vertex_handle}. Therefore, also \cppLstinline{Arrangement_2::Vertex_iterator} is converted to the python class \pyLstinline{Vertex}. Every converted iterator is supplied with both magic functions \pyLstinline{__iter__} and \pyLstinline{__next__}, thus made a \emph{Python iterator}.\footnote{For more information about Python iterators, see, e.g., \url{https://docs.python.org/3/c-api/iterator.html}.} The \pyLstinline{__iter__} function simply returns the input object, and the \pyLstinline{__next__} function returns the next object in the traversal order of the corresponding iterator.

Similar to the \cppLstinline{Iterator} concept and its refinements, the \cppLstinline{Circulator} concept and its refinements, also provided by the \HandlesandCirculatorsPackage{} package, describe types that can be used to identify and traverse the elements of container that contains circular data, for example, the halfedges incident to a face or the halfedges incident to a vertex in an arrangement.  A circulator object does not have a past-the-end value. Instead, the range $[a, b)$ of two circulators $a$ and $b$ denotes either the empty range, or the sequence of all elements in the container. nanobind supports several features that aid in the wrapping if iterators. Circulators are artificially converted to iterators to leverage on those features. In addition, while traversing a circulator, an stop-iteration exception is thrown when a circular refers to an item that has been traversed already.

% -----------------------------------------------------------------------------
\subsubsection{Functions Accepting Collections of Elements}
\label{sec:code:input}
% -----------------------------------------------------------------------------
The Standard Template Library (STL) provides various type-safe containers for storing collections of related objects. The \cppLstinline{Container} concept describes types of objects that store collections of other objects (their elements).\footnote{For more   information about STL containers, see, e.g.,   \url{https://en.cppreference.com/w/cpp/container}.} Container models are implemented as class templates, which enables great flexibility in the elements types and in the implemented algorithms that operate on container objects. Traditionally, implementations of algorithms in C++ manipulate iterators pointing into the containers they operate on. To date it is popular to pass a collection of elements to a function via two generic input iterators, where the first points at the first element of a container and the second points past the end of the container. An input iterator object, the type of which is a model of the the \cppLstinline{InputIterator} concept,\footnote{For more information about input iterators, see, e.g.,   \url{https://en.cppreference.com/w/cpp/iterator/input_iterator}.} supports reading from the location obtained via the dereference operator and can be pre- and post incremented.  The Boost.Range library provides an alternative modern method, which uses \emph{ranges}, for applying algorithms on collections of objects. In particular, it utilizes a new family of concepts that refine a basic concept called \cppLstinline{Range}. The \cppLstinline{Range} concept is similar to the \cppLstinline{Container} concept; it requires the provision of iterators for accessing a half-open range of elements and provides information about the number of elements in the range. The new method results in code that is more efficient and more expressive (and thus more comprehensible).\footnote{See   \url{https://www.boost.org/doc/libs/1_78_0/libs/range/doc/html/index.html} for the documentation of Boost.Range.}  Soon, (C++20) ranges will become part of the standard.\footnote{For more information about STL ranges, see, e.g., \url{https://en.cppreference.com/w/cpp/ranges}.} The \HandlesandCirculatorsPackage{} package of \cgal{} provides a variant of a \cppLstinline{Range} concept suitable for \cgal{}. Many functions in \cgal{} accept as input collections of elements, and some of them already use this concept. The use of ranges is expected to grow. For every function in \cgal{} that operates on one or more collections of elements, and regardless of the interface of the function, that is, whether a pair of generic input iterators, or a range, is used to represent each collection, we introduce and expose a wrapper function that accepts a Python list (of type \cppLstinline{py::list}) as an argument;\footnote{For more information on Python lists, see, e.g.,   \url{https://docs.python.org/3/tutorial/datastructures.html\#more-on-lists}.} the wrapper function calls the original function, properly passing the input collection; see Listing~\ref{lst:insert} for an example of such a function that uses iterators; the wrapper function is shown in Listing~\ref{lst:insertWrapper}; an excerpt code in Python that exploits the above is shown in Listing~\ref{lst:insertPython}. The function \cppLstinline{insert_curves()} (in Listing~\ref{lst:insertWrapper}) exploits an iterator adapter called \cppLstinline{stl_input_iterator}

\begin{lstfloat}
  \begin{lstlisting}[style=cppFramedStyle,basicstyle=\normalsize,%
                     label={lst:insert},%
                     caption={The signature of the free function that inserts a collections of curves into an arrangement.}]
template <typename InputIterator>
CGAL::insert(Arrangement_2& arr, InputIterator first, InputIterator past_the_end)
  \end{lstlisting}
\end{lstfloat}

\begin{lstfloat}
  \begin{lstlisting}[style=cppFramedStyle,basicstyle=\normalsize,%
                     label={lst:insertWrapper},%
                     caption={A function that wraps the function shown in Listing~\ref{lst:insert} and accepts a Python list that contains the input curves.}]
void insert_curves(Arrangement_2& arr, py::list& lst) {
  if (! lst) return;
  if (py::isinstance<X_monotone_curve_2>(lst[0]).check()) {
    auto begin = stl_input_iterator<X_monotone_curve_2>(lst);
    auto end = stl_input_iterator<X_monotone_curve_2>();
    CGAL::insert(arr, begin, end);
  }
  else if (py::isinstance<Curve_2>(lst[0]).check()) {
    auto begin = stl_input_iterator<Curve_2>(lst);
    auto end = stl_input_iterator<Curve_2>();
    CGAL::insert(arr, begin, end);
  }
}
  \end{lstlisting}
\end{lstfloat}

\begin{lstfloat}
  \begin{lstlisting}[style=pythonFramedStyle,basicstyle=\normalsize,%
                     label={lst:insertPython},%
                     caption={A code sample in Python that inserts three curves into an arrangement.}]
arr = Arrangement_2()
c1 = X_monotone_curve_2(Point_2(0, 0), Point_2(1, 0))
c2 = X_monotone_curve_2(Point_2(1, 0), Point_2(0, 1))
c3 = X_monotone_curve_2(Point_2(0, 1), Point_2(1, 1))
Aos2.insert(arr, [c1, c2, c3])
  \end{lstlisting}
\end{lstfloat}

% -----------------------------------------------------------------------------
\subsubsection{Functions Resulting in Collections of Elements}
\label{sec:code:output}
% -----------------------------------------------------------------------------
Functions may compute collections of elements as results. Using output iterators enables an efficient and flexible method for passing output collections from functions. An output iterator object, the type of which is a model of the the \cppLstinline{OutputIterator} concept,\footnote{For more information of output iterators, see, e.g.,   \url{https://en.cppreference.com/w/cpp/iterator/output_iterator}.} supports writing to the location obtained by the dereference operator and can be pre- and post-incremented. A function that computes a collection of elements accepts an output iterator as an argument and populates the underlying collection. Modern C++ compilers support the efficient transfer of resources from one object of a certain type to another object of the same type, referred to as \emph{move semantics}.\footnote{For more information of move semantics, see, e.g.,   \url{https://en.cppreference.com/w/cpp/language/move_constructor}.} If the underlying collection supports move semantics, a function that computes a collection of elements can simply return the collection, resulting in an elegant yet efficient code. For every function in \cgal{} that results in a collection of elements, and regardless of the interface of the function, that is, whether an output iterator is used or the collection is returned, we introduce and expose a wrapper function that returns a Python list (of type \cppLstinline{py::list}); the wrapper function calls the original function, properly populating a Python list with the output collection, and finally it returns the list.

The free function template \cppLstinline{CGAL::decompose()} accepts an arrangement $\mathcal{A}$ and computes a collection of polymorphic elements via an output iterator. For each vertex $v$ of $\mathcal{A}$ the output collection contains a pair of features---one that directly lies below $v$ and another that directly lies above $v$. Let $v$ be a vertex of $\mathcal{A}$. The feature above (respectively below) $v$
may be one of the following:
\begin{compactitem}
\item Another vertex $u$ having the same $x$-coordinate as $v$.
\item An arrangement edge associated with an $x$-monotone curve that contains $v$ in its $x$-range.
\item An unbounded face in case $v$ is incident to an unbounded face, and there is no curve lying above (respectively below) it.
\item An empty object, in case $v$ is the lower (respectively upper) endpoint of a vertical edge in the arrangement.
\end{compactitem}
Listing~\ref{lst:decompose} shows the signature of the function. Dereferencing the output iterator must yield an object the type of which is \cppLstinline{Decompose_result} shown in Line~2 of Listing~\ref{lst:decomposeWrapper}.

\begin{lstfloat}
  \begin{lstlisting}[style=cppFramedStyle,basicstyle=\normalsize,%
                     label={lst:decompose},%
                     caption={The signature of the free function that decomposes an arrangement into pseudo trapezoids.}]
template <typename OutputIterator>
CGAL::decompose(Arrangement_2& arr, OutputIterator oi);
  \end{lstlisting}
\end{lstfloat}

The wrapper of the \cppLstinline{CGAL::decompose()} template function, shown in Listing~\ref{lst:decomposeWrapper}, returns a list of Python elements; each element is a Python tuple of two items; the first wraps a C++ vertex and the second wraps another Python tuple of two items; the first is the cell above the vertex or \pyLstinline{py::Object} if none exists and the second is the cell below the vertex or \pyLstinline{py::Object} is none exists. Implementing such a wrapper can be done, for example, by introducing an intermediate standard container, say \cppLstinline{container}, and populating it with the elements, of the \cppLstinline{Decompose_result} polymorphic type, computed by the \cppLstinline{CGAL::decompose()} function in C++ (by passing an object of type \cppLstinline{std::back_insert_iterator<Decompose_result>} obtained by the call \cppLstinline{std::back_inserter(container)}, as the output iterator argument of the function), and then transforming the container elements into Python elements. Clearly, this is inefficient. Instead, we use the Boost c++ \pyLstinline{function_output_iterator<>} adapter to directly populate a Python list instead of an intermediate container. The adapter (i) utilizes a helper function called \cppLstinline{decompose_helper()} (not listed), which is applied to every resulting element transforming it to the corresponding bounded Python object, and (ii) appends the latter to the output Python list.

\begin{lstfloat}
  \begin{lstlisting}[style=cppFramedStyle,basicstyle=\normalsize,%
                     label={lst:decomposeWrapper},%
                     caption={A function that wraps the function shown in Listing~\ref{lst:decompose} and returns a Python list that represents a vertical decomposition.}]
py::list decompose(Arrangement_2& arr) {
  using Decompose_result = std::pair<Arrangement_2::Vertex_const_handle,
                                     std::pair<boost::optional<variant>,
                                               boost::optional<variant>>>;
  // The argument type of boost::function_output_iterator (UnaryFunction) must
  // be Assignable and Copy Constructible; hence the application of std::ref().
  auto it = boost::make_function_output_iterator(std::ref(op));
  CGAL::decompose(arr, it);
  return lst;
}
  \end{lstlisting}
\end{lstfloat}

\begin{figure}[!htp]
  \centerline{%
    \begin{tikzpicture}
      \draw node[point=white] (x) at (4,2) {};
      \draw node[point] (u1) at (0,0) {};
      \draw node[point=cyan] (u2) at (2,2) {};
      \draw node[point] (u3) at (3,3) {};
      \draw node[point=cyan] (u4) at (5,1) {};
      \draw node[point] (u5) at (6,0) {};
      \draw[blue,thick] (u1)--(u2)--(u3)--(x)--(u4)--(u5);
      \draw node[point] (v1) at (2,0) {};
      \draw node[point=cyan] (v2) at (3,1) {};
      \draw node[point] (v3) at (5,3) {};
      \draw node[point=cyan] (v4) at (6,2) {};
      \draw node[point] (v5) at (8,0) {};
      \draw[blue,thick] (v1)--(v2)--(x)--(v3)--(v4)--(v5);
      \draw[dots=10 per 1cm,very thick] (0,-1)--(u1)--(0,4);
      \draw[dots=10 per 1cm,very thick] (2,-1)--(v1)--(u2)--(2,4);
      \draw[dots=10 per 1cm,very thick] (3,-1)--(v2)--(u3)--(3,4);
      \draw[dots=10 per 1cm,very thick] (4,-1)--(x)--(4,4);
      \draw[dots=10 per 1cm,very thick] (5,-1)--(u4)--(v3)--(5,4);
      \draw[dots=10 per 1cm,very thick] (6,-1)--(u5)--(v4)--(6,4);
      \draw[dots=10 per 1cm,very thick] (8,-1)--(v5)--(8,4);
    \end{tikzpicture}}
  \caption{An arrangement of four line segments and its vertical decomposition into pseudo trapezoids, as constructed by the Python code shown in Listing~\ref{lst:py:decompose}.  The segments of the arrangement are drawn in solid blue lines and the segments of the vertical decomposition are drawn in dark dotted lines.}
  \label{fig:decompose}
\end{figure}
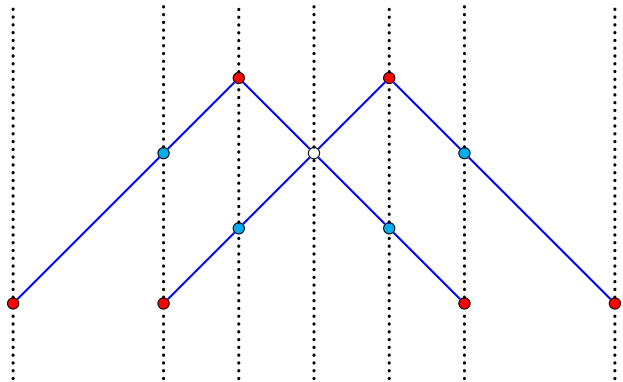

The Python code example shown in Listing~\ref{lst:py:decompose} exploits the above and computes the vertical decomposition shown in Figure~\ref{fig:decompose}.

%% The output:
%% 0 0
%%   unbounded face
%%   unbounded face
%% 2 0
%%   unbounded face
%%    0 0 3 3
%% 3 3
%%    2 0 4 2
%%   unbounded face
%% 4 2
%%   unbounded face
%%   unbounded face
%% 5 3
%%    4 2 6 0
%%   unbounded face
%% 6 0
%%   unbounded face
%%    5 3 8 0
%% 8 0
%%   unbounded face
%%   unbounded face

\begin{lstfloat}
  \def\filename{vertical_decomposition.py}%
  \lstinputlisting[style=pythonFramedStyle,basicstyle=\normalsize,%
                   label={lst:py:decompose},%
                   firstline=15,
                   caption={A code sample in Python that decomposes an
                     arrangement.}]
                  {\pythonDir\filename}
\end{lstfloat}

%% % -----------------------------------------------------------------------------
%% \subsection{\myLstinline{Python_distance}}
%% \label{sec:code:distance}
%% % -----------------------------------------------------------------------------
%% Similarly to overlay traits, a user may want to define a custom
%% distance class to use with search queries. Here too, as defining a new
%% C++ class is not possible from python, we provide a similar solution:
%% A \myLstinline{Python_distance} class is exposed, which is a model of
%% \myLstinline{GeneralDistance}
%% concept. \url{https://doc.cgal.org/5.2/Spatial_searching/classGeneralDistance.html}
%% It expects 5 functions in its constructor, and those are saved as
%% class members and are used for the respective 5 operations [the
%%   optional are not included] of the concept.

%% Making some things more pythonic:

%% (the extended arrangement data type) Those 10 functions are saved as
%% class members and are then called in the respective 10 [overloaded]
%% versions of \myLstinline{create_vertex}, \myLstinline{create_edge},
%% \myLstinline{create_face object}) (the matching is done by the order
%% of the passed functions and the order of the [overloaded] versions in
%% the official concept reference).

%% This way the user can still define the way the result face/edge/vertex
%% data will be calculated.

% =============================================================================
\section{Concepts-Binding Coupling}
\label{sec:concepts}
% =============================================================================
Generic programming enables the implementation of generic algorithms, which work on collections of different types, can be easily maintained, extended, and customized, and are type safe and easier to read. As mentioned in the introduction, \cgal{} rigorously adheres to the generic programming paradigm. As a consequence, most components of \cgal{} are either class or function templates. Many of the parameters of these templates are described in terms of models of concepts. When a class or a function template is instantiated, each one of its template parameters is substituted with a model of one or more concepts associated with the template parameter. Close to~750 concepts can be identified in \cgal{} at the time this article is written. Most hierarchy graphs of concepts are small. Few graphs, such as the graph of concepts of the geometry traits of the \iiDArrangementsPackage{} package, are quite large with intricate refinement relations. We use clusters of closely related concepts to describe the refinement relations among them; see, e.g., Figure~\ref{fig:atc}. We have introduced tight coupling between concepts and (i) binding generations and (ii) type annotation. We describe these relations in the following sections.

% -----------------------------------------------------------------------------
\subsection{Following Generic Concepts}
\label{sec:code:concepts}
% -----------------------------------------------------------------------------
For each concept we have introduced a template function that generates bindings for all the type and function members required by the concept. We call this function for every model of the concept. This systematic approach guaranties that all the documented functions and types, but nothing else, are exposed. Consider for example the cluster of geometry traits concepts depicted in Figure~\ref{fig:atc:central}. The template functions \cppLstinline{export_AosBasicTraits\_2()}, \cppLstinline{export_AosXMonotoneTraits\_2()}, and \cppLstinline{export_AosTraits\_2()} accept binding class objects\footnote{A binding class object of type   \cppLstinline{py:class_<T>} is used to expose the C++ type \cppLstinline{T} to Python.} for particular geometry traits models and populate it with all attributes that correspond to the requirements of the concepts \cppLstinline{AosBasicTraits\_2}, \cppLstinline{AosXMonotoneTraits\_2}, and \cppLstinline{AosTraits\_2}, respectively. If a concept \cppLstinline{B} refines a concept \cppLstinline{A}, then the function \cppLstinline{export_B()} calls the function \cppLstinline{export_A()}. The bindings of different traits models are generated in separate compilation units. As nodes in the concept refinement graphs may have more than a single parent, we ensure that a function that corresponds to a concept is not invoked more than once in a given compilation unit; see Lines~3--4 in the definition of the function \cppLstinline{export_AosBasicTraits\_2()}
in Listing~\ref{lst:cpp:exportBasic}.

\begin{lstfloat}
  \begin{lstlisting}[style=cppFramedStyle,basicstyle=\normalsize,%
                     label={lst:cpp:exportBasic},%
                     caption={Exposing attributes that correspond to the concept \cppLstinline{AosBasicTraits_2}.}]
template <typename GeometryTraits_2, typename ClassObject, typename Concepts>
void export_AosBasicTraits_2(ClassObject c, Concepts& concepts) {
  // Sentinel
  static bool exported = false;
  if (exported) return;

  // Expose attributes that correspond to the traits
  auto& classes = concepts.m_basic_traits_classes;
  classes.m_point_2 = py::class_<Point_2>(m, "Point_2")
    .def(py::init<>())
    .def(py::init<const Point_2&>());
  ~~$\cdots$~~
}
  \end{lstlisting}
\end{lstfloat}

Typically, a concept in our case requires the provision of nested types that are themselves models of other concepts. For example, the concept \cppLstinline{AosBasicTraits\_2} requires the provision of the nested type \cppLstinline{Point_2} used to represent a two-dimensional point. The type \cppLstinline{Point_2} must be \cppLstinline{DefaultConstructible} and \cppLstinline{CopyConstructible}. Let \cppLstinline{p} be an object of type \cppLstinline{Point_2} required by the \cppLstinline{AosBasicTraits\_2} concept. Most traits models also support the calls \cppLstinline{p.x()} and \cppLstinline{p.y()}; they return the $x$ and $y$ coordinates of the points $p$, respectively. The number type used to represent the coordinates depends on the traits instance; for example, the type \cppLstinline{Kernel::FT} is used for the \cppLstinline{Arr_segment_traits<Kernel>} traits instance. However, the \cppLstinline{Point_2} type nested in the traits model \cppLstinline{Arr_Bezier_curve_traits\_2} does not maintain an exact representation of the coordinates. Thus, it instead supports the call \cppLstinline{p.approximate()}, which returns a pair of floating point numbers that only approximate the real coordinates. The function \cppLstinline{export_AosBasicTraits\_2()} constructs a class object for the \cppLstinline{Point\_2} type, but adds nothing but Python attributes that expose the default and copy constructors of this type; see Lines~10--11 in the definition of the function \cppLstinline{export_AosBasicTraits\_2()} in the listing. It also saves the class object of the \cppLstinline{Point_2} type (see Line~8), so that later on the functions responsible for the bindings of the various traits models can add additional Python attributes. For example, the function that exposes the  \cppLstinline{Arr_segment_traits<Kernel>} instance adds the attributes \pyLstinline{x} and \pyLstinline{y}, which expose the member functions \cppLstinline{x()}, \cppLstinline{y()}, respectively, to the Python object that exposes the type \cppLstinline{Point_2}. Similarly, the function that exposes the \cppLstinline{Arr_Bezier_curve_traits\_2} instance adds the attribute \pyLstinline{approximate}, which exposes the member function \cppLstinline{approximate()} to the Python object that exposes the type \cppLstinline{Point_2}.

The concept \cppLstinline{AosHorizontalSideTraits\_2} requires the provision of the \cppLstinline{Compare\_x\_on\_boundary\_2} functor\footnote{A functor is a class that has one or more \cppLstinline{operator()} members; thus, it acts as a function.} that has two implementations of \cppLstinline{operator()} as follows; one overload accepts two ends of open or unbounded curves (refer to the manual for the precise encoding of a curve end) and another overload accepts a point and an end of a curve that is open or unbounded at that end. The former compares the $x$-coordinates of the curves at their respective limits or ends and the latter compares the $x$-coordinates of the point and the $x$-coordinates of the curve at its respective limit or end.
The concept \cppLstinline{AosIdentifiedHorizontalTraits\_2} indirectly refines the concept \cppLstinline{AosHorizontalSideTraits\_2} mentioned above. (The former would be used to define a traits model that handles arcs on a sphere with poles on the left and right sides; Figure~\ref{fig:atc:sphericalBoundary} depicts a cluster of concepts for models that handle arcs on a sphere with poles on the top and bottom sides.) In addition to the two implementations of the overloaded members \cppLstinline{operator()} of the functor \cppLstinline{Compare\_x\_on\_boundary\_2} it requires the provision of a third implementation that accepts two points and compares their $x$ coordinates. Similar to the case described in the previous paragraph, where the class object that handles the bindings for the \cppLstinline{Point_2} type must be available in two separate functions, also here the class object that handles the bindings for the functor \cppLstinline{Compare\_x\_on\_boundary\_2} must be available in both functions \cppLstinline{export_AosHorizontalSideTraits\_2()} and \cppLstinline{export_AosIdentifiedHorizontalTraits\_2()}. Saving the class object in one function so that it can be reused latter on in the other addresses this necessity.

% -----------------------------------------------------------------------------
\subsection{Type Annotation}
\label{ssec:concepts:annotation}
% -----------------------------------------------------------------------------
\cgal{} is a large library to start with. It is divided into approximately~150 packages with countless function and class templates. The vast number of function and type instances that can be defined using these templates can be overwhelming. The development of non critical code that is based on \cgal{} can be expedited using Python and \cgal{} Python bindings.  Using code completion (which is based on type annotations) offered by several Python IDEs can accelerate the development process much further. Type annotation in Python refers to annotations of Python functions, arguments, and variables in a way that can be used by various tools. It is an optional feature of Python that has been introduced in Version~3.5 and enhanced in successive versions. With this feature implemented a type checker, a tool separate from the Python interpreter, can be used to statically analyze code base to find bugs during early development stages and spelunk code of large projects. Naturally, it can be used by IDEs to interactively provide hints and suggestions to programmers and type check the code as it is being developed. Annotations can and should be placed near the code if possible. Observe that the Python interpreter ignores these annotations. Annotations can also be placed in external files (which have the extension \lstinline{.pyi}), referred to as stubs, for cases where the code base is inaccessible, such as the case with bindings. Stub files contain the signatures of the annotated functions with the body of the functions discarded. They also list the attributes of annotated types. In our case they contain the annotated interface of the \cgal{} bindings. Stub files have the same syntax as regular Python modules. There is one feature of the typing module that is different in stub files, namely, the \pyLstinline{@overload} decorator. This decorator allows describing functions that have different combinations of argument types. Consider, for example, the overloaded free C++ functions \cppLstinline{overlay()}; see Section~\ref{sec:code:overlay}. The annotation for the corresponding Python function \pyLstinline{overlay()} in the stub file is shown in
Listing~\ref{lst:py:overloading}.

\begin{lstfloat}
\begin{lstlisting}[style=pythonFramedStyle,basicstyle=\normalsize,%
                   label={lst:py:overloading},%
                   caption={Type annotation of the overloaded overlay function.}]
@overload
def overlay(r:Arrangement_2, b:Arrangement_2, o:Arrangement_2) -> None: pass
@overload
def overlay(r:Arrangement_2, b:Arrangement_2, o:Arrangement_2, t:Arr_overlay_traits) -> None: pass
@overload
def overlay(r:Arrangement_2, b:Arrangement_2, o:Arrangement_2, t:Arr_overlay_function_traits) -> None: pass
\end{lstlisting}
\end{lstfloat}

Annotating a Python type that wraps a C++ type, which is a model of some concepts is guided by the concepts. Similar to the C++ code that generates the bindings, we have introduced a tight coupling between concepts and Python type annotations; see
Section~\ref{sec:code:concepts}. As opposed to the C++ binding code, we have introduced a framework for the automatic generation of the annotation stubs. The framework includes a Python script called \bashLstinline{generate.py} that accepts as input static data files in json format that describe the concepts, their refinement relations, the exposed C++ classes, and their modeling relations with concepts. The script generates annotation stubs for the bindings of modules selected by the user. As mentioned in Section~\ref{sec:gen}, \cmake{} is used to generate the native build environment. We have augmented the scope of \cmake, and now \cmake{} is also used to generate the annotation stubs. Before we delve into the details of our annotation stub generation system, we describe in the following two paragraphs an unsuccessful attempt, which resembles the C++ binding code. It reached a dead-end because of limitations of Python.

Imagine that for each C++ concept we introduce a Python annotation class that annotates all the requirements of the concept. The annotated Python classes do not exist; these annotation classes are introduced to facilitate the creation of the annotation classes of real Python classes. Every annotation-class that annotates a certain model type, say \cppLstinline{T}, inherits from all the annotation-classes that correspond to the concepts that \cppLstinline{T} models. For example, the type annotation for the concept \cppLstinline{AosBasicTraits_2} contains the annotations the nested type \cppLstinline{Point_2}, which in turn, contains annotations for the default constructor and copy constructor of the type \cppLstinline{Point_2} defined by every model of the concept \cppLstinline{AosBasicTraits_2}; see Listing~\ref{lst:py:aosBasicTraitsErr} for partial annotations. Observe that the annotated Python class does not exist in reality, thus the \pyLstinline{_} prefix in the name of the annotation class \pyLstinline{_AosBasicTraits_2} in the listing. Assume that bindings are generated for the \cppLstinline{Arrangement_2<GeometryGtaits_2, Dcel>} type instantiated with a geometry traits type that is an instance of the \cppLstinline{Arr_segment_traits_2} class template.  This type models several concepts; see, e.g., Figure~\ref{fig:atc:central}. Partial
erroneous Python type annotation for the type \pyLstinline{Arr_segment_traits_2} is shown in Listing~\ref{lst:py:aosSegmentTraitsErr}.

\begin{lstfloat}
  \begin{draft}
  \begin{lstlisting}[style=pythonFramedStyle,basicstyle=\scriptsize,%
                     %backgroundcolor=\color{red!20},
                     label={lst:py:aosBasicTraitsErr},%
                     caption={Erroneous partial type annotation for the attributes that correspond to the concept \cppLstinline{AosBasicTraits_2}.}]
class _AosBasicTraits_2():
  class Point_2():
    def __init__(): ...
    def __init__(const Point_2&): ...
  class Equal_2():
    @overload
    def __call__(self, p: _AosBasicTraits_2.Point_2, q: _AosBasicTraits_2.Point_2) -> bool: ...
    @overload
    def __call__(self, p: _AosBasicTraits_2.X_monotone_curve_2, q: _AosBasicTraits_2.X_monotone_curve_2) -> bool: ...
  def equal_2_object(self) -> _AosBasicTraits_2.Equal_2: ...
  \end{lstlisting}
  \end{draft}
\end{lstfloat}

\begin{lstfloat}
  \begin{draft}
  \begin{lstlisting}[style=pythonFramedStyle,basicstyle=\scriptsize,%
                     %backgroundcolor=\color{red!20},
                     label={lst:py:aosSegmentTraitsErr},%
                     caption={Erroneous partial Python type annotation for the type \pyLstinline{Arr_segment_traits_2}.}]
class Arr_segment_traits_2(AosBasicTraits_2._AosBasicTraits_2,
                          AosXMonotoneTraits_2._AosXMonotoneTraits_2,
                          AosTraits_2._AosTraits_2,
                          AosLandmarkTraits_2._AosLandmarkTraits_2,
                          AosVerticalSideTraits_2._AosVerticalSideTraits_2,
                          AosVerticalSideTraits_2._AosHorizontalSideTraits_2) :
  class Point_2(AosBasicTraits_2._AosBasicTraits_2.Point_2):
    def x(self) -> Ker.FT: ...
    def y(self) -> Ker.FT: ...
  \end{lstlisting}
  \end{draft}
\end{lstfloat}

Unfortunately, the code presented in Listing~\ref{lst:py:aosBasicTraitsErr} and
Listing~\ref{lst:py:aosSegmentTraitsErr} does not work. Python and static type checkers for Python (e.g., mypy)\footnote{See, e.g.,   \url{https://mypy.readthedocs.io/en/stable/}.} have made significant progress since Python was conceived. However, even with the introduction of the \pyLstinline{Typing} module, and in particular the \pyLstinline{Generic} class within,\footnote{See   \url{https://docs.python.org/3/library/typing.html\#module-typing}.} there are still barriers that cannot be bridged. The barrier that we encountered is related to deep nesting. The culprit lies in Line~7 and Line~9 of Listing~\ref{lst:py:aosBasicTraitsErr}. In Line~7, for example, the class \pyLstinline{Point_2} nested in the utility class \pyLstinline{_AosBasicTraits_2} is refered to from a method of the class \pyLstinline{Equal_2}, which is nested in \pyLstinline{_AosBasicTraits_2}. An object of type \pyLstinline{Arr_segment_traits_2.Equal_2} was not recognized (probably because \pyLstinline{Arr_segment_traits_2.Equal_2} is not substituted by \pyLstinline{_AosBasicTraits_2.Equal_2} even though \pyLstinline{Arr_segment_traits_2} is derived from \pyLstinline{_AosBasicTraits_2}.) and our ambitious attempt failed flat.

Assume that two users select the \iiDArrangementsPackage{} module for binding generation. In particular, one user chooses to generate bindings for the \cppLstinline{Arrangement_2<GeometryGtaits_2, Dcel>} type instantiated, where the template parameter \cppLstinline{GeometryGtaits_2} is substituted with an instance of the \cppLstinline{Arr_segment_traits_2} class template and the other user chooses to generate bindings for the \cppLstinline{Arrangement_2<GeometryGtaits_2, Dcel>} type, where \cppLstinline{GeometryGtaits_2} is substituted with an instance of the \cppLstinline{Arr_Bezier_curve_traits_2} class template; see Listing~\ref{lst:py:aosSegmentTraits} and Listing~\ref{lst:py:aosBezierTraits} for the corresponding partial valid Python type annotations generated by our system. The input data files describe the concepts and the classes. More specifically, for each concept $C$, there exists a record that lists the requirements of the concept $C$ and the concepts the $C$ refines. For each type $T$, there exists a record that lists the concepts that $T$ models and the additional members and types nested in $T$ that are not covered by the concepts. Type annotations are, naturally, generated only for the types. An important feature of the system is the ability to merge methods of a class that appears both as a nested type of another class and as a type required by a concept. An example of such a scenario is the \cppLstinline{Point_2} type required by the concept
\cppLstinline{Aosbasictraits_2} and nested in every geometry traits class, e.g., \cppLstinline{Arr_segment_traits_2} and \cppLstinline{Arr_Bezier_curve_2}.

\begin{lstfloat}
  \begin{lstlisting}[style=pythonFramedStyle,basicstyle=\normalsize,%
                     %backgroundcolor=\color{red!20},
                     label={lst:py:aosSegmentTraits},%
                     caption={Valid Python type annotation for the type \pyLstinline{Arr_segment_traits_2}.}]
from typing import Iterator, overload
class Arr_segment_traits_2():
  class Point_2():
    @overload
    def __init__(self) -> None: ...
    @overload
    def __init__(self, p: Point_2) -> None: ...
    @overload
    def __init__(self, x: float, y: float) -> None: ...
    def x(self) -> FT: ...
    def y(self) -> FT: ...

  class Equal_2():
    def __call__(self, p: Point_2, Q: Point_2) -> Boolean: ...

  def equal_2_object(self) -> Equal_2: ...
  \end{lstlisting}
\end{lstfloat}

\begin{lstfloat}
  \begin{lstlisting}[style=pythonFramedStyle,basicstyle=\normalsize,%
                     %backgroundcolor=\color{red!20},
                     label={lst:py:aosBezierTraits},%
                     caption={Valid Python type annotation for the type \pyLstinline{Arr_segment_traits_2}.}]
from typing import Iterator, overload
class Arr_Bezier_curve_traits_2():
  class Point_2():
    @overload
    def __init__(self) -> None: ...
    @overload
    def __init__(self, p: Point_2) -> None: ...
    def approximate(self) -> list[float]: ...

  class Equal_2():
    def __call__(self, p: Point_2, Q: Point_2) -> Boolean: ...

  def equal_2_object(self) -> Equal_2: ...
  \end{lstlisting}
\end{lstfloat}

While using a code-generation approach adds a level of complexity to our binding system, it achieves the goal, it is flexible, and it posses an additional advantage. Once the input data files for specific concepts and models are in place, they can be used to automatically generate the documentation for those models and concept; we certainly plan to pursue this goal in the future. The use of the type-annotation subsystem, and in particular the \bashLstinline{generate.py} script or the script that generates the documentation that we intend to develop are not limited to \cgal{}---they can be used by any binding generation system that generates Python bindings for generic C++ code.

% =============================================================================
\section{Applications}
\label{sec:app}
% =============================================================================
The \cgal{} library, and in particular the \cgal{} arrangement
packages have been used intensively at the Computational Geometry Lab
at Tel Aviv University for over twenty years now, not only for
research, where the researchers typically (though not always---see
Section~\ref{ssec:neural}) have good knowledge of generic programming
in C++, but also for teaching. We have been using \cgal{} for projects
in the Computational Geometry course, for assignments and final
projects in the courses ``Algorithmic Robotics and Motion Planning''
and ''Algorithms for 3D printing and other Manufacturing Processes'',
and for the implementation of a variety of robot algorithms in guided
software workshops. Using \cgal{} for teaching has persistently been a
pain-point in these courses especially due to the intricate
utilization of C++ in the implementation of the library. This has
recently dramatically changed with the introduction of the Python
bindings presented in this paper, as we describe in two examples below.

% -----------------------------------------------------------------------------
\subsection{A Multi Robot Motion Planning Platform}
\label{ssec:discopygal}
We developed a program called \disco{} written in Python that provides visualization and verification for multi-robot motion planning in 2D. The program uses Python bindings of \cgal{} arrangements, \cgal{}
Minkowski sums, and other components of \cgal{} needed for the implementation of fundamental motion-planning algorithms and enables the dynamic loading of motion-planning algorithms in the form of Python scripts during run time. It allows users to easily swap and compare between different motion-planning algorithms, and even modify an existing implementation of an algorithm while the program is running. (There is no need to recompile the code or restart the program.) 
% The precompiled library for the \cgal{} bindings used by the program can be supplied with it, such that even % the installation of \cgal{}, boost, or a C++ compiler is not required in order to use the program. 
We were invited to give a crash course in multi-robot motion planning in a summer school geared primarily toward engineering doctoral students,\footnote{\url{http://disc.tudelft.nl/education/summer-school/disc-summer-school-2021/}}
where the organizers particularly asked for hands-on algorithm implementation and experimentation session. After two hours of lecture including fifteen minutes of presentation of \disco, students installed it and were able to interact with it. Testimonies of students and organizers affirmed that the system was friendly and
easy-to-use, and fruitful in assisting the students to absorb the algorithmic ideas. The students only needed to access and write Python code, while still employing heavy-duty \cgal{} procedures in the background.

In last year's Algorithmic Robotics course, the experimental assignments are based on \disco{}. This was the first time that the introduction of the software into the assignment passed completely smoothly, without compromising the required level of algorithmic sophistication. The system allowed the students to focus on algorithmic issues using Python, a programming language they are all familiar with, freeing them from many of the past points of hardship raised by working directly with \cgal{}, and in particular with the advanced generic-programming mechanisms that \cgal{} uses.

\begin{figure}[!htp]
  \centering%
  \includegraphics[width=\linewidth]{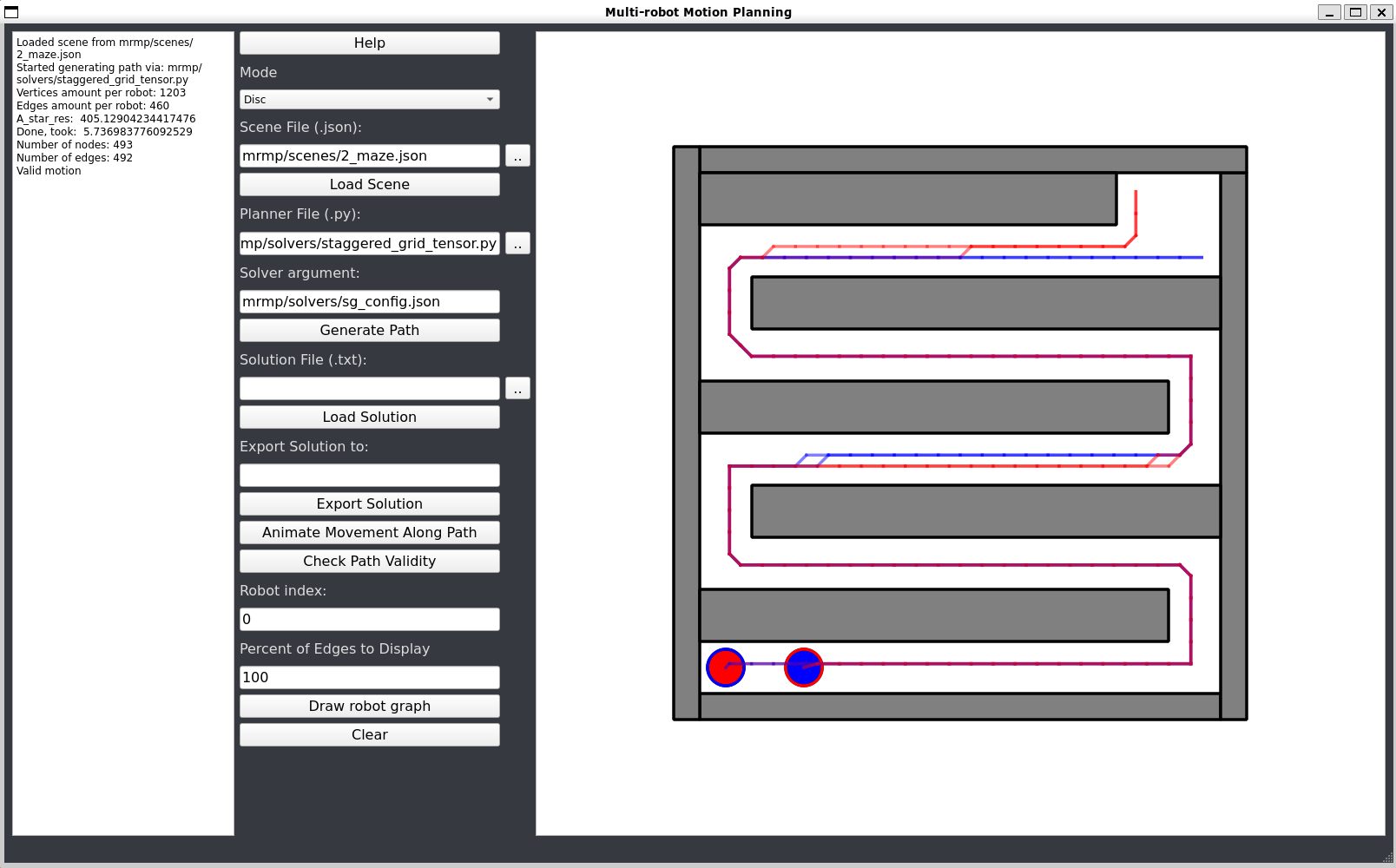}
  \caption{A snapshot of the \disco{} application window.}
  \label{fig:disco}
\end{figure}

A snapshot of the \disco{} application window is shown in Figure~\ref{fig:disco}. The window is divided into three parts; informative messages are displayed on the console on the left, a rich menu resides to the right of the console; the workspace is drawn on the right. The particular workspace in the figure consists of two disc robots. The objective is to swap their positions. The path of each robot planned by the system is drawn in the color of the robot. For more information visit the page \url{http://acg.cs.tau.ac.il/discopygal}.

% -----------------------------------------------------------------------------
\subsection{Neural Collision Detection}
\label{ssec:neural}
% -----------------------------------------------------------------------------
In collaboration with neuroscientists, we helped developing a simulation system for the interaction between neurons and blood vessels in the brain~\cite{Har-Gil2021.07.20.452894}. The need arose to use three-dimensional alpha shapes. The \cgal{} offerings in this respect looked a perfect fit, with the caveat that the main tools of the simulation are developed in Python and the most convenient programming language for that part of the project is Python. We developed Python bindings for the \cgal{} alpha-shape package, and their usage was immediately and almost effortlessly integrated into the overall system by the neuroscientists.

% \bibliographystyle{abbrv}
% \bibliography{abrev,cgal,cgal_python_bindings}

\begin{thebibliography}{10}

\bibitem{bhsbp-ag-03}
D.~Abrahams and R.~W. Grosse-Kunstleve.
\newblock Building hybrid systems with boost.python.
\newblock {\em C/C++ Users Journal}, pages 31--39, 2003.

\bibitem{a-gps-99}
M.~H. Austern.
\newblock {\em Generic Programming and the {\slshape\scshape Stl}}.
\newblock Addison-Wesley, Boston, MA, USA, 1999.

\bibitem{bfhhm-epsph-15}
A.~Baram, E.~Fogel, D.~Halperin, M.~Hemmer, and S.~Morr.
\newblock Exact {M}inkowski sums of polygons with holes.
\newblock {\em Computational Geometry: Theory and Applications}, 1:01--01,
  2015.

\bibitem{bbcds-cbbw-11}
S.~Behnel, R.~Bradshaw, C.~Citro, L.~Dalc{\'{\i}}n, D.~S. Seljebotn, and
  K.~Smith.
\newblock Cython: The best of both worlds.
\newblock {\em Computational Sciences Engineering}, 13(2):31--39, 2011.

\bibitem{bfhks-apsca-10}
E.~Berberich, E.~Fogel, D.~Halperin, M.~Kerber, and O.~Setter.
\newblock Arrangements on parametric surfaces {II}: Concretizations and
  applications.
\newblock {\em Mathematics in Computer Science}, 4:67--91, 2010.

\bibitem{bfhmw-apsgf-10}
E.~Berberich, E.~Fogel, D.~Halperin, K.~Mehlhorn, and R.~Wein.
\newblock Arrangements on parametric surfaces {I}: General framework and
  infrastructure.
\newblock {\em Mathematics in Computer Science}, 4:45--66, 2010.

\bibitem{by-ag-98}
J.-D. Boissonnat and M.~Yvinec.
\newblock Triangulations.
\newblock In {\em Algorithmic Geometry}. Cambridge University Press, 1998.
\newblock Translated by Herv{\'e} Br{\"o}nnimann.

\bibitem{cgal:bfghhkps-lgk23-22}
H.~Br{\"o}nnimann, A.~Fabri, G.-J. Giezeman, S.~Hert, M.~Hoffmann, L.~Kettner,
  S.~Pion, and S.~Schirra.
\newblock {2D} and {3D} linear geometry kernel.
\newblock In {\em \fontfamily{cmr} {\slshape\scshape Cgal} User and Reference
  Manual}. \textsc{Cgal} Editorial Board, {5.5} edition, 2022.

\bibitem{cgal:ct-pt3-22}
M.~Caroli, A.~Pell{\'e}, M.~Rouxel-Labb{\'e}, and M.~Teillaud.
\newblock {3D} periodic triangulations.
\newblock In {\em \fontfamily{cmr} {\slshape\scshape Cgal} User and Reference
  Manual}. \textsc{Cgal} Editorial Board, {5.5} edition, 2022.

\bibitem{ct-cpt-09}
M.~Caroli and M.~Teillaud.
\newblock Computing 3d periodic triangulations.
\newblock In {\em Proceedings of the 17{th} Annual European Symposium on
  Algorithms (ESA)}, volume 5757 of {\em LNCS}, pages 37--48. Springer-Verlag,
  2009.

\bibitem{cgal:d-as2-22}
T.~K.~F. Da.
\newblock {2D} alpha shapes.
\newblock In {\em \fontfamily{cmr} {\slshape\scshape Cgal} User and Reference
  Manual}. \textsc{Cgal} Editorial Board, {5.5} edition, 2022.

\bibitem{cgal:dy-as3-22}
T.~K.~F. Da, S.~Loriot, and M.~Yvinec.
\newblock {3D} alpha shapes.
\newblock In {\em \fontfamily{cmr} {\slshape\scshape Cgal} User and Reference
  Manual}. \textsc{Cgal} Editorial Board, {5.5} edition, 2022.

\bibitem{cgal:dksy-hc-22}
O.~Devillers, L.~Kettner, S.~Pion, M.~Seel, and M.~Yvinec.
\newblock Handles and circulators.
\newblock In {\em \fontfamily{cmr} {\slshape\scshape Cgal} User and Reference
  Manual}. \textsc{Cgal} Editorial Board, {5.5} edition, 2022.

\bibitem{DBLP:journals/jea/FlatoHHNE00}
E.~Flato, D.~Halperin, I.~Hanniel, O.~Nechushtan, and E.~Ezra.
\newblock The design and implementation of planar maps in {CGAL}.
\newblock {\em {ACM} J. Exp. Algorithmics}, 5:13, 2000.

\bibitem{fhw-caass-12}
E.~Fogel, D.~Halperin, and R.~Wein.
\newblock {\em {\slshape\scshape Cgal} Arrangements and Their Applications, A
  Step by Step Guide}.
\newblock Springer-Verlag, Berlin Heidelberg, Germany, 2012.

\bibitem{cgal:fwzh-rbso2-22}
E.~Fogel, O.~Setter, R.~Wein, G.~Zucker, B.~Zukerman, and D.~Halperin.
\newblock {2D} regularized boolean set-operations.
\newblock In {\em \fontfamily{cmr} {\slshape\scshape Cgal} User and Reference
  Manual}. \textsc{Cgal} Editorial Board, {5.5} edition, 2022.

\bibitem{geos-manual}
{GEOS contributors}.
\newblock {\em {GEOS} coordinate transformation software library}.
\newblock Open Source Geospatial Foundation, 2021.

\bibitem{hlw-gfapm-00}
D.~Halperin, J.-C. Latombe, and R.~H. Wilson.
\newblock A general framework for assembly planning: {T}he motion space
  approach.
\newblock {\em Algorithmica}, 26:577--601, 2000.

\bibitem{my-hs-atarr-95}
D.~Halperin and M.~Sharir.
\newblock Arrangements and their applications in robotics: Recent developments.
\newblock In K.~Goldbergs, D.~Halperin, J.-C. Latombe, and R.~Wilson, editors,
  {\em International Workshop on Algorithmic Foundations of Robotics}, pages
  495--511. A.K. Peters, Ltd., Boston, MA, 1995.

\bibitem{hs-a-18}
D.~Halperin and M.~Sharir.
\newblock Arrangements.
\newblock In J.~E. Goodman, J.~O'Rourke, and C.~T\'oth, editors, {\em Handbook
  of Computational Geometry}, chapter~28. Chapman \& Hall/CRC, 3rd edition,
  2017.

\bibitem{Har-Gil2021.07.20.452894}
H.~Har-Gil, Y.~Jacobson, A.~Pr{\"o}nneke, J.~F. Staiger, O.~Tomer, D.~Halperin,
  and P.~Blinder.
\newblock Neural collision detection: an open source library to study the
  three-dimensional interactions of neurons and other tree-like structures.
\newblock {\em bioRxiv}, 2021.

\bibitem{nanobind}
W.~Jakob.
\newblock nanobind -- seamless operability between c++17 and python, 2022.
\newblock https://github.com/wjakob/nanobind.

\bibitem{cgal:pt-t3-22}
C.~Jamin, S.~Pion, and M.~Teillaud.
\newblock {3D} triangulations.
\newblock In {\em \fontfamily{cmr} {\slshape\scshape Cgal} User and Reference
  Manual}. \textsc{Cgal} Editorial Board, {5.5} edition, 2022.

\bibitem{cgal:k-pt2-13-22}
N.~Kruithof.
\newblock {2D} periodic triangulations.
\newblock In {\em \fontfamily{cmr} {\slshape\scshape Cgal} User and Reference
  Manual}. \textsc{Cgal} Editorial Board, {5.5} edition, 2022.

\bibitem{ld-hppcn-16}
W.~Lavrijsen and A.~Dutta.
\newblock High-performance python-c++ bindings with pypy and cling.
\newblock In {\em Proceedings of the 6{th} Workshop on Python}, pages 27--35.
  IEEE Computer Society Press, 2016.

\bibitem{DBLP:journals/jcb/MartinYBZD11}
J.~W. Martin, A.~K. Yan, C.~Bailey{-}Kellogg, P.~Zhou, and B.~R. Donald.
\newblock A geometric arrangement algorithm for structure determination of
  symmetric protein homo-oligomers from noes and rdcs.
\newblock {\em Journal of Computational Biology}, 18(11):1507--1523, 2011.

\bibitem{mh-mc-08}
K.~Martin and B.~Hoffman.
\newblock {\em Mastering \textsc{CMake}}.
\newblock Kitware, Inc, 4th edition, 2008.

\bibitem{m-rgece-06}
M.~Meyerovitch.
\newblock Robust, generic and efficient construction of envelopes of surfaces
  in three-dimensional space.
\newblock In {\em Proceedings of the 14{th} Annual European Symposium on
  Algorithms (ESA)}, volume 4168 of {\em LNCS}, pages 792--803.
  Springer-Verlag, 2006.

\bibitem{cgal:s-gkd-22}
M.~Seel.
\newblock {dD} geometry kernel.
\newblock In {\em \fontfamily{cmr} {\slshape\scshape Cgal} User and Reference
  Manual}. \textsc{Cgal} Editorial Board, {5.5} edition, 2022.

\bibitem{cgal:tf-ssd-22}
H.~Tangelder and A.~Fabri.
\newblock {dD} spatial searching.
\newblock In {\em \fontfamily{cmr} {\slshape\scshape Cgal} User and Reference
  Manual}. \textsc{Cgal} Editorial Board, {5.5} edition, 2022.

\bibitem{cgal:eb-22}
{The \textsc{Cgal} Project}.
\newblock {\em \fontfamily{cmr} {\slshape\scshape Cgal} User and Reference
  Manual}.
\newblock \textsc{Cgal} Editorial Board, {5.5} edition, 2022.

\bibitem{DBLP:conf/gis/KreveldSW04}
M.~J. van Kreveld, {\'{E}}.~Schramm, and A.~Wolff.
\newblock Algorithms for the placement of diagrams on maps.
\newblock In {\em Proceedings of the 12{th} ACM International Workshop
  Geographic Information Systems}, pages 222--231, Washington, DC, USA, 2004.
  PUB-ACM.

\bibitem{cgal:w-rms2-22}
R.~Wein, A.~Baram, E.~Flato, E.~Fogel, M.~Hemmer, and S.~Morr.
\newblock {2D} minkowski sums.
\newblock In {\em \fontfamily{cmr} {\slshape\scshape Cgal} User and Reference
  Manual}. \textsc{Cgal} Editorial Board, {5.5} edition, 2022.

\bibitem{cgal:wfzh-a2-22}
R.~Wein, E.~Berberich, E.~Fogel, D.~Halperin, M.~Hemmer, O.~Salzman, and
  B.~Zukerman.
\newblock {2D} arrangements.
\newblock In {\em \fontfamily{cmr} {\slshape\scshape Cgal} User and Reference
  Manual}. \textsc{Cgal} Editorial Board, {5.5} edition, 2022.

\bibitem{DBLP:journals/comgeo/WeinFZH07}
R.~Wein, E.~Fogel, B.~Zukerman, and D.~Halperin.
\newblock Advanced programming techniques applied to \textsc{Cgal}'s
  arrangement package.
\newblock {\em Computational Geometry: Theory and Applications},
  38(1--2):37--63, 2007.

\bibitem{cgal:y-t2-22}
M.~Yvinec.
\newblock {2D} triangulations.
\newblock In {\em \fontfamily{cmr} {\slshape\scshape Cgal} User and Reference
  Manual}. \textsc{Cgal} Editorial Board, {5.5} edition, 2022.

\end{thebibliography}
\appendix
% =============================================================================
\section{Extenders}
\label{sec:extenders}
% =============================================================================
In several places in our binding code we are given a type that models certain concepts and based on several conditions, this type must be replaced with a new type that models additional concepts. Instead of using old style code in the form of \cppLstinline{#if,#elif,#else,#endif} directives, we exploit generic programming and implement these extenders using modern code that is type safe. In the following we present four such scenarios.

% -----------------------------------------------------------------------------
\subsection{2D Arrangement Traits Extender}
\label{sec:extenders:arrTraits}
% -----------------------------------------------------------------------------
When the user enables the generation of bindings for the \iiDArrangementsPackage{} package, the binding for an instance of the class template \cppLstinline{Arrangement_2<Traits, Dcel>} is generated. When this class template is instantiated the \cppLstinline{Traits} template parameter must be substituted with a specific traits model. By default an instance of the \cppLstinline{Arr_segment_traits_2<Kernel>} is used; however, the user can override this selection and choose any one of the supported geometry traits. We refer to this traits as the basic geometry traits. If the user also enables the generation of bindings for the
\iiDRegularizedBooleanSetOperationsPackage{} package, the basic traits must be extended, because it must also model the \cppLstinline{GeneralPolygonSetTraits_2} concept (see Figure~\ref{fig:atc:gps}); see Section~\ref{ssec:modules:aos2}. The extension however differs for different basic traits. It is automatically done via a dedicated class template and four specializations shown in Listing~\ref{lst:arrTraitsExtender}. The final traits is obtained using this extender as shown in Listing~\ref{lst:arrFinalTraits}. Here,
\cppLstinline{boolean_set_operations_2_bindings()} is a constant binary function (evaluated at compile time) that determines whether the generation of bindings for the
\iiDRegularizedBooleanSetOperationsPackage{} package is enabled, and
\cppLstinline{BGT} is the basic geometry traits.

\begin{lstfloat}
  \begin{lstlisting}[style=cppFramedStyle,basicstyle=\normalsize,%
                     label={lst:arrTraitsExtender},%
                     caption={2D Arrangement Traits Extender.}]
// Traits extender
template <bool b, typename Base> struct Tr {};

// Specialization for the case where bindings for Bso2 are not generated
template <typename Base>
struct Tr<false, Base> { typedef Base type; };

// Specialization for the general case where bindings for Bso2 are generated
template <typename Base>
struct Tr<true, Base> { typedef CGAL::Gps_traits_2<Base> type; };

// Specialization for the segment traits
template <>
struct Tr<true, CGAL::Arr_segment_traits_2<Kernel>> {
  typedef CGAL::Gps_segment_traits_2<Kernel, Point_2_container> type;
};

// Specialization for the circle segment traits
template <>
struct Tr<true, CGAL::Arr_circle_segment_traits_2<Kernel>> {
  typedef CGAL::Gps_circle_segment_traits_2<Kernel> type;
};
  \end{lstlisting}
\end{lstfloat}

\begin{lstfloat}
  \begin{lstlisting}[style=cppFramedStyle,basicstyle=\normalsize,%
                     label={lst:arrFinalTraits},%
                     caption={The final 2D Arrangement traits type definition.}]
typedef Tr<boolean_set_operations_2_bindings(), BGT>::type      Geometry_traits;
  \end{lstlisting}
\end{lstfloat}

% -----------------------------------------------------------------------------
\subsection{2D Arrangement DCEL Cell Extenders}
\label{sec:extenders:arrDcel}
% -----------------------------------------------------------------------------
When the user enables the generation of bindings for the \iiDArrangementsPackage{} package, the binding for an instance of the class template \cppLstinline{Arrangement_2<Traits, Dcel>} is generated. When this class template is instantiated the \cppLstinline{Dcel} template-parameter should be substituted with a type that models the \cppLstinline{ArrangementDcel} concept. We substitute this parameter with an instance of the template \cppLstinline{CGAL::Arr_dcel_base<V, H, F>}, where \cppLstinline{V}, \cppLstinline{H}, and \cppLstinline{F} must model the concepts \cppLstinline{ArrangementDcelVertex}, \cppLstinline{ArrangementDcelHalfedge}, and
\cppLstinline{ArrangementDcelFace}, respectively. By default, we substitute these parameters with the instances \cppLstinline{Arr_vertex_base<Traits::Point_2>},
\cppLstinline{Arr_halfedge_base<Traits::X_monotone_curve_2>}, and \cppLstinline{Arr_face_base}, respectively, referred to as the basic types of cells. If the user chooses to extend the vertex type, then we substitute \cppLstinline{V} with the instance \cppLstinline{Arr_extended_ vertex<Vb, py::object>}, where \cppLstinline{Vb}
is the basic vertex type above. This conditional extension is carried out via a dedicated class template and two specializations shown in Listing~\ref{lst:arrDcelVertexExtender}. The final vertex type is obtained using this extender as shown in Listing~\ref{lst:arrFinalVertex}. Here, \cppLstinline{is_vertex_extended()} is a constant binary function that determines whether the user extension of vertices is enabled.

\begin{lstfloat}
  \begin{lstlisting}[style=cppFramedStyle,basicstyle=\normalsize,%
                     label={lst:arrDcelVertexExtender},%
                     caption={2D Arrangement DCEL user extended vertex type extender.}]
// Vertex extender
template <bool b, typename Vb> struct Vertex_extender {};

// Specialization
template <typename Vb> struct Vertex_extender<false, Vb> { typedef Vb type; };

// Specialization
template <typename Vb> struct Vertex_extender<true>
{ typedef CGAL::Arr_extended_vertex<Vb, py::object> type; };
  \end{lstlisting}
\end{lstfloat}

\begin{lstfloat}
  \begin{lstlisting}[style=cppFramedStyle,basicstyle=\normalsize,%
                     label={lst:arrFinalVertex},%
                     caption={The final 2D Arrangement vertex type definition.}]
CGAL::Arr_vertex_base<Geometry_traits::Point_2>                         Vb;
typedef Vertex_extended<is_vertex_extended(), Vb, py::object>::type     V;
  \end{lstlisting}
\end{lstfloat}

If the user also enables the generation of bindings for the \iiDRegularizedBooleanSetOperationsPackage{} package, the template parameters \cppLstinline{H} and \cppLstinline{F} must be substituted with types that also model the
\cppLstinline{GeneralPolygonSetDcelHalfedge} and \cppLstinline{GeneralPolygonSetDcelFace} concepts, respectively; see Section~\ref{ssec:modules:bso2}. In addition, similar to the case of vertex extension, if the user chooses to extend the halfedge or face types, then we substitute \cppLstinline{V} with the instances \cppLstinline{Arr_extended_halfedge<Hb, py::object>} or \cppLstinline{Arr_extended_ face<Fb, py::object>}, respectively, where
\cppLstinline{Hb} and \cppLstinline{Fb} are the types above. The conditional extensions of the halfedge type are carried out via dedicated class templates and corresponding specializations shown in Listing~\ref{lst:arrDcelBasicHalfedgeExtender} and
Listing~\ref{lst:arrDcelHalfedgeExtender}. The final halfedge type is obtained using these extenders as shown in Listing~\ref{lst:arrFinalHalfedge}. Here, \cppLstinline{boolean_set_operations_2_bindings()} is a constant binary function that determines whether the generation of bindings for the \iiDRegularizedBooleanSetOperationsPackage{} package is enabled, and \cppLstinline{is_halfedge_extended()} is a constant binary function that determines whether the user extension of halfedges is enabled.

\begin{lstfloat}
  \begin{lstlisting}[style=cppFramedStyle,basicstyle=\normalsize,%
                     label={lst:arrDcelBasicHalfedgeExtender},%
                     caption={2D Arrangement DCEL Halfedge extender for 2D Regularized Boolean set operations.}]
// Basic halfedge extender
template <bool b> struct Halfedge_gps {};

// Specialization
template <> struct Halfedge_gps<false>
{ typedef CGAL::Arr_halfedge_base<Geometry_traits::X_monotone_curve_2> type; };

// Specialization
template <> struct Halfedge_gps<true>
{ typedef CGAL::Gps_halfedge_base<Geometry_traits::X_monotone_curve_2> type; };
  \end{lstlisting}
\end{lstfloat}

\begin{lstfloat}
  \begin{lstlisting}[style=cppFramedStyle,basicstyle=\normalsize,%
                     label={lst:arrDcelHalfedgeExtender},%
                     caption={2D Arrangement DCEL user extended halfedge type extender.}]
// Halfedge extender
template <bool b, typename Hb> struct Halfedge_extender {};

// Specialization
template <typename Hb> struct Halfedge_extender<false, Hb> { typedef Hb type; };

// Specialization
template <typename Hb> struct Halfedge_extender<true, Hb>
{ typedef CGAL::Arr_extended_halfedge<Hb, py::object> type; };
  \end{lstlisting}
\end{lstfloat}

\begin{lstfloat}
  \begin{lstlisting}[style=cppFramedStyle,basicstyle=\normalsize,%
                     label={lst:arrFinalHalfedge},%
                     caption={The final 2D Arrangement DCEL halfedge type definition.}]
typedef Halfedge_gps<boolean_set_operations_2_bindings()>::type         Hb;
typedef Halfedge_extended<is_halfedge_extended(), Hb, py::object>::type H;
  \end{lstlisting}
\end{lstfloat}

Similarly, the conditional extensions of the face type are carried out via dedicated class templates and corresponding specializations shown in Listing~\ref{lst:arrDcelBasicFaceExtender} and Listing~\ref{lst:arrDcelFaceExtender}. The final face type is obtained using these extenders as shown in Listing~\ref{lst:arrFinalFace}.
\begin{lstfloat}
  \begin{lstlisting}[style=cppFramedStyle,basicstyle=\normalsize,%
                     label={lst:arrDcelBasicFaceExtender},%
                     caption={2D Arrangement DCEL Face extender for 2D Regularized Boolean set operations.}]
// Basic face extender
template <bool b> struct Face_gps {};

// Specialization
template <> struct Face_gps<false> { typedef CGAL::Arr_face_base type; };

// Specialization
template <> struct Face_gps<true> { typedef CGAL::Gps_face_base type; };
  \end{lstlisting}
\end{lstfloat}

\begin{lstfloat}
  \begin{lstlisting}[style=cppFramedStyle,basicstyle=\normalsize,%
                     label={lst:arrDcelFaceExtender},%
                     caption={2D Arrangement DCEL user extended face type extender.}]
// Face extender
template <bool b, typename Fb> struct Face_extender {};

// Specialization
template <typename Fb> struct Face_extender<false, Fb> { typedef Fb type; };

// Specialization
template <typename Fb> struct Face_extender<true, Fb>
{ typedef CGAL::Arr_extended_face<Fb, py::object> type; };
  \end{lstlisting}
\end{lstfloat}

\begin{lstfloat}
  \begin{lstlisting}[style=cppFramedStyle,basicstyle=\normalsize,%
                     label={lst:arrFinalFace},%
                     caption={The final 2D Arrangement DCEL face type definition.}]
typedef Face_gps<boolean_set_operations_2_bindings()>::type             Fb;
typedef Face_extended<is_face_extended(), Fb, py::object>::type         F;
  \end{lstlisting}
\end{lstfloat}

% -----------------------------------------------------------------------------
\subsection{2D Triangulation Cell Extenders}
\label{sec:extenders:iitri}
% -----------------------------------------------------------------------------
When the user enables the generation of bindings for the \iiDTriangulationsPackage{} package, the binding for an instance of one of the 2D triangulation class templates is generated. Each of these class templates has a template parameter that must be substituted with a model of the concept \cppLstinline{TriangulationDataStructure_2} when the triangulation class template is instantiated. We substitute this parameter with an instance of the class template \cppLstinline{Triangulation_data_structure_2<V, F>}.

The type that substitutes the \cppLstinline{V} template parameter when the class template \cppLstinline{Triangulation_data_structure_ 2<V, F>} is instantiated must model the \cppLstinline{TriangulationDSVertexBase_2} concepts. If the binding is generated for a periodic triangulation, this parameter must be substituted with a type that also models the concept \cppLstinline{Periodic_2TriangulationDSVertexBase_2}. The extension is
automatically done via a dedicated class template and three specializations shown in Listing~\ref{lst:iitriBaseVertexExtender}. Here \cppLstinline{Traits} is the geometry traits type; see Section~\ref{ssec:modules:tri2} for an explanation on how this type is
determined.

\begin{lstfloat}
  \begin{lstlisting}[style=cppFramedStyle,basicstyle=\normalsize,%
                     label={lst:iitriBaseVertexExtender},%
                     caption={The 2D Triangulation base vertex type extender.}]
// Vertex base selection
template <bool b, int i> struct Vertex_base_name {};

// Specialization
template <int i> struct Vertex_base_name<false, i>
{ typedef CGAL::Triangulation_vertex_base_2<Traits> type; };

// Specialization
template <> struct Vertex_base_name<false, CGALPY_TRI2_REGULAR>
{ typedef CGAL::Regular_triangulation_vertex_base_2<Traits> type; };

// Specialization
template <int i> struct Vertex_base_name<true, i>
{ typedef CGAL::Periodic_2_triangulation_vertex_base_2<Traits> type; };
  \end{lstlisting}
\end{lstfloat}

If the user chooses to extend the vertex type, the type that substitutes the \cppLstinline{V} template parameter is further extended. This is automatically done via yet another dedicated class template and two specializations shown in Listing~\ref{lst:iitriExtendedVertexExtender}.

\begin{lstfloat}
  \begin{lstlisting}[style=cppFramedStyle,basicstyle=\normalsize,%
                     label={lst:iitriExtendedVertexExtender},%
                     caption={The 2D Triangulation user extended vertex type extender.}]
// Vertex with info
template <bool b, typename Vb> struct Vertex_with_info {};

// Specialization
template <typename Vb> struct Vertex_with_info<false, Vb> { typedef Vb type; };

// Specialization
template <typename Vb> struct Vertex_with_info<true, Vb>
{ typedef CGAL::Triangulation_vertex_base_with_info_2<py::object, Traits, Vb> type; };
  \end{lstlisting}
\end{lstfloat}

If the triangulation is used to define an alpha shape type, the template
parameter \cppLstinline{V} must be substituted with a type that also models
the concept \cppLstinline{AlphaShapeVertex_2}; see
Section~\ref{ssec:modules:as2}. This is automatically done via yet another
class template shown in Listing~\ref{lst:iitriASVertexExtender}.

\begin{lstfloat}
  \begin{lstlisting}[style=cppFramedStyle,basicstyle=\normalsize,%
                     label={lst:iitriASVertexExtender},%
                     caption={The 2D Triangulation for alpha shapes vertex type extender.}]
// Vertex alpha shape
template <bool b, typename Vb, typename ExactComparison>
struct Vertex_alpha_shape {};

// Specialization
template <typename Vb, typename ExactComparison>
struct Vertex_alpha_shape<false, Vb, ExactComparison> { typedef Vb type; };

// Specialization
template <typename Vb, typename ExactComparison>
struct Vertex_alpha_shape<true, Vb, ExactComparison>
{ typedef CGAL::Alpha_shape_vertex_base_2<Traits, Vb, ExactComparison> type; };
  \end{lstlisting}
\end{lstfloat}

Finally, if the binding is generated for a hierarchy triangulation, e.g., an instance of the template parameter \cppLstinline{Triangulation_hierarchy_2<Triangulation_2>}, the
\cppLstinline{V} template parameter must be substituted with a type that also models the concept \cppLstinline{TriangulationHierarchyVertexBase_2}. (Observe that there is no special requirements on the type that substitutes the \cppLstinline{F} parameter in this case.) This is automatically done via yet another class template shown in Listing~\ref{lst:iitriHierarchyVertexExtender}.

\begin{lstfloat}
  \begin{lstlisting}[style=cppFramedStyle,basicstyle=\normalsize,%
                     label={lst:iitriHierarchyVertexExtender},%
                     caption={The 2D Triangulation hierarchy vertex type extender.}]
// Vertex triangulation hierarchy
template <bool b, typename Vb> struct Vertex_hierarchy {};

// Specialization
template <typename Vb> struct Vertex_hierarchy<false, Vb> { typedef Vb type; };

// Specialization
template <typename Vb> struct Vertex_hierarchy<true, Vb>
{ typedef CGAL::Triangulation_hierarchy_vertex_base_2<Vb> type; };
  \end{lstlisting}
\end{lstfloat}

% Basic
%%%%%%%
The type \cppLstinline{Vb} is defined as the vertex base type.  If bindings are generated for a periodic triangulation, the type \cppLstinline{Vb} is defined as \cppLstinline{Periodic_2_triangulation_vertex_base_2<Traits>}; otherwise, if bindings are generated for an instance of \cppLstinline{Regular_triangulation_2<Traits, Tds>}, the type \cppLstinline{Vb} is defined as \cppLstinline{Regular_triangulation_ vertex_base_2<Traits>}; otherwise, the type \cppLstinline{Vb} is defined as
\cppLstinline{Triangulation_vertex_ base_2<Traits>};
% Extended
%%%%%%%%%%
If the user enables vertex-type extension, the type \cppLstinline{Vbi} is defined as
\cppLstinline{Triangulation_vertex_base_with_ info_2<py::object, Kernel, Vb>}; otherwise the type \cppLstinline{Vbi} is defined as \cppLstinline{Vb}.
% Alpha Shape
%%%%%%%%%%%%%
If the triangulation is used to define an alpha shape type, the type \cppLstinline{Vbia} is defined as \cppLstinline{Alpha_shape_vertex_base_2<Traits, Vb, ExactComparison>},
where \cppLstinline{ExactComparison} is substituted with either \cppLstinline{true} or \cppLstinline{false} based on the setting of the \cmake{} flag \cmakeLstinline{CGALPY_AS2_EXACT_COMPARISON}; see Table~\ref{tab:as2}. Otherwise, \cppLstinline{V} is defined as \cppLstinline{Vbi}.
% Hierarchy
%%%%%%%%%%%
If the binding is generated for a hierarchy triangulation, the final vertex type \cppLstinline{V} is defined as \cppLstinline{Triangulation_hierarchy_vertex_ base_2<Vbi>}; otherwise the final vertex type \cppLstinline{V} is defined as \cppLstinline{Vbia}; see Listing~\ref{lst:iitriFinalVertex}. Here, \cppLstinline{is_periodic()}, \cppLstinline{vertex_with_ info()}, \cppLstinline{alpha_shape_2_bingings()}, and
\cppLstinline{hierarchy()} are constant binary functions. \cppLstinline{is_periodic()} that determines whether the triangulation, binding for which is generated, is periodic.
\cppLstinline{vertex_with_info()} determines whether the user extension of vertices is enabled. \cppLstinline{alpha_shape_2_bingings()} indicates whether bindings for the \iiDAlphaShapesPackage{} package. \cppLstinline{hierarchy()} indicates whether the binding is generated for a hierarchy triangulation.

\begin{lstfloat}
  \begin{lstlisting}[style=cppFramedStyle,basicstyle=\normalsize,%
                     label={lst:iitriFinalVertex},%
                     caption={The final 2D Triangulation vertex type definition.}]
typedef Vertex_base_name<is_periodic(), CGALPY_TRI2>::type           Vb;
typedef Vertex_with_info<vertex_with_info(), Vb>::type               Vbi;
typedef Vertex_alpha_shape<alpha_shape_2_bindings(), Vbi, Ec>::type  Vbia;
typedef Vertex_hierarchy<hierarchy(), Vbia>::type                    V;
  \end{lstlisting}
\end{lstfloat}

The final face type is determined in a similar fashion. The only difference between the selections of the vertex and face types is that the setting of the \cmake{} flag \cmakeLstinline{CGALPY_TRI2_HIERARCHY} does not have an effect on the face type selection.

% -----------------------------------------------------------------------------
\subsection{3D Triangulation Cell Extenders}
\label{sec:extenders:iiitri}
% -----------------------------------------------------------------------------
When the user enables the generation of bindings for the \iiiDTriangulationsPackage{} package, the binding for an instance of one of the 3D triangulation class templates is generated; see Section~\ref{ssec:modules:tri3}. The scenario here is similar to the scenario of the \iiDTriangulationsPackage{} package described in Section~\ref{sec:extenders:iitri}. We skip the description of the extenders and jump to the description of their application.

If bindings are generated for a periodic triangulation, the type \cppLstinline{Vbp} is defined as \cppLstinline{Periodic_3_triangulation_ ds_vertex_base_3<>}; otherwise, \cppLstinline{Vbp} is defined as \cppLstinline{Triangulation_ds_vertex_base_3<>}.
If bindings are generated for an instance of either \cppLstinline{Regular_triangulation_vertex_base_3<Kernel>} or \cppLstinline{Periodic_3_ Regular_triangulation_vertex_base_3<Kernel>}, the type \cppLstinline{Vb} is defined as \cppLstinline{Regular_triangulation_ vertex_base_3<Traits, Vbp>}; otherwise, the type \cppLstinline{Vb} is defined as \cppLstinline{Triangulation_vertex_base_3<Traits, Vbp>}.
If the vertex type is extended, the type \cppLstinline{Vbi} is defined as
\cppLstinline{Triangulation_vertex_base_ with_info_3<py::object, Traits, Vb>}; otherwise, the type \cppLstinline{Vbi} is defined as \cppLstinline{Vb}.
If the binding is generated for a hierarchy triangulation, the type \cppLstinline{Vbih} is defined as \cppLstinline{Triangulation_hierarchy_vertex_base_3<Vbi>}; otherwise the type \cppLstinline{Vbih} is defined as \cppLstinline{Vbi}.
If the triangulation is used to define an alpha shape type, the final type \cppLstinline{V} (that substitutes the \cppLstinline{V} parameter) is defined as either
\cppLstinline{Alpha_shape_vertex_base_3<Traits, Vbih, ExactComparison>}
or
\cppLstinline{Fixed_Alpha_shape_vertex_base_3<Traits, Vbih, ExactComparison>} based on the selection of the alpha shape class (either fixed or plain); see Table~\ref{tab:as3}; \cppLstinline{ExactComparison} is substituted with either \cppLstinline{true} or \cppLstinline{false} based on the user selection of exact comparisons; see Table~\ref{tab:as3}. Otherwise, \cppLstinline{V} is defined as \cppLstinline{Vbih}; see Listing~\ref{lst:iiitriFinalVertex}. (Observe that while 3D periodic regular triangulations are supported in the form of the class template \cppLstinline{Periodic_3_regular_triangulation_3}, corresponding 2D triangulations are not; thus the difference between the structures of the code in Listing~\ref{lst:iitriFinalVertex} and Listing~\ref{lst:iiitriFinalVertex}.

\begin{lstfloat}
  \begin{lstlisting}[style=cppFramedStyle,basicstyle=\normalsize,%
                     label={lst:iiitriFinalVertex},%
                     caption={The final 3D Triangulation vertex type definition.}]
typedef Vertex_periodic<is_periodic()>::type                         Vb;
typedef Vertex_regular<is_regular(), Vb>::type                       Vbr;
typedef Vertex_with_info<vertex_with_info(), Vbr>::type              Vbri;
typedef Vertex_alpha_shape<alpha_shape_3_bindings(), Vbri, Ec>::type Vbria;
typedef Vertex_hierarchy<hierarchy(), Vbria>::type                   V;
  \end{lstlisting}
\end{lstfloat}

The final cell type is determined in a similar fashion. The only difference between the selections of the vertex and cell types is that the setting of the \cmake{} flag \cmakeLstinline{CGALPY_TRI3_HIERARCHY} does not have an effect on the cell type selection.

%%%%%%%%%%%%%%%%%%%%%%%%%%%%%%%%%%%%%%%%%%%%%%%%%%%%%%%%%%%%%%%%%%%%%%%%%%%%%%
%  appendices
%%%%%%%%%%%%%%%%%%%%%%%%%%%%%%%%%%%%%%%%%%%%%%%%%%%%%%%%%%%%%%%%%%%%%%%%%%%%%%
\end{document}
\documentclass{article}